\documentclass[12pt]{article}  
\usepackage{jheppub}\usepackage{titlesec}
\usepackage{amsmath}\usepackage{amsfonts}\usepackage{amssymb}\usepackage{mathtools}
\usepackage{mathrsfs}\usepackage{bbm} 
\usepackage{xcolor}\usepackage{color}
\usepackage{hyperref}  
 \usepackage{bm}

\title{\Large On underestimation of the inelastic interactions \\ in the direct dark matter search}
\author[]{V.A.~Bednyakov}\emailAdd{bedny@jinr.ru}
\affiliation[]{Dzhelepov Laboratory of Nuclear Problems, JINR, 141980, Dubna, Russia}
\abstract{In the paper explicit expressions are obtained for the event rates expected in experiments aimed at direct detection of dark matter (DM) particles. These expressions allow one to estimate the rates taking into account  simultaneously elastic (coherent) and inelastic (incoherent) channels of DM particle interaction with target nuclei.
The nonzero nuclear excitation energies are used in the calculation of the inelastic scattering contributions.
A strong correlation between the excitation energy and the recoil energy of the excited nucleus limits the possibility of the inelastic channel detection with a number of nuclei. Together with the standard model of the DM distribution in the Galaxy some other models are considered, which allow higher velocities of the DM particle.
As the recoil energy of the nucleus, $T_A$, increases, the dominance of the elastic interaction channel is smoothly replaced by the  dominance of the inelastic one. Therefore, if a DM detector is set up to detect only elastic scattering events, it starts to lose capability of “seeing” anything. 
 The only way to "notice"\/ the interaction remains the $\gamma$ radiation from the deexcitation of the target nucleus. In the case of spin-independent DM interaction, as $T_A$ increases, the inelastic contribution quickly dominates, while  the observable differential event rate decreases quite insignificantly.
If the DM particle interacts with (zero-spin) nuclei only spin-dependently,  the detectors focused on registration of the elastic spin-dependent DM signal will see nothing, since the entire signal "goes  through"\/ the inelastic channel. It looks like the desired DM interaction could have a noticeable intensity,  but the DM detector is unable to detect it.
Therefore, a setup aimed at the direct DM detection should register two signals.
The first is the nuclear recoil energy and the second is the $\gamma$-quanta with a certain 
energy from the target nucleus deexitation.
The experiment will provide the complete information about the DM interaction.}
\begin{document}
\maketitle
\section{Direct detection of galactic dark matter particles}\label{1DM-IntroDuction} 
Over almost a century, the presence of dark matter (DM) in the outer space around us not only has been 
well confirmed by a variety of astrophysical observations
\cite{Bringmann:2012ez,Sofue:2015xpa,Feldmann:2013hqa,Madhavacheril:2014slf,Feng:2010gw,Famaey:2015bba,Safdi:2023wsk},
but also has become one of the most intriguing problem of fundamental physics
\cite{Cebrian:2022brv,Bernabei:2020mon,Bernabei:2014maa,ArkaniHamed:2008qn,Hoeneisen:2015rva,Livio:2014gda}.
According to modern science, this extra, non-luminous and non-baryonic matter, manifesting itself 
only gravitationally so far, consists of particles yet unknown in the Standard Model (SM)
\cite{Bernabei:2022loo,Slatyer:2021qgc,Cooley:2021rws,Bertone:2004pz,Drukier:1986tm,Gelmini:2015zpa}.
For a long time one of the best candidates for the DM particle  was the so-called Weakly Interacting Massive Particle (WIMP), since, rather weakly interacting and  having a mass in the range from 1 GeV$/c^2$ to 1 TeV$/c^2$, it naturally satisfied the cosmological requirements of the early universe.
The search for WIMPs was carried out  in three main directions.
The first is the direct detection, which is aimed at registration of the DM particle interaction
with ordinary matter (on the Earth).
The second is the indirect detection through detecting products  of the DM particle annihilation inside the outer space objects.
The third is the search for appearance of the DM candidates at modern accelerators with sufficiently high energy and luminosity
\cite{Bednyakov:2015uoa}.
\par
In the light of overwhelming recognition of the DM existence,  the results of all above-mentioned  DM searches look very inconsistent.
One the one hand, one believes that WIMPs were not yet detected in these experiments
\cite{Cebrian:2022brv,Livio:2014gda}.
Indeed, during several decades intensive searches for the WIMPs were carried out.
Many restrictions were obtained on the cross section of their interaction with nucleons,
which were brought rather close to the absolute  minimum limit for GeV$/c^2$ DM masses,  
 the so-called neutrino flour \cite{Boehm:2018sux,Papoulias:2018uzy,Dai:2022lsj}.
 \par 
On the other hand, no one has yet been able to refute the results of the DAMA/LIBRA collaboration. 
This  is {\em the only} collaboration that has measured the annual modulation signal 
corresponding to the presence of DM particles in our Galaxy at the 13$\sigma$ confidence level
\cite{Bernabei:2022loo,Bernabei:2020mon,Bernabei:2014maa,Zurowski:2020dxe}.
The rotation of the Earth (with a DM detector) around the Sun,
when both of them move through the galactic DM environment,
causes very specific annual modulation of the intensity of the DM interaction with nuclei
\cite{Aboubrahim:2022lwb,Kahlhoefer:2018knc,Bednyakov:2012cu,Freese:1987wu}.
When the movement of the Earth is opposite to the direction of the Sun, 
the relative speed of the DM particle and the detector (and the intensity of their interaction) 
reaches a minimal value.  
When the Earth moves in the same direction as the Sun, the relative speed and the intensity of the DM-nucleus interaction are maximal.
This phenomenon is almost independent of the nature of the interaction.
 If the interaction is possible, it is mainly determined  by the distribution of the DM-particle velocities in the Galaxy 
 (more precisely, near the Earth).
Although the modulation component of the signal is noticeably smaller than the 
time-averaged total signal of the interaction, its value can hardly be overestimated.
Nowadays it is the only available positive signature that is caused by the galactic nature of the WIMPs.
Only this signature and only experiments {\em aimed at the direct DM detection}
allow one to prove the presence of  DM  in the space around us
\cite{Bednyakov:2012cu}.
\par
In connection with the DAMA/LIBRA results, the following should be emphasized.
The modulation observed has the phase, amplitude and period very well corresponding to our Galaxy DM.
There is no other collaboration which is able to observe this modulation at the same confidence level.
The other above-mentioned searches for DM particles are only able of discover a potential {\em DM candidate}.
To prove its belonging to the galactic DM population, one has to observe the modulation signal induced right by {\em this DM candidate}.
\par
Therefore, despite significant technical difficulties and serious systematic uncertainties
\footnote{One needs very accurate low-background and low-threshold detectors, and strong 
protection against a variety of background processes. One has very small interaction probability, and the
low event rate. The statistics must be collected over years. 
There are a lot of uncertainties in the DM distribution both in the Galaxy and near the Earth. 
The interaction of DM particles with one type of target nuclei does not guarantee 
any interaction with other nuclei, for example, due to different nucleon and spin composition, etc.},
direct DM experiments are inevitably important, because one is unable 
to understand the nature of  DM without them
\cite{Spooner:2007zh,Bednyakov:2015uoa,Bednyakov:2020njj,Slatyer:2021qgc,Cebrian:2022brv,Aboubrahim:2022lwb,Lawrence:2022niq,Cushman:2013zza,Saab:2012th}.
  \par
As a response to the long-term lack of positive results from the existing experiments
searching for the WIMP dark matter (excluding the DAMA/LIBRA), many new, sometimes very exotic, alternative models and suggestions appeared in the literature. 
They are concerning with compositions of  DM
\cite{Nakayama:2019rhg,Hurtado:2020vlj,Du:2020ldo,Baryakhtar:2020rwy,Majumdar:2021vdw,Afek:2021vjy,Lebedev:2023uzp,Mohapatra:2023aei,Olle:2019kbo,Belanger:2022gqc,Dvali:2023xfz,Kitajima:2023fun,Friedlander:2023jzw,Mistele:2023wao,Chiueh:2022krl,Abbas:2023ion,Ralegankar:2023pyx,Bai:2023mfi,Marfatia:2022jiz,Biswas:2023azl,Chu:2023zbo,Arguuelles:2023ktd,DeLuca:2023laa,Hong:2022gzo,Roy:2023zar,Biggio:2023gtm,Carney:2022gse}, 
 DM properties
\cite{Giudice:2017zke,Zurowski:2020dxe,Wang:2021jic,Feng:2021hyz,Emken:2021vmf,Granelli:2022ysi,Filimonova:2022pkj,Bell:2022yxn,Freese:2023fcr,Baryakhtar:2022hbu,DeRocco:2022rze,Emken:2021lgc,Proukakis:2023nmm,Biswas:2023eju,Garcia:2023awt,Das:2023vsq,Maldonado:2023alg,Lee:2023hrg,Arellano-Celiz:2019pax,Wang:2023xgm},
as well as very new and sophisticated methods of the DM detection
\cite{Tsuchida:2019hhc,Coskuner:2021qxo,Hamaide:2021hlp,Boos:2022gtt,Flambaum:2022zuq,Fan:2022uwu,Blanco:2022cel,Billard:2022cqd,Araujo:2022wjh,Li:2022acp,Badurina:2022ngn,Endorsers:2022aie,Chigusa:2023hms,Ruzi:2023cvp,DarkSide-50:2022hin,DarkSide:2022dhx,Albakry:2023aby,Saez:2023sjt,Kryemadhi:2022vuk,Su:2022wpj,Das:2022srn,PALEOCCENE:2023rjj,ALPHA:2022rxj,Delos:2023exh,Catena:2023qkj,Catena:2023awl,Ruzi:2023mxp,Ray:2023auh,DuttaBanik:2023yxj,Dhakal:2022rwn,Kim:2020bwm,Leane:2020wob,DarkSide-50:2023fcw}.
\par
In fact, the main interest of the DM-search community was shifted towards "light"\/ DM particles,
whose masses are comparable with the masses of electrons and/or nucleons
\cite{Bardhan:2022ywd,Krnjaic:2022ozp}.
The sensitivity to light DM particles of detectors aimed at direct searches for WIMPs 
strongly decreases with decreasing DM mass, because they can measure only the nuclear recoil energy.
For example, a DM particle with a mass of about 1 GeV/$c^2$ and a kinetic energy in the 
local Galaxy halo is not able to cause a nuclear recoil above the threshold of 1 keV. 
Therefore one has the rapid decrease in sensitivity
\cite{Granelli:2022ysi}.
This problem with the light DM particles, first, required new ideas and technologies for  more sensitive detectors, 
at least with significantly lower energy thresholds
\cite{Billard:2022cqd,Proceedings:2022hmu,Du:2020ldo,EDELWEISS:2022ktt,Edelweiss:2022bzh,EDELWEISS:2019vjv}.
Second, since one believes that the light DM particles can effectively produce the recoil electrons
with energy above their detection threshold
(about 0.186 keV for xenon \cite{XENON:2019gfn}),
it was proposed to register the light DM particles via their interactions with the atomic electrons
\cite{XENON:2019gfn,PandaX-II:2021nsg,Granelli:2022ysi}.
Third, this problem has stimulated searches for possible ways to increase 
the kinetic energy of the light DM particles.
\par
One variant of the light DM energy increase is based on the accelerators with high intensities,  
at which a sufficiently abundant production of the light DM particles could be expected, 
for example, directly in beams of leptons or protons,  in the decays of mesons and baryons, 
by bremsstrahlung, etc. \cite{Krnjaic:2022ozp,Boos:2022gtt,Kim:2017qdi}.
The other way is based on "natural"\/ mechanisms of almost relativistic DM particle 
production in the {\em modern}\/ universe.
As a rule, one considers the streams of DM particles accelerated in such a way  
as less intense, but more energetic. 
It became clear that the light DM particles can be accelerated to (almost) relativistic 
velocities due to their elastic interaction with the cosmic rays in the halo of the Milky Way
\cite{PhysRevLett.122.171801,Lei:2020mii,Xia:2020apm,Cappiello:2019qsw,Ema:2018bih,Dent:2019krz,PandaX-II:2021kai,Alvey:2022pad}.
A new DM particle acceleration mechanism, called the blazar-boosted DM was proposed,  
where the DM particles can be significantly accelerated  and can have sufficiently large local densities
due to their interaction with high-energy proton jets of a blazar 
\cite{Wang:2021jic,Granelli:2022ysi,Gondolo:1999ef}.
There are other mechanisms of the DM particle acceleration, see for example,  
\cite{Agashe:2014yua,Kouvaris:2015nsa,An:2017ojc,Emken:2017hnp,Granelli:2022ysi,Xia:2022tid,CDEX:2022fig,Bhowmick:2022zkj,Bardhan:2022bdg,Maity:2022exk,Wang:2019jtk,Elor:2021swj,Hu:2016xas,TEXONO:2018nir}.
This allows one  to overcome the sensitivity limitations of the 
low-energy threshold detectors \cite{Berger:2022cab}.
\par
Furthermore, a new mechanism was proposed to explain how the light nonrelativistic DM particles
can produce detectable (nuclear) recoil energy. 
It concerns the  models of the dark sector with large enough mass difference between 
the DM particles in the initial and final state
\cite{Giudice:2017zke,Cui:2009xq,Kang:2019uuj,Bell:2021xff,Bell:2022yxn,Filimonova:2022pkj,Feng:2021hyz,Baryakhtar:2020rwy,Zurowski:2020dxe,Smith-Orlik:2023kyl,Adams:2023pws}.
One believes that these energetic "secondary"\/ DM particles
will transfer sufficient energy to the  nuclei  and produce  signals that exceed the detection threshold. 
On this way one expects to lower the DM-mass limit to a level of about 100 keV$/c^2$, 
which can be registered in the direct DM experiment  \cite{CDEX:2022fig}.
\par
In addition, it was found \cite{Lawrence:2022niq} that the velocity distribution in the galacto-centric reference frame has strong deviation from the Maxwell-Boltzmann form.
This indicates the presence of some high-speed DM substructure with particle velocities as large as 
800--1000 km/s with a noticeable increase in the local density.
\par
Summarizing, we emphasize the critical significance of the direct DM search experiments.
Let us stress that the range of kinetic energy of the DM particles (with both small and large masses) 
is significantly expanded today beyond limits of the standard DM halo model, 
which is based on the Maxwell-Boltzmann distribution with  the galaxy escape velocity of 540 km/s.
Therefore, taking into account the high importance of the topic 
\cite{Bednyakov:2015uoa,Bednyakov:2020njj,Bednyakov:2012cu,Feng:2022rxt}, 
it seems  premature to consign to oblivion the traditional way of the direct DM detection 
without a critical analysis of the generally accepted assumptions.
\par
To this end,  the first analysis was started in \cite{Bednyakov:2022dmc}
based on the approach \cite{Bednyakov:2018mjd,Bednyakov:2019dbl,Bednyakov:2021ppn,Bednyakov:2021bty}.
It was shown that as the nuclear recoil energy $T_A$ increases,  in the massive nonrelativistic $\chi$ particle scattering on a nucleus a transition occures from dominance of the elastic $\chi A$ interaction 
to the dominance of the inelastic one.
In this situation, as $T_A$ increases, a device tuned only to the elastic $\chi A$ scattering  
begins "to go blind"\/, since the number of the elastic events strongly decreases. 
They become gradually replaced by the inelastic events that the device is not able to see.
It was shown \cite{Bednyakov:2022dmc}  that  this "phenomenon"\/ can 
strongly manifest itself in the direct DM search experiments, 
whose results are interpreted in terms of the {\em spin-independent}\/ 
and {\em spin-dependent}\/ interaction of the DM particle with nucleons.
It sould be stressed that both of these interactions are beyond the scope of the SM and
may have a completely unexpected character \cite{Bednyakov:2022dmc}.
\par
In addition, if the recoil energy of the nucleus, after the $\chi A$ scattering, 
turns out to be below the detection threshold, i.e., $T_A< T_A^{\min}$, then the elastic signal cannot be detected at all.
For this "invisible"\/ $T_A$, the only evidence of the $\chi A$ interaction is the radiation from the nuclear deexcitation, i.e., the inelastic signal, although its intensity at $T_A< T_A^{\min}$
can be an order of magnitude less than the intensity of the elastic one
\cite{Bednyakov:2022dmc}.
In general,  it is not possible to understand which process, elastic or inelastic, took place
when only the nuclear recoil energy $T_A$ is measured
\cite{Bednyakov:2015uoa,Papoulias:2018uzy,Boehm:2018sux,Cooley:2021rws,Cebrian:2022brv}.
In other words, it is unclear whether the elastic or inelastic formula
should be used in data analysis.
\par
Thus, the goal of this work is to study within formalism of 
\cite{Bednyakov:2018mjd,Bednyakov:2019dbl,Bednyakov:2021ppn,Bednyakov:2022dmc}
some parameter space (kinematic and physical)
where the expected DM event rate caused by
the inelastic (incoherent) process $\chi A\to \chi A^{*}$ can compete with
the event rate caused by elastic (coherent) $\chi A\to \chi A^{}$ scattering.
Because of  the {\em beyond}\/-the-Standard-Model nature of the $\chi A$ interaction, 
this parameter space could explain the "blindness" of the DM detectors,
intended only for detection of the elastic DM scattering events.

\section{Kinematics and cross section of ${\chi A}$ scattering}\label{2chiA-Kinematics} 
\begin{figure}[h] 
\includegraphics[scale=1.3]{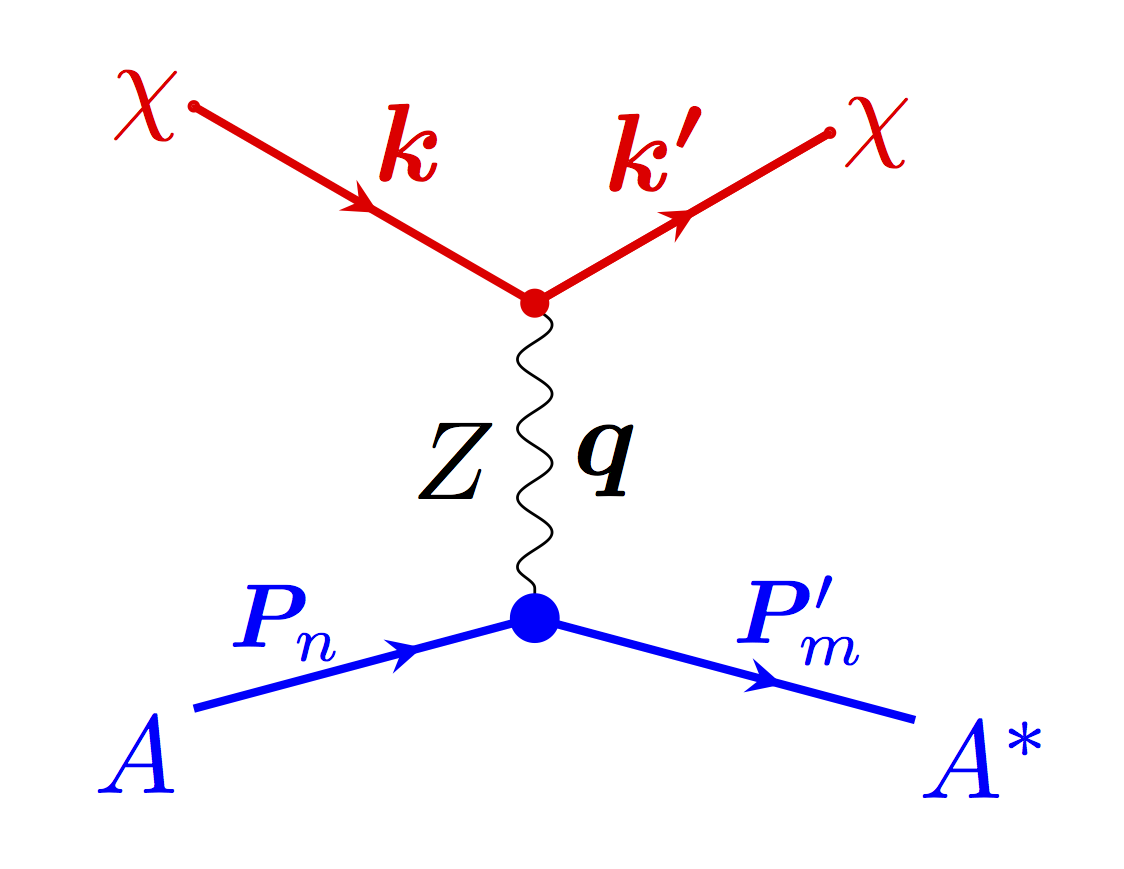}
\includegraphics[scale=0.3]{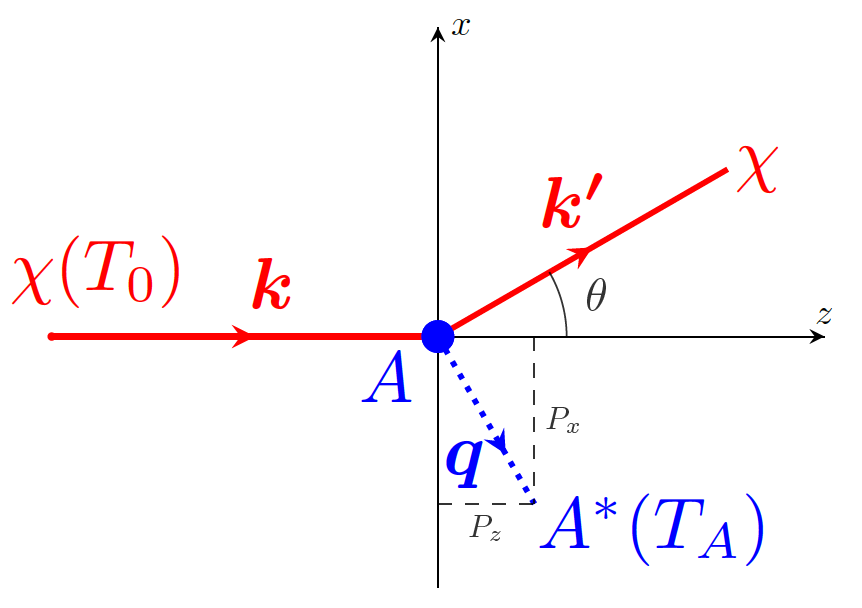}
\caption{\small An example of $\chi A$-interactions due to the exchange of the neutral $Z$-boson (left). Kinematics of this process in the laboratory frame where the nucleus $A$ is at rest (right). } \label{fig:DiagramCENNS}
\end{figure}

In the case of the interaction of two particles with the formation of two particles,  \linebreak
$ \chi(k)+A( P_n)\rightarrow \chi(k' )+A^{(*)}(P'_m),$ 
the 4-momenta of the incoming and the outgoing neutral massive lepton ($\chi$ particle) are
denoted as $k=(k_0=E_\chi,\bm{k})$ and $k'=(k'_0=E_\chi',\bm{k}')$, and the
4-momenta of the initial and the final state of the nucleus are denoted, respectively, as $P_n=(P^0_n,\bm{P}_n)$ and $P'_m=(P^0_m, \bm{P}_m)$ (Fig.~\ref{fig:DiagramCENNS}, left).
The total energy of the state $|P_n\rangle$ is $P_n^0 = E_{\bm{P}}+\varepsilon_n$,
where $\varepsilon_n$ is the internal energy of the $n$th quantum level of the nucleus.
If the $\chi$ particle with mass $m_\chi$ and momentum $\bm{k}$ hits the nucleus $A$ at rest
along the $z$-axis and flies away at angle $\theta$ to the $x$-axis with momentum $\bm{k}'$, 
the 4-momenta can be written as
\par
\begin{eqnarray*}
k&=&\big(k_0=\sqrt{m^2_\chi+|\bm{k}|^2}, 0, 0, k_z=|\bm{k}|\big), \qquad
P_n = \big(P_n^0=m_A +\varepsilon_n, 0, 0, 0\big), \\
k'&=&\big(k'_0=\sqrt{m^2_\chi + |\bm{k}'|^2}, k'_x= |\bm{k}'|\sin\theta, 0, k'_z = |\bm{k}'|\cos\theta\big), \\
P'_m&=&\big(P_m^0=\varepsilon_m + \sqrt{m_A^2 + (P_m^x)^2 + (P_m^z)^2}, P_m^x, 0, P_m^z\big )= \\ &=& \big(P_m^0=\varepsilon_m + \sqrt{m_A^2 + \bm{q}^2}, - |\bm{k}'|\sin\theta, 0, |\bm{k }|- |\bm{k}'|\cos\theta\big),
\end{eqnarray*}
where $m_A$ is the mass of the target nucleus, and $\varepsilon_m$ is the excitation energy of the 
$m$th level of the nucleus.
The 4-momentum $q =(q_0, \bm{q})$ transferred to the nucleus is related to these quantities as follows:
\begin{eqnarray} \nonumber
\label{eq:2DM-Kinematics-momentum-transfer}
q^2&\equiv &(k-k')^2 = 2\big(m^2_\chi- (kk')\big) =
2 \big(m^2_\chi -\sqrt{(m^2_\chi + |\bm{k}'|^2)(m^2_\chi + |\bm{k}|^2)} + |\bm{k}| |\bm{k}'|\cos\theta\big),
\\ q_0 &=& k_0 -k'_0 = P^0_m - P^0_n = \Delta\varepsilon_{mn}+ T_A,
\\ \bm{q}^2 &=& (\bm{k}-\bm{k}')^2 = (-|\bm{k}'|\sin\theta)^2+(|\bm{k}|- |\bm{k}'|\cos\theta)^2=
  |\bm{k}|^2 + |\bm{k}'|^2 - 2 |\bm{k}| |\bm{k}'|\cos\theta.
\nonumber\end{eqnarray}
The kinetic energy of the motion of the recoil nucleus is defined as
\begin{equation}\label{eq:2DM-Kinematics-Recoil-kinetic-energy}
T_A=\sqrt{m^2_A+\bm{q}^2}-m_A
.\end{equation}
The energy conservation law, i.e., $k_0 + P_n^0=k'_0 + P_m^0$, can be rewritten in the form
\begin{equation}\label{eq:2DM-Kinematics-Kinematics-EnergyConservation}
k_0+m_A-\Delta\varepsilon_{mn}=\sqrt{m^2_\chi + |\bm{k}'|^2}+ \sqrt{m^2_A+|\bm{k}'|^2+ |\bm{k}|^2 - 2 |\bm{k}| |\bm{k}'|\cos\theta}
, \end{equation} 
where the difference between the energies of the nuclear $|m\rangle$ and $|n\rangle$ states is denoted as
\begin{equation}\label{eq:2DM-Kinematics-delta_Epsilon}
\Delta\varepsilon_{mn} \equiv \varepsilon_m-\varepsilon_n.
\end{equation}
From expression (\ref{eq:2DM-Kinematics-Kinematics-EnergyConservation}) it follows that
  \begin{equation}\label{eq:2chiA-Kinematics-T_A-vs-T_0-etc}
T_A(|\bm{k}'|,\cos\theta)=k_0-k'_0-\Delta\varepsilon_{mn}
=\sqrt{m^2_\chi + |\bm{k}|}-\sqrt{m^2_\chi + |\bm{k}'|^2}-\Delta\varepsilon_{mn}.
\end{equation}
\par 
In the Standard Halo Model \cite{Bertone:2004pz,Drukier:1986tm,Freese:1987wu,Lewin:1996rx}, 
the DM-particle velocity near the Earth is estimated at approximately $10^{-3}$ of the speed of light, 
$ |\bm{v}|={|\bm{k}|}/{m_\chi} \simeq 10^{-3}c\simeq 350~$km/s, with the maximal (escape) 
velocity to be about 540~km/sec.
As mentioned in the introduction, the maximal DM velocity can be as large as 1000~km/sec.
Nevertheless, the kinetic energy of DM particles incident on the target nucleus at rest
\begin{equation}\label{eq:2chiA-Kinematics-T_0}
T_0\equiv \dfrac{|\bm{k}|^2}{2 m_\chi} =\dfrac{|\bm{v}|^2 m_\chi}{2 }
\simeq 10^{-(6\div 4)} \dfrac{m_\chi c^2}{2 }
\end{equation}
still remains in the region of tens of MeV, which is noticeably smaller than the mass of the DM particle $m_\chi$,
although, according to other estimations, this energy can reach the GeV level.
In the direct DM experiments one usually has $m_A\ge 10$ GeV$/c^2$,
the recoil energies of nuclei do not exceed 200 keV,
the typical difference between the excitation energies of nuclei is in the region of
several hundred keV. 
The characteristic scales of these quantities are
\begin{eqnarray}\nonumber
\label{eq:2DM-Kinematics-DDMD-scales}
&\quad& 10 \le m_A \le 100 \text{~GeV}/c^2 , \quad 1<m_\chi< 10^4 \text{~GeV}/c^2
, \quad 1 \text{~keV} \le T_0 \le 1 \text{~GeV}
,\\ &&  T_A \le 200 \text{~keV}, \quad |\bm{q}| \le 0.2 \div 0.3 \text{~GeV}/c, \quad
  \Delta\varepsilon_{mn}\le 1 \text{~MeV}.\end{eqnarray}
Therefore, in what follows, it is enough  to use the {\em nonrelativistic}  approximation
$$k_0=(m^2_\chi + |\bm{k}|^2)^{1/2}\simeq m_\chi + \dfrac{|\bm{k}|^2}{2 m_\chi} = m_\chi + T_0,
\quad \sqrt{m^2_\chi + |\bm{k}'|^2}\simeq m_\chi + \dfrac{|\bm{k}'|^2}{2 m_\chi},\quad \text{and}$$
\begin{equation}\label{eq:2DM-Kinematics-Kinetic-energy-of-nucleus-via-k-k'-nonrel}
T_A(|\bm{k}'|, \cos\theta) \simeq \dfrac{\bm{q}^2}{2 m_A} =
\dfrac{|\bm{k}'|^2+|\bm{k}|^2 - 2 |\bm{k}| |\bm{k}'|\cos\theta}{2m_A}
. \end{equation}
\par 
The observable differential cross section of the process $\chi A\to \chi A^{(*)}$
is the sum of  the incoherent and coherent terms
\cite{Bednyakov:2022dmc},
which are expressed via scalar products $(l_{s's},h^{p/n}_{r'r})$ of  the lepton and nucleon currents
\begin{eqnarray}
\label{eq:41chiA-CrossSection-Coh-vs-InCoh-CrossSection-TwoTerms-with-s-sprime}
\frac{d\sigma_{s's}}{d T_A}(\chi A\to \chi A^{(*)})&=&
 \frac{d\sigma^{s's}_\text{inc}}{d T_A}(\chi_s A\to \chi_{s'} A^{*}) + \frac{d\sigma^{s's}_\text{coh}}{d T_A}(\chi_s A\to \chi_{s'} A),
\quad \text{where} 
\end{eqnarray}\begin{eqnarray}  \nonumber
\label{eq:41chiA-CrossSection-Coh-vs-InCoh-CS-with-s-sprime-for-proton-neutron}
\frac{d\sigma^{s's}_\text{inc}}{d T_A} &=& c_A(T_0, m_A, m_\chi) g_\text{i} \sum_{f=p,n}[1-|F_f(\bm{q})|^2]
\Big[A^f_+  \sum^{}_{r'=\pm}|(l_{s's}, h^{\eta,f}_{r'+})|^2+A^f_- \sum^{}_{r'=\pm}|(l_{s's},h^{\eta,f}_{r'-})|^2 \Big], \qquad 
\\  \frac{d\sigma^{s's}_\text{coh}}{d T_A} &=&c_A(T_0, m_A, m_\chi) g_\text{c}
\Big| \sum_{f=p,n}F_f(\bm{q})  [A^f_+(l_{s's},h^{\eta,f}_{++})+ A^f_-(l_{s's},h^{\eta,f}_{--})] \Big|^2
, \end{eqnarray}
and $A^f_\pm$ is the number of $f$-type nucleons ($f=p,n$) with the spin projection $\pm 1$ on
some direction (for example, the arrival of the $\chi$ particle).
In the experimental situation (or in calculations), the initial external quantity 
is the recoil energy $T_A$. 
The $\chi$-particle emission angle in the laboratory system
as a function of $T_A$, $\Delta\varepsilon_{mn}$ and $T_0$ has the form
\cite{Bednyakov:2022dmc}
\begin{eqnarray}\label{eq:2DM-Kinematics-CosT-from-k2prime-and-T_A}
  \cos\theta(T_A) &=&
  \dfrac{m_\chi( 2 T_0- \Delta\varepsilon_{mn}) - T_A (m_\chi +m_A)}{2 m_\chi\sqrt{ T_0 (T_0-\Delta\varepsilon_{mn}-T_A) }}
.\end{eqnarray}
In formulas (\ref{eq:41chiA-CrossSection-Coh-vs-InCoh-CS-with-s-sprime-for-proton-neutron})
 the universal common factor is introduced
\begin{eqnarray}
\label{eq:43chiA-CrossSection-via-ScalarProducts-c_A-definition}
c_A \equiv c_A(T_0, m_A, m_\chi) &\equiv& \dfrac{G^2_F m_A}{2^6\pi m^2 (2T_0 m_\chi)}
=\dfrac{G^2_F m_A}{4\pi }\dfrac{1}{2^4 m^2 |\bm{k}^l_\chi|^2 },\quad
\end{eqnarray}
where $m$ is the nucleon mass 
and 
the dependence on the initial energy $T_0$ of the $\chi$ particle incident on the nucleus at rest 
is explicitly distinguished, and the intensity of the hypothetical interaction of the $\chi$ particle with nucleons
(proportional to $G^2_F$) is "hidden"\/ in the corresponding scalar products.
For all spin projections of the nucleon ($r',r=\pm1$) and the $\chi$ particle ($s',s=\pm1$), 
the scalar products of the $\chi$ particle and nucleon weak currents 
are defined as follows \cite{Bednyakov:2021pgs}:
\begin{eqnarray}
\label{eq:42chiA-CrossSection-ScalarProducts-ChiEta-All-WeakGeneralCurrent-definition}
(l^w_{s's}, h^{w,f}_{r'r}) &=& \alpha_f (l^v_{s's}\, h^v_{r'r}) + \beta_f (l ^v_{s's}\, h^a_{r'r})
+ \gamma_f (l^a_{s's}\, h^v_{r'r}) + \delta_f (l^a_{s's}\, h^a_{r'r})
.\end{eqnarray}
In the nonrelativistic approximation, these scalar products look like
\begin{eqnarray} \nonumber
\label{eq:42chiA-CrossSection-ScalarProducts-ChiEta-All-Weak-Nonrel}
(l^w_{\pm\pm}, h^{w,f}_{\pm\pm}) &=&   m^2_c (\alpha_f - \delta_f)
, \quad (l^w_{\pm\pm}, h^{w,f}_{\mp\mp}) =  m^2_c (\alpha_f + \delta_f)
,\quad  (l^w_{\pm\mp}, h^{w,f}_{\mp\pm}) = - 2 m^2_c \delta_f
,\\ (l^w_{\pm\mp}, h^{w,f}_{\mp\mp}) &= &  \pm m^2_s e^{\mp i\phi} (\alpha_f-\delta_f)  
, \quad  (l^w_{\pm\mp}, h^{w,f}_{\pm\pm}) = \pm m^2_s e^{\mp i\phi} (\alpha_f+\delta_f)
, \\  (l^w_{\pm\pm}, h^{w,f}_{\pm\mp})  &= &  \mp 2m^2_s e^{\mp i \phi} \delta_f 
,\quad (l^w_{\pm\pm}, h^{w,f}_{\mp\pm}) =  (l^w_{\pm\mp}, h^{w,f}_{\pm\mp})^{}\simeq  0
.  \nonumber  \end{eqnarray}
Here the notations are introduced
\begin{equation}
\label{eq:42chiA-CrossSection-ScalarProducts-ChiEta-All-nonrel-parameters}
\displaystyle m^2_c \equiv 4 m_\chi m \cos\frac{\theta}{2}, \quad m^2_s \equiv 4 m_\chi m \sin\frac{\theta}{2}.
\end{equation}
It is convenient to rewrite 
formulas (\ref{eq:41chiA-CrossSection-Coh-vs-InCoh-CS-with-s-sprime-for-proton-neutron})
in terms of the total number of $f$-type nucleons, 
$A_f$, and the difference in the number of nucleons, 
$\Delta A_f $, with the positive and negative spin projections
\begin{equation}
\label{eq:43chiA-CrossSection-via-ScalarProducts-to-A_f-and-DeltaA_f}
  A^f_\pm =\dfrac12 (A_f\pm \Delta A_f),
\quad\text{where}\quad A_f\equiv A^f_+ + A^f_- \quad\text{and}\quad \Delta A_f \equiv A^f_+ - A^f_-
.\end{equation}
The cross sections for the nonrelativistic {\em coherent}\/ $\chi A$ interaction have the form
\begin{eqnarray}\nonumber
\label{eq:43chiA-CrossSections-via-ScalarProducts-Weak-nonrel-CohSC-general}
\frac{d\sigma^{\mp\mp}_\text{coh}(\bm{q})}{g_\text{c} d T_A}&=&  \cos^2\frac{\theta}{2} 
\dfrac{G^2_F  m_A}{4\pi}\dfrac{ m^2_\chi }{|\bm{k}^l_\chi|^2 } 
 \Big[ \sum^{}_{f=p,n} F_{f}(\bm{q}) {A_f} \Big(\alpha_f \pm   \delta_f \dfrac{\Delta A_f}{A_f} \Big)\Big]^2
,\\\frac{d\sigma^{\mp\pm}_\text{coh}(\bm{q})}{ g_\text{c} d T_A}&=&\sin^2\frac{\theta}{2}
\dfrac{G^2_F  m_A}{4\pi} \dfrac{m^2_\chi}{ |\bm{k}^l_\chi|^2 }    
 \Big[ \sum^{}_{f=p,n} F_{f}(\bm{q}) A_f \Big( \alpha_f  \mp \delta_f \dfrac{\Delta A_f}{A_f} \Big)  \Big]^2
,\\\dfrac{d\sigma^{\text{total}}_\text{coh}(\bm{q})}{ g_\text{c} d T_A}&=&  \dfrac12 \sum^{}_{s's}\dfrac{d\sigma^{s's}_\text{coh}}{ g_\text{c} d T_A} = 
  \dfrac{G^2_F  m_A}{4\pi}\dfrac{m^2_\chi}{|\bm{k}^l_\chi|^2 }  \Big(\big[\sum^{}_{f=p,n}\alpha_f A_f F_{f}(\bm{q}) \big]^2+ \big[\sum^{}_{f=p,n}\delta_f {\Delta A_f} F_{f}(\bm{q}) \big]^2  \Big)
 \nonumber
.\end{eqnarray}
The cross sections for the  nonrelativistic {\em incoherent}\/ $\chi A$ interaction are the following:
\begin{eqnarray} \nonumber 
\label{eq:43chiA-CrossSections-via-ScalarProducts-Weak-nonrel-InCohSC-general}
  \frac{d\sigma^{\mp\mp}_\text{inc}(\bm{q})}{g_\text{i} d T_A}&=&
\dfrac{ G^2_F  m_A}{4\pi}\dfrac{m^2_\chi }{|\bm{k}^l_\chi|^2} \sum_{f=p,n}\big[1-F^2_f(\bm{q})\big] A_f 
\times\\&&\times  \nonumber 
\Big\{\cos^2\frac{\theta}{2}\alpha^2_f + (1+\sin^2\frac{\theta}{2})\delta^2_f
\pm\dfrac{2\Delta A_f }{A_f}\delta_f\Big[\cos^2\frac{\theta}{2} \alpha_f +\sin^2\frac{\theta}{2}\delta_f \Big]\Big\}
, \\ 
 \frac{d\sigma^{\mp\pm}_\text{inc}(\bm{q})}{g_\text{i} d T_A}&=&
\dfrac{ G^2_F m_A}{4\pi}\dfrac{m^2_\chi }{|\bm{k}^l_\chi|^2}\sum_{f=p,n}\big[1-F^2_f(\bm{q})\big]  A_f 
\times\\&&\times \nonumber
\Big\{\sin^2\frac{\theta}{2}\alpha^2_f + (1+\cos^2\frac{\theta}{2} )\delta^2_f\mp\dfrac{2\Delta A_f}{A_f }\delta_f\Big[\sin^2\frac{\theta}{2} \alpha_f+ \cos^2\frac{\theta}{2}\delta_f\Big]\Big\}
,\\ 
\frac{d\sigma^{\text{total}}_\text{inc}(\bm{q})}{g_\text{i} d T_A} &=&\frac12 \sum^{}_{s's} \frac{d\sigma^{s's}_\text{inc}}{g_\text{i} d T_A}   = \dfrac{ G^2_F m_A}{4\pi}\dfrac{m^2_\chi }{|\bm{k}^l_\chi|^2}
 \sum_{f=p,n} A^f \big[1-F^2_f(\bm{q})\big]  \big[\alpha^2_f + 3 \delta^2_f \big]
 \nonumber   .\end{eqnarray}
The measurable total cross section of the $\chi A$ interaction in the nonrelativistic 
approximation has the form
 \begin{eqnarray}\label{eq:43chiA-CrossSections-via-ScalarProducts-Weak-nonrel-Full}
\dfrac{d\sigma^{\text{weak}}_{\text{nonrel}}}{d T_A}(\chi A \to \chi A^*)&=&
\dfrac{G^2_F  m_A }{4\pi}  \dfrac{m^2_\chi  }{|\bm{k}^l_\chi|^2}
\Big\{g_\text{i} \sum_{f=p,n} A^f \big[1-F^2_f(\bm{q})\big]  \big[\alpha^2_f + 3 \delta^2_f \big]
\\&& +g_\text{c} \Big(\big[\sum^{}_{f=p,n}\alpha_f A_f F_{f}(\bm{q}) \big]^2+ \big[\sum^{}_{f=p,n}\delta_f {\Delta A_f} F_{f}(\bm{q}) \big]^2  \Big)\Big\} 
\nonumber. \end{eqnarray}
Here following the notation is used:
\begin{equation}\label{43chiA-CrossSection-via-ScalarProducts-c_A+}
c_A (4m_\chi m)^2 = (4m_\chi m)^2 \dfrac{ G^2_F m_A}{2^6\pi m^2 |\bm{k}^l_\chi|^2 }
  = \dfrac{ G^2_F m_A}{4\pi}\dfrac{m^2_\chi }{|\bm{k}^l_\chi|^2 }
.\end{equation} 

\section{Expected event  counting rate}\label{44DM-CrossSection-via-ScalarProducts-EventRates} 
In an experiment aimed at the direct DM detection (see, for example, 
\cite{Lewin:1996rx,Freese:1987wu,Bednyakov:2015uoa,Vergados:1996hs,Baudis:2012ig,Tanabashi:2018oca}), one tries to measure the so-called event counting rate, i.e., some number of useful events per unit of time.
This value is proportional to the cross section of the particle interaction
multiplied by their relative velocity, $ v \sigma (v, ...)$. 
It is  further integrated over the relative velocity $v$ with a distribution function $f(v)$,  
which determines the probability to have velocity $v$ at the moment of the interaction.
\par
A "traditional estimate"\/ of the event rate 
 is based on a rather substantial expected flow of  the DM particles through the Earth
\cite{Baudis:2012ig}, 
which for the DM mass $ m_{\rm{DM}} = 100$ GeV/$c^2$ is  
$${\Phi}_{\rm{DM}} = \frac{ {\rho}_{\rm{DM}} }{ m_{\rm{DM}} } \langle v \rangle\cong \dfrac{6.6 \times 10^{4}}{{\rm{sm^{2}~{\rm sec} }}}\Big[ \frac{ {\rho}_{\rm{DM}} }{ 0.3~{ \rm{GeV/sm^{3} } } }\Big]\Big[ \frac{ 100~{\rm{GeV}} }{ m_{\rm{DM}} } \Big]\Big[ \frac{ \langle v \rangle }{ 220~{ \rm{km/sec} } } \Big]
.$$
Therefore, for a target with the atomic number $A = $ 100 and typical cross section at a level of 
$10^{-38}~\rm {sm^{2}}$ the rate has the following rather small value:
 {\small
 $$  R=\sigma \frac{ N_{\rm{Av}} }{ A } {\Phi}_{\rm{DM}} \cong 0.13~\frac{ \rm{events} }{ \rm{kg\, year} }
  \Big[ \frac{ 100~{\rm{g/mol}} }{ A } \Big] \Big[ \frac{ {\rho}_{\rm{DM}} }{ 0.3~{ \rm{GeV/sm^{3} } } }  \Big]
  \Big[ \frac{ 100~{\rm{GeV}} }{ m_{\rm{DM}} } \Big] \Big[ \frac{ \langle v \rangle }{ 220~{ \rm{km/sec} } } \Big] 
  \Big[ \frac{\sigma}{ 10^{-38}~{\rm{cm^{2}}} }\Big]. $$}%
Here $ \langle v \rangle $ and $ {\rho}_{\rm{DM}}$ are the averaged velocity and density of 
the DM particles near the Earth, $ {\sigma}$ is the cross section for the 
DM particle--nucleon interaction.
$ N_{\rm{Av}} = 6.02 \times 10^{23}~{\rm{mol}}^{-1} $ is the Avogadro number 
(the number of protons in one gram of protons), $A$ is the mass of one mole, i.e., it is a weight of $6.02 \times 10^{23}$ nuclei of the target material. 
\par 
In fact, the integral counting event rate is determined in the form 
\begin{equation}\label{eq:44DM-CrossSection-via-ScalarProducts-EventRates-Rate-vs-T_A}
R(T_A^{\min})=   n_\chi \, N_A \int_{v_{\min}}^{v_{\max}}\!\! f(v)dv  \int_{T_A^{\min}}^{T_A^{\max}}\!\frac{v \,d\sigma}{dT_A}(v,T_A) \, dT_A 
. \end{equation}
As a function of the nuclear recoil energy, $T_A \in (T_A^{\min},T_A^{\max})$,  the 
differential event rate is 
\begin{equation}
\label{eq:44DM-CrossSection-via-ScalarProducts-EventRates-dRate/dT_A-vs-T_A}
\frac{d R(T_A^{\min})}{d T_A} = n_\chi \, N_A  \int_{v_{\min}}^{v_{\max}}\!\! f(v) \,  \frac{v\,  d\sigma}{d T_A}(v,T_A)\, dv
 .\end{equation} 
Here $N_A$ is the number of scattering centers (the number of $A$-type nuclei) in the target.
If the target mass is one gram, $N_A= \dfrac{ N_{\rm{Av}} }{ A }$.
Furthermore, 
 $n_\chi \equiv \dfrac{\rho_\chi}{m_\chi}$ is the number density of the $\chi$ particles, 
 where $\rho_\chi \simeq 0.35 \, \text{GeV}/$cm$^3$ is the density of the halo
 (assumed to consist entirely of the $\chi$ particles)   around the Earth, 
 $T_A^{\min}$ is the minimal recoil energy of the nucleus (the lower energy threshold), 
 $v_{\min}$ is the minimal relative velocity of the $\chi$ particle and nucleus 
 at which the recoil energy of the $T_A^{\min}$ is still possible
 \begin{equation}  
\label{eq:44DM-CrossSection-via-ScalarProducts-EventRates-Vmin}
v^{}_{\min} = \sqrt{T_A^{\min}  \dfrac{(m_A+m_\chi)^2}{ 2 m_A   m^2_\chi}} 
= \sqrt{ \dfrac{2T_A^{\min} }{m_\chi  \mu_A} }    , \qquad \text{where}\qquad 
 \mu_A  \equiv \frac{4 m_\chi m_A}{(m_\chi+m_A)^2}  
 .\end{equation}
When the relative velocity $v< v_{\min}$, any event registration is not possible.
Further, $v_{\max} = v_{\text{esc}}$ is the maximal velocity of the $\chi$ particle for a given DM distribution.
Then the maximal kinetic energy of the $\chi$ particle with mass $m_\chi$  and the maximal kinetic 
energy (recoil) of the nucleus with mass $m_A$, caused by 
the {\em elastic}\/ $\chi$ particle--nucleus collision are
\begin{equation} \label{eq:3DM-Elastic-Tmax}
T_\chi^{\max} = \dfrac{m_\chi  v^2_{\max}} {2} \quad \text{and}\quad
 T_A^{\max} =  \dfrac{q^2_{\max}}{2 m_A} =   \dfrac{ 2 m_A  v^2_{\max} m^2_\chi}{(m_A+m_\chi)^2} 
= T_\chi^{\max} \mu_A 
.  \end{equation} 
In the Standard Halo Model (SHM) \cite{Kang:2019uuj,Evans:2018bqy,Zurowski:2020dxe}
the  $\chi$-particle velocities {\em near the detector}\/ are distributed as follows:
\begin{eqnarray}
\label{eq:3DM-fSHM}
 f_{\rm SHM}(\bm{w})&=&\frac{1}{(\pi v_0^2)^{3/2}}\exp\Big[\frac{-(\bm{w}-\bm{v}_\oplus)^2}{v_0^2}\Big] \text{~~and~~}
 \\  v_{\rm esc} &=& |\bm{v}|_{\max} = |\bm{w}+\bm{v}_\oplus|_{\max}  \simeq 540~\frac{\rm km}{\rm sec}
 \nonumber , \end{eqnarray}
where $v_0 \simeq 220$~km/sec is the circular velocity in the halo. 
It is usually assumed \cite{Freese:1987wu}
 that the speed of the Sun $v_\odot \equiv v_{\rm Sun} \simeq v_{\text{circ}} \equiv v_0 = \bar{v} \sqrt{\frac{2}{3}} \simeq 232\pm 20$~km/sec, where $\bar{v}$ is the velocity dispersion in the halo ($\simeq 270\pm 25$ km/s).
The Earth's velocity relative to the Galactic halo is $\bm{v}_\oplus\ne 0$.
 \par
For the one-dimensional DM velocity 
relative to the Earth (and relative to the detector) $w=|\bm{w}|$  one has the following 
distribution function \cite{Freese:1987wu}: 
\begin{eqnarray}
\label{eq:3DM-SHM-speeds-in-Earth-system}
f_{\rm SHM}(w)dw \equiv f_{\rm SHM}(x) dx,  \text{~where~}
f_{\rm SHM}(x) = \frac{4 x^2 }{\sqrt{\pi}} e^{-(x^2+\eta^2)}\frac{\mbox{sinh}(2x\eta)}{2x\eta}, \ \int\!\! f_{\rm SHM}(x)dx = 1.\qquad
\end{eqnarray}
Here $x = \dfrac{w}{v_0}$ is the dimensionless velocity of the DM particle relative to the Earth,
$\eta = \dfrac{v_\oplus}{v_0}$ is the dimensionless velocity of the Earth relative to the Galactic halo.
It depends on the time $\eta=\eta(t)$.
Since the Earth has its annual revolution around the Sun, 
one has  $\langle \eta(t) \rangle_{\text{year}} = \eta_0 \simeq 1.05. $
\par
It is currently believed that the SHM distribution (\ref{eq:3DM-fSHM}) does not accurately describe 
the DM distribution in the Galaxy.
There are anisotropic structures, such as DM particle flows,
due to impact of nearby galaxies \cite{OHare:2018trr}.
In what follows, we will use the following
"simplified-generalized"\/ expression from \cite{OHare:2018trr}:
\begin{equation}
f_{\rm str}(\bm{w})=\frac{1}{\pi^{3/2} (\epsilon v_0)^{3}}\exp\Big[-\frac{\big(\bm{w}-(\bm{v}_\oplus+\bm{v}_{\rm str})\big)^2}{(\epsilon v_0)^2}\Big], \quad \text{where}\quad |\bm{v}_{\rm str}|\simeq 300 \text{~km/sec}
\label{eq:3DM-stream}
.\end{equation}
In this "external-DM-flow" distribution the vector sum of the Earth velocity and the flow velocity 
relative to the Galactic halo
\footnote{We neglect here their time dependences.}, 
$\bm{v}_\oplus+\bm{v}_{\rm str}$, plays 
the role of the Earth's velocity $ \bm{v}_\oplus$ from  (\ref{eq:3DM-fSHM}).
Therefore, by analogy with formula (\ref{eq:3DM-SHM-speeds-in-Earth-system}) one can write
\begin{eqnarray}\label{eq:3DM-stream-speeds-in-Earth-system}
f_{\rm str}(w)dw \equiv f_{\rm str}(y) dy,  \text{~where~} 
f_{\rm str}(y) = \frac{4}{\sqrt{\pi}}  y^2 e^{-(y^2+\xi^2)} \frac{\mbox{sinh}(2y\xi)}{2y\xi} \text{~and~~} \int f_{\rm str}(y)dy = 1
, \qquad \end{eqnarray}
where $y = \dfrac{w}{\epsilon v_0}$ is the dimensionless velocity of the DM particle (belonging to the external flow) relative to the Earth, and  ${\xi}\equiv \dfrac{|\bm{v}_\oplus+ \bm{v}_{\rm str}|}{\epsilon v_0 }$ 
is the dimensionless sum of the Earth and flow velocities. 
For simplicity, we assume that
${\xi}=\dfrac{{v}_\oplus+{v}_{\rm str}}{\epsilon v_0 } $.
The complete (one-dimensional) {\em normalized to unity}\/ DM distribution (of $\chi$ particles)
in terms of the dimensionless velocity relative to the Earth has the form
\begin{eqnarray*}
f_{\rm SHM+str}(w)dw  &=& \Big[\Big(1-\frac{\rho_{\rm str}}{\rho}\Big) f_{\rm SHM}(w) +\frac{\rho_{\rm str}}{\rho} f_{\rm str}(w)\Big] dw  \equiv \Big(1-\frac{\rho_{\rm str}}{\rho}\Big) f_{\rm SHM}(x) dx + \frac{\rho_{\rm str}}{\rho} f_{\rm str}(y) dy
.\end{eqnarray*}
With allowance for $y=\dfrac{x}{\epsilon}$ and $\xi\equiv \dfrac{\eta_{\rm str}}{\epsilon} $, 
this expression can be rewritten in the final form
\begin{eqnarray}
\label{eq:3DM-both-speeds-in-Earth-system}
 f(x)dx &\equiv&f_{\rm SHM+str}(w)dw =\gamma^{}_{1}f^{}_{\rm SHM}(x)dx + \gamma^{}_{\eta} f^{}_{\rm str}(x)dx 
, \text{~~where~now} \\  f_{\rm str}(x) &=& \frac{4x^2 }{\sqrt{\pi}} 
e^{-\dfrac{x^2 + \eta^2_{\rm str}}{\epsilon^2}} \frac{\mbox{sinh}(2x\eta_{\rm str}/\epsilon^2)}{2x\eta_{\rm str}}
\qquad \text{and}\qquad \gamma^{}_{1}+\gamma^{}_{\eta}=1
\nonumber .\end{eqnarray}
In formula (\ref{eq:3DM-both-speeds-in-Earth-system}), one has a small generalization by means 
of weights of the distributions, the sum of which is equal to one.
In what follows, we will use expression (\ref{eq:3DM-both-speeds-in-Earth-system})
for the function $f(v)$ from the counting event rate 
(\ref{eq:44DM-CrossSection-via-ScalarProducts-EventRates-dRate/dT_A-vs-T_A}).
\par
The total (summed over all spin indices of the $\chi$ particle) differential cross section included in
(\ref{eq:44DM-CrossSection-via-ScalarProducts-EventRates-Rate-vs-T_A}) and
(\ref{eq:44DM-CrossSection-via-ScalarProducts-EventRates-dRate/dT_A-vs-T_A}),
according to (\ref{eq:41chiA-CrossSection-Coh-vs-InCoh-CrossSection-TwoTerms-with-s-sprime})
and (\ref{eq:43chiA-CrossSections-via-ScalarProducts-Weak-nonrel-Full}),
is the sum of the coherent and incoherent contributions to the $\chi A$ cross section
$$ \frac{d\sigma}{d T_A}(v,T_A)\equiv  \frac{d\sigma_\text{i(nc)}}{d T_A}(\chi_s A\to \chi_{s'} A^{*}) + \frac{d\sigma_\text{c(oh)}}{d T_A}(\chi_s A\to \chi_{s'} A). $$
Then the total event rate, 
integrated over the nuclear recoil energy  $T_A $ in the interval $T_A^{\min}\le T_A \le T_A^{\max}$, 
can be given as the sum of the {\em coherent} and {\em incoherent} event rates
\begin{eqnarray} \label{eq:3DM-EventRates-Coh+Inc}
R(T_A^{\min},T_A^{\max}) &\equiv& R_\text{c(oh)}(T_A^{\min}, T_A^{\max}) +  R_\text{i(nc)}(T_A^{\min}, T_A^{\max}),  \quad \text{where}
\\ R_\text{c/i}(T_A^{\min}, T_A^{\max}) &\equiv& \int_{T_A^{\min}}^{T_A^{\max}} dT_A\frac{d R(T_A)_\text{c/i} }{d T_A}  ,\text{~~and~~} \nonumber \\ \nonumber
 \frac{d R(T_A)_\text{c/i} }{d T_A}&=&
 n_\chi v_0 N_A \int_{x_{\min}}^{x_{\max}} f(x) x dx\frac{d\sigma^{}_\text{c/i}(xv_0 ,T_A)}{dT_A}.
\end{eqnarray}
For the differential event counting rates one has
\begin{eqnarray} \label{eq:3DM-Diff-EventRates-Coh+Inc}
 \frac{d R(T_A)_\text{c/i} }{n_\chi v_0 N_A d T_A} &=& \int_{x_{\min}}^{x_{\max}}
 \Big[ \gamma^{}_{1}  f^{}_{\rm SHM}(x)+ \gamma^{}_{\eta} f^{}_{\rm str}(x) \Big]
 x dx\frac{d\sigma^{}_\text{c/i}(xv_0 ,T_A)}{dT_A}
 =\\&=&  \gamma^{}_{1}\int_{x_{\min}}^{x^{\max}_1}f_{\rm SHM}(x)dx\frac{x d\sigma^{}_\text{c/i}(x,T_A)}{dT_A} 
 + \gamma^{}_{\eta} \int_{x_{\min}}^{x^{\max}_\eta}  f_{\rm str}(x)dx\frac{x d\sigma^{}_\text{c/i}(x,T_A)}{dT_A} 
\nonumber
.\end{eqnarray}
There are the following relations according to (\ref{eq:44DM-CrossSection-via-ScalarProducts-EventRates-Vmin}):
\begin{eqnarray}\nonumber
\label{eq:3DM-xmin-xmax}
x^{}_{\min} &=&  \dfrac{w^{}_{\min}}{v_0} =  \Big[\dfrac{T_A^{\min}}{T_A^{0}}\Big]^{1/2}
, \text{~~where~~} T_A^{0} \equiv   \mu_A\dfrac{m_\chi  v_0^2}{2}
,\\ x^{\max}_{1} &=& \dfrac{w^{\rm SHM}_{\max}}{v_0} = \dfrac{v_{\rm esc} + v^{}_{\oplus}}{v_0} 
= \dfrac{v_{\rm esc}}{v_0}+\eta_1, \text{~~where~~} \eta_1= \dfrac{v^{}_{\oplus}}{v_0}\simeq 1 
\text{~~and~~}
\\  x^{\max}_{\eta} &=& \dfrac{w^{\rm str}_{\max}}{v_0}= \dfrac{v_{\rm esc}+ {v}_\oplus+{v}_{\rm str} }{v_0}
 =\dfrac{v_{\rm esc}}{v_0} +\eta_{}, \text{~~where~~} \eta= \dfrac{v^{}_{\oplus}+{v}_{\rm str} }{v_0}.
\nonumber\end{eqnarray} 
Here $x^{\max}_{1,\eta}$ corresponds to the maximal collision velocity between  the $\chi$ particle and the target, i.e., when the escape velocity and the velocity of the Sun (together with the external DM stream velocity)
fold in the opposite direction.
\par
If one assumes that the interaction of the massive $\chi$ lepton with the nucleons has the form of 
 the {\em weak interactions}
(\ref{eq:42chiA-CrossSection-ScalarProducts-ChiEta-All-WeakGeneralCurrent-definition}),
the total cross section of the $\chi A$ scattering
in the {\em nonrelativistic approximation}\/ is given by formula
(\ref{eq:43chiA-CrossSections-via-ScalarProducts-Weak-nonrel-Full}).
It explicitly contains  the coherent and incoherent contributions to the $\chi A$ cross section as follows:
 \begin{eqnarray}\label{eq:3DM-Diff-EventRates-Coh+Inc+1}
\dfrac{d\sigma^{\text{weak}}_{\text{i(nc)}}(x,T_A)}{d T_A}&=&\dfrac{G^2_F  m_A }{4\pi}  \dfrac{1}{ x ^2 v_0^2} 
\Big\{g_\text{i}\sum_{f=p,n} A^f \big[1-F^2_f(\bm{q})\big]  \big[\alpha^2_f + 3 \delta^2_f \big]\Big\}_{\text{i}}
,\\ \dfrac{d\sigma^{\text{weak}}_{\text{c(oh)}}(x,T_A)}{d T_A}&=&\dfrac{G^2_F  m_A }{4\pi}\dfrac{1}{ x ^2 v_0^2} 
\Big\{g_\text{c}  \big[\sum^{}_{f=p,n}\alpha_f A_f F_{f}(\bm{q}) \big]^2+ \big[\sum^{}_{f=p,n}\delta_f {\Delta A_f} F_{f}(\bm{q}) \big]^2 \Big\}_{\text{c}} 
\nonumber. \end{eqnarray}
Here it is used that $ \dfrac{m^2_\chi }{|\bm{ k}^l_\chi|^2} = \dfrac{1}{ x ^2 v_0^2} $, since
in the nonrelativistic approximation  the momentum of the $\chi$ lepton incident on the nucleus at rest is
 $|\bm{k}^l_\chi|^2 = m^2_\chi w^2$. 
As a result, the differential rates of the coherent and incoherent events
 (\ref{eq:3DM-Diff-EventRates-Coh+Inc}) take the form
\begin{eqnarray*} \nonumber
 \frac{d R(T_A)_\text{c/i} }{n_\chi v_0 N_A d T_A}
&=&  \gamma^{}_{1} \int_{x_{\min}}^{x^{\max}_1}f_{\rm SHM}(x)dx\frac{x d\sigma^{}_\text{c/i}(x,T_A)}{dT_A} 
 + \gamma^{}_{\eta}  \int_{x_{\min}}^{x^{\max}_\eta}  f_{\rm str}(x)dx\frac{x d\sigma^{}_\text{c/i}(x,T_A)}{dT_A} 
=\\&=&  \dfrac{G^2_F  m_A }{4\pi  v_0^2}  \Big\{..\Big\}_{\text{c/i}} \Big[
 \gamma^{}_{1}\int_{x_{\min}}^{x^{\max}_1}f_{\rm SHM}(x)\dfrac{dx}{x}
 +  \gamma^{}_{\eta}  \int_{x_{\min}}^{x^{\max}_\eta}  f_{\rm str}(x) \dfrac{dx}{x} \Big]
\nonumber
.\end{eqnarray*}
Finally, taking into account $\big\{...\big\}$-parentheses from (\ref{eq:3DM-Diff-EventRates-Coh+Inc+1})
and notation (\ref{eq:3DM-xmin-xmax}),  one arrives at
\begin{eqnarray}  \label{eq:3DM-Diff-EventRates-Coh+Inc+2}
\frac{d R(T_A)_\text{c/i} }{d T_A}&=& 
R_0(A,\chi) \Big\{g_{\text{c/i}}...\Big\}_{\text{c/i}} \Big[\gamma^{}_{1}  S^{}_{1}(x^{}_{\min}, x^{\max}_{1})+
\gamma^{}_{\eta}  S^{}_{\eta}(x^{}_{\min}, x^{\max}_{\eta})  \Big]
.\end{eqnarray}
Here, the auxiliary notation is introduced 
\begin{eqnarray}
\label{eq:44DM-CrossSection-via-ScalarProducts-EventRates-Rate_0}
R_0(A,\chi) &\equiv& \dfrac{ G^2_F  m_A  }{4 \pi}\dfrac{  n_\chi  N_A }{ v_0 }, 
\end{eqnarray}
which sets the scale of the event rate, 
and the DM halo contributions are given as
\begin{eqnarray}
S_{1}^{}(x^{}_{\min}, x^{\max}_1) &\equiv& \int_{x_{\min}}^{x^{\max}_1}f_{\rm SHM}(x)\frac{dx}{x}
= S_{}(x^{}_{\min},\eta_1,1)-S_{}(x^{\max}_1,\eta_1,1)
,\\ S^{}_{\eta}(x^{}_{\min}, x^{\max}_\eta) &\equiv& \int_{x_{\min}}^{x^{\max}_\eta}f_{\text{str}}(x)  \frac{dx}{x} 
=S_{}(x_{\min},\eta_{},\epsilon)-S_{}(x^{\max}_\eta,\eta_{},\epsilon)
. \end{eqnarray}
Here the common distribution function is introduced
\begin{equation}  \label{eq:4-S-def-main}
S_{}(x,\eta,\epsilon)\equiv\dfrac{\epsilon}{2\eta} 
\Big\{\text{erf}\big(\dfrac{x+\eta}{\epsilon}\big)-\text{erf}\big(\dfrac{x-\eta}{\epsilon}\big)\Big\}
=\dfrac{\epsilon }{\sqrt{\pi}\eta} \int_{\frac{x-\eta}{\epsilon}}^{\frac{x+\eta}{\epsilon}} e^{-t^2}d t.
\end{equation}
 The common value of $x^{}_{\min} $ from (\ref{eq:3DM-xmin-xmax}) is proportional to the square root of
the recoil energy threshold $T_A^{\min} $.
 In the  {\em elastic}\/  $\chi A$ interaction, according to (\ref{eq:3DM-Elastic-Tmax}) and (\ref{eq:3DM-xmin-xmax}), $x^{\max}_{\eta}$
determines the maximal nuclear recoil energy as a function of the maximal DM velocity
\begin{equation}
\label{eq:3DM-TAmax+TAmax}
T_{A,\eta}^{\max}  =   \mu_A\dfrac{m_\chi ( v_0 x^{\max}_{\eta})^2 }{2}
= \mu_A T_{\chi,\eta}^{\max} 
.\end{equation}
In the {\em inelastic}\/ $\chi A$ collision (when $\Delta\varepsilon_{mn}>0$)
the maximal nuclear recoil energy can be obtained by substituting the maximal
energy of the DM particle (\ref{eq:3DM-Elastic-Tmax}) into
 formula (\ref{eq:2DM-Kinematics-CosT-from-k2prime-and-T_A}) 
 and taking into account that the $\chi$ particle, incident on the nucleus, 
"is reflected"\/ strictly in the opposite direction, i.e., from the following condition:
$$ \cos\theta(T_{A^*}) =  \dfrac{m_\chi( 2 T_\chi^{\max} - \Delta\varepsilon_{mn})  - T_{A^*} (m_\chi +m_A)}{2 m_\chi\sqrt{ T_\chi^{\max} (T_\chi^{\max}-\Delta\varepsilon_{mn}-T_{A^*})}} =-1
.$$ 
The solution to this equation is
\begin{equation}\label{eq:3DM-Inelastic-TAmax}
T_{A^*,\eta}^{\max}(r,\Delta\varepsilon_{mn})  =  T_{\chi,\eta}^{\max} \dfrac{\mu_A}{2} 
\Big[1+\sqrt{1 -\dfrac{( r  +1)  \Delta\varepsilon_{mn}}{ T_{\chi,\eta}^{\max}} }\Big]
-\dfrac{{r} \Delta\varepsilon_{mn}}{r+1},  \quad \text{where~~}  r\equiv \dfrac{m_\chi}{m_A} 
 .\end{equation}
If  $T_{\chi,\eta }^{\max} \gg (r+1)\Delta\varepsilon_{mn}$ 
(or $\Delta\varepsilon_{mn}=0$), then $T_{A^*,\eta }^{\max}(r,\Delta\varepsilon_{mn})$ transforms back in 
(\ref{eq:3DM-TAmax+TAmax})
\footnote{With $\mu_A = \dfrac{4 r}{(1+r)^2}$, one has 
$T_{A^*}^{\max} = T_\chi^{\max} \mu_A-\dfrac{2 r  \Delta\varepsilon_{mn}}{1+r}
 =T_\chi^{\max} \mu_A  \Big(1- \dfrac{\Delta\varepsilon_{mn}}{T_\chi^{\max}}
\dfrac{(r+1)}{2 }\Big) \simeq T_\chi^{\max} \mu_A.$}. 
Solution (\ref{eq:3DM-Inelastic-TAmax}) exists only when
\begin{equation}\label{eq:3DM-Inelastic-condition}
 T_{\chi,\eta }^{\max} \ge \Big(\dfrac{m_\chi}{m_A}+1\Big) \Delta\varepsilon_{mn}
\quad \text{or}\quad  ( x^{\max}_\eta)^2 \dfrac{v_0^2 }{2}
 \ge \Big(1 +\dfrac{m_A}{m_\chi }\Big)\dfrac{\Delta\varepsilon_{mn}}{m_A}
.\end{equation}
If $T_{\chi,\eta}^{\max} = (r+1)\Delta\varepsilon_{mn}$, 
the "minimal"\/ value of the recoil energy is reached 
$$T_{A^*,\eta}^{\max}(r,\Delta\varepsilon_{mn})  =  T_{\chi,\eta}^{\max}\dfrac{\mu_A}{4}
,$$
which is a {\em unique} value of the nuclear recoil energy,
 at which excitation of the nucleus with energy $\Delta\varepsilon_{mn}$ is still possible.
In other words, if for a nucleus  with mass $m_A$ and excitation energy $\Delta\varepsilon_{mn}^{}$  
and a $\chi$ particle  with mass $m_\chi$ and maximal kinetic energy $T_ {\chi,\eta}^{\max}$
condition (\ref{eq:3DM-Inelastic-condition}) is not satisfied, the inelastic process is impossible.
In this situation, the DM particle does not have enough energy to excite the nucleus and 
 cause its  {\em motion}. 
Because of the momentum conservation law,  excitation of the nucleus 
by impact must be accompanied by some motion of the excited nucleus, i.e.,  
by the nuclear recoil kinetic energy.
\par
Considering  (\ref{eq:3DM-Diff-EventRates-Coh+Inc+2}),
the coherent and incoherent event rates have two terms corresponding to different DM distributions 
with different maximal nuclear recoils
\begin{eqnarray} \nonumber
\label{eq:3DM-DiffEventRates-Coh+Inc-SHM+str}
\frac{d R(T_A)_\text{c} }{g_\text{c} R_0()  d T_A}&=& 
\gamma^{}_{1} S^{}_{1}(x^{}_{\min}, x^{\max}_{1})\Big\{\big[\sum^{}_{f=p,n}\alpha_f A_f F_{f}(\bm{q})\big]^2+\big[\sum^{}_{f=p,n}\delta_f{\Delta A_f}F_{f}(\bm{q})\big]^2\Big\}\Theta(T_{A,\eta_1}^{\max}-T_A) 
\\&& \nonumber
+\gamma^{}_{\eta} S^{}_{\eta}(x^{}_{\min}, x^{\max}_\eta)  \Big\{\big[\sum^{}_{f=p,n}\alpha_f A_f F_{f}(\bm{q})\big]^2+\big[\sum^{}_{f=p,n}\delta_f{\Delta A_f}F_{f}(\bm{q})\big]^2\Big\} \Theta(T_{A,\eta}^{\max}-T_A)
,\\[-5pt] \\ \nonumber
\frac{d R(T_A)_\text{i} }{g_\text{i}R_0()  d T_A}&=&
\gamma^{}_{1}  S^{}_{1}(x^{}_{\min}, x^{\max}_1)\Big\{\sum_{f=p,n} A^f \big[1-|F_f(\bm{q})|^2\big]  \big[\alpha^2_f + 3 \delta^2_f \big]\Big\}\Theta(T_{A^*,\eta_1}^{\max}-T_A)
\\&&+  \nonumber
\gamma^{}_{\eta}  S^{}_{\eta}(x^{}_{\min}, x^{\max}_\eta)\Big\{\sum_{f=p,n} A^f \big[1-|F_f(\bm{q})|^2\big]  \big[\alpha^2_f + 3 \delta^2_f \big]\Big\}\Theta(T_{A^*,\eta}^{\max}-T_A)
. \end{eqnarray}
Here $T_{A^{(*)},\eta_1,\eta}^{\max}$
are defined by formula (\ref{eq:3DM-Inelastic-TAmax}), and $T_A$ is the measurable 
recoil energy for both excited and unexcited nuclei.
It is related to the momentum transferred to the nucleus by
 formulas (\ref{eq:2chiA-Kinematics-T_A-vs-T_0-etc}) and 
 (\ref{eq:2DM-Kinematics-Kinetic-energy-of-nucleus-via-k-k'-nonrel}), or $ \bm{q}^2 \simeq 2 m_A T_A $.
Recall that any detector aimed at registration of the nuclear recoil energy is 
unable to distinguish the "elastic"\/ recoil $T_{A}$ from the "inelastic"\/ recoil $T_{A^{*}}$.
  \par
Formulas (\ref{eq:3DM-DiffEventRates-Coh+Inc-SHM+str})
are {\em the most general expressions} for differential event rates
in the case of the {\em nonrelativistic} weak $\chi A$ interaction and  
two different DM distributions.
\par
Let us assume (for illustration) that the correction factors are 
$g_\text{c} = g_\text{i} \simeq 1$
\cite{Bednyakov:2023bbg}
and the nuclear form factors are the same for protons and neutrons  \cite{Bednyakov:2022dmc}
\begin{equation}\label{eq:3DM-gi=gc+Fp=Fn}F(\bm{q})  \equiv F_p(\bm{q}) =F_n(\bm{q})
.\end{equation}
Then after introducing generalized (in)coherent  weak charges of nuclei
\begin{equation}
\label{eq:3DM-DiffEventRates-Qc-Qi}
Q_c(A)\equiv \big[\sum^{}_{f=p,n}\alpha_f A_f \big]^2+\big[\sum^{}_{f=p,n}\delta_f{\Delta A_f}\big]^2
\quad\text{and}\quad 
Q_i(A)\equiv \sum_{f=p,n} A^f \big[\alpha^2_f + 3 \delta^2_f \big]
\end{equation} 
and $T_A$-dependent (in)coherent form factor functions
\begin{equation}
\label{eq:3DM-DiffEventRates-T_A-formfactors}
\Phi(T_A)_c\equiv |F_{}(\bm{q})|^2 \text{~~and~~} \Phi(T_A)_i\equiv 1- |F_{}(\bm{q})|^2 
,\end{equation}
general expression (\ref{eq:3DM-DiffEventRates-Coh+Inc-SHM+str})
for the event counting rates takes the form
\begin{eqnarray} 
\label{eq:3DM-DiffEventRates-Coh+Inc-SHM+str+Fp=Fn-Qci}
\frac{ d R(T_A)_\text{c/i} }{d T_A}&=& R_0(A,\chi)  Q_{\rm c/i}(A)\Phi(T_A)_{\rm c/i} 
C_{\rm c/i}(x_{\min},\eta_{},\epsilon,T_A), \quad \text{where}
\\  C_{\rm c/i}(x_{\min},\eta_{},\epsilon,T_A)&\equiv & \nonumber  
\gamma^{}_{1}S^{}_{1}(x^{}_{\min}, x^{\max}_1) \Theta(T_{A/A^{*},1}^{\max}-T_A) 
+\gamma^{}_{\eta}S^{}_{\eta }(x^{}_{\min}, x^{\max}_\eta) \Theta(T_{A/A^{*},\eta}^{\max}-T_A) 
. \end{eqnarray}
Since $x^{\max}_\eta > x^{\max}_1$ and $T_{A,\eta }^{\max} > T_{A,1}^{\max}$, 
then,  as the recoil energy $T_A$ increases, the role of the second term increases, 
and for $T_{A,1}^{\max}< T_A< T_{A,\eta }^{\max}$, one will have the only
DM stream distribution (the second term).
Futhermore, the incoherent-to-coherent event rate ratio  is as follows:
\begin{eqnarray}
\label{eq:3DM-Diff-EventRates-Ratio}
{\dfrac{d R_\text{i} }{d T_A}}/{\dfrac{d R_\text{c} }{d T_A}}
=\dfrac{Q_i(A) \Phi(T_A)_{\rm i}}{Q_c(A)\Phi(T_A)_{\rm c} }
\dfrac{C_{\rm i}(x_{\min},\eta_{},\epsilon,T_A)}{C_{\rm c}(x_{\min},\eta_{},\epsilon,T_A)}
.\end{eqnarray}
The total event rates (\ref{eq:3DM-EventRates-Coh+Inc}), after integration of the differential  rates 
(\ref{eq:3DM-DiffEventRates-Coh+Inc-SHM+str}),  take the form
\begin{eqnarray} \nonumber
\label{eq:3DM-EventRates-Coh+Inc-SHM+str}
\frac{R(T_A^{\min})_\text{c} }{g_\text{c} R_0(A,\chi)}&=& \gamma_1S_{1}(x^{}_{\min}, x^{\max}_1)
\int_{T_A^{\min}}^{T_{A,1}^{\max}}\!\! dT_A  \Big\{\big[\sum^{}_{f=p,n}\alpha_f A_f F_{f}(\bm{q})\big]^2+\big[\sum^{}_{f=p,n}\delta_f{\Delta A_f}F_{f}(\bm{q})\big]^2\Big\}
\\&& \nonumber
+\gamma_\eta S_{\eta }(x^{}_{\min}, x^{\max}_\eta)  \int_{T_A^{\min}}^{T_{A,\eta }^{\max}}\!\! dT_A 
 \Big\{\big[\sum^{}_{f=p,n}\alpha_f A_f F_{f}(\bm{q})\big]^2+\big[\sum^{}_{f=p,n}\delta_f{\Delta A_f}F_{f}(\bm{q})\big]^2\Big\}
,\\[-5pt] \\ \nonumber
\frac{R(T_A^{\min})_\text{i} }{g_\text{i}R_0(A,\chi)}&=& \gamma_1S_{1}(x^{}_{\min}, x^{\max}_1)
\int_{T_A^{\min}}^{T_{A^*,1}^{\max}}\!\! dT_A  \Big\{\sum_{f=p,n} A^f \big[1-F^2_f(\bm{q})\big]  \big[\alpha^2_f + 3 \delta^2_f \big]\Big\}
\\&&+  \nonumber \gamma_\eta S_{\eta }(x^{}_{\min}, x^{\max}_\eta )
\int_{T_A^{\min}}^{T_{A^*,\eta }^{\max}}\!\! dT_A 
\Big\{\sum_{f=p,n} A^f \big[1-F^2_f(\bm{q})\big]  \big[\alpha^2_f + 3 \delta^2_f \big]\Big\}
. \end{eqnarray}
Since $T_{A^*,\eta}^{\max} < T_{A,\eta}^{\max}$, 
for sufficiently large nuclear excitation energies $\Delta\varepsilon_{mn}$, i.e., when
the condition $T_{\chi,\eta }^{\max} \gg (r+1)\Delta\varepsilon_{mn}$ is  no more satisfied,  the
integration of the incoherent contributions is carried out over a narrower interval of the
nuclear recoil energy.
\par
With allowance for approximations 
(\ref{eq:3DM-gi=gc+Fp=Fn})--(\ref{eq:3DM-DiffEventRates-T_A-formfactors}), formulas 
(\ref{eq:3DM-EventRates-Coh+Inc-SHM+str}) are simplified
\begin{eqnarray} \nonumber
\label{eq:3DM-EventRates-Coh+Inc-SHM+str-Fp=Fn+}
R(T_A^{\min})_\text{c/i} &=&  R_0(A,\chi)
\gamma_1S_{1}(x^{}_{\min}, x^{\max}_1) Q_{\rm c/i}(A) \int_{T_A^{\min}}^{T_{A^{(*)},1}^{\max}}\!\! dT_A \Phi(T_A)_{\rm c/i} 
\\&&+  R_0(A,\chi) \gamma_\eta S_{\eta }(x^{}_{\min}, x^{\max}_\eta) 
Q_{\rm c/i}(A) \int_{T_A^{\min}}^{T_{A^{(*)},\eta}^{\max}}\!\! dT_A  \Phi(T_A)_{\rm c/i}
. \end{eqnarray}
The total incoherent-to-cohernet rate ratio as a function of the threshold $T_A^{\min}$ is
{\small
\begin{eqnarray} \label{eq:3DM-EventRates-Ratio-SHM+str-Fp=Fn+}
\dfrac{R_\text{i} }{R_\text{c} }(T_A^{\min})=\dfrac{Q_i(A)}{Q_c(A)}\!
\dfrac{\gamma_1S_{1}(x^{}_{\min}, x^{\max}_1)\int_{T_A^{\min}}^{T_{A^*,1}^{\max}}\!\! dT_A  \Phi(T_A)_{\rm i}  
+\gamma_\eta S_{\eta }(x^{}_{\min}, x^{\max}_\eta)\int_{T_A^{\min}}^{T_{A^*,\eta }^{\max}}\!\! dT_A  \Phi(T_A)_{\rm i} }
{\gamma_1 S_{1}(x^{}_{\min}, x^{\max}_1)\int_{T_A^{\min}}^{T_{A,1}^{\max}}\!\! dT_A  \Phi(T_A)_{\rm c} 
+\gamma_\eta S_{\eta} (x^{}_{\min}, x^{\max}_\eta) \int_{T_A^{\min}}^{T_{A,\eta }^{\max}}\!\! dT_A  \Phi(T_A)_{\rm c}}.
\qquad  \end{eqnarray}}%
One can notice (from these formulas) that both coherent and incoherent event rates depend on 
the same energy threshold $T_A^{\min}$.
It is the minimal recoil nuclear kinetic energy,
at which a measurable effect (ionization or heat release) is still caused in the detector.
When $T_A < T_A^{\min}$, this effect is  not visible at the background level, 
and no event can be registered.
\par 
This ``logic'' works well for the coherent $\chi A$ interaction, 
when the recoil energy of the nucleus, $T_A$, is the only detectable
"signal"\/ that the  (elastic) interaction occurred.
This energy coincides with the energy, $\Delta E_\chi$, transferred to the nucleus 
by the incident $\chi$ particle (lost by the particle). 
Therefore, if $T_A < T_A^{\min}$,  one always has the coherent rate $R_\text{c}(T_A^{\min})=0$. 
This is exactly the meaning of the value  $T_A^{\min }$.
\par
However, in the case of the incoherent rate $R_\text{i}(T_A^{\min})$, in addition to the recoil nuclear energy, $T_A$, which is registered {\em or even not registered} above the threshold $T_A^{\min}$, the other mandatory  
signature of the incoherent (inelastic) event should be the nuclear deexcitation 
\cite{Bednyakov:2021ppn,Bednyakov:2021bty,Bednyakov:2018mjd}. 
It can manifest itself as emission of $\gamma$-quanta or other (massive) particles with the total  energy $\Delta\varepsilon_{mn}$, which is the difference between the $m$th and $n$th energy levels of the target nucleus.
\par
This deexitation signature enhances  the role of the incoherent rate $R_\text{i}(T_A^{\min})$.
The requirement for the rate absence,  $R_\text{i}(T_A < T_A^{\min})=0$, 
when the nuclear recoil energy is below the threshold $T_A < T_A^{\min}$, is removed, 
since detection of the deexcitation energy $\Delta\varepsilon_{mn}$
is independent of the recoil energy $T_A$.
The set of the deexcitation energies is fixed for the individual nucleus and, as a rule, 
these energies are large enough.
Therefore, any detection of deexcitation energy specific for a given nucleus, $\Delta\varepsilon_{mn}$, 
{\em outside} the recoil energy range $T_A^{\min}< T_A \le 100\div 200 \text{~keV}$
\footnote{Where the expected coherent rate, $R_\text{coh}$,  vanishes in sufficiently massive nuclei.}, 
can be considered as a new possibility for the direct DM detection at the Earth.
To this end, a traditional low-background and low-threshold DM detector should be surrounded by detector(s) aimed at registration of $\gamma$-quanta, whose energy spectrum coincides with the deexcitation spectrum of the target nucleus.
It is obvious that there should be no other external sources of these nuclei excitation
(such as an electron beam or cosmic rays).
Furthermore, the annual variation (modulation) of this deexcitation signal
\cite{Sahu:2020kwh,Sahu:2020cfq}, would be an (extra) important 
signature of the (incoherent) interaction of DM particles with nuclear matter.

\section{Numerical estimates and discussions} \label{4DM-ResultsAndDiscussion}
\subsection*{\em Dark matter distributions and kinematics}
One can make numerical estimates of values from (\ref{eq:3DM-xmin-xmax}).
It is always $x^{}_{\min} >0$.
Further
\begin{eqnarray} \label{eq:3DM-xmin-xmax-numbers}
x^{\max}_1 = \dfrac{v_{\rm esc}}{v_0}+\eta_1 \simeq  \dfrac{550 + 220}{220} = 3.5,
\text{~~and~~}  x^{\max}_\eta = \dfrac{v_{\rm esc}}{v_0}+\eta_1+\eta_{\rm str}  
= 3.5+ \dfrac{300 }{220} \simeq 4.9,\quad 
\end{eqnarray}
where $\eta_1 =\dfrac{v^{}_{\oplus}}{v_0}  \simeq 1$ 
and $\eta_{\rm str} = \xi \epsilon=\dfrac{{v}_{\rm str} }{v_0} \simeq 1.4$.
Then
\footnote{Due to $\text{erf}(z+1)-\text{erf}(z-1)=\dfrac{2}{\sqrt{\pi}} \big[\int_0^{z+1} e^{-t^2}d t- \int_0^{z-1} e^{-t^2}d t\big]=\dfrac{2}{\sqrt{\pi}}\int_{z-1}^{z+1} e^{-t^2}d t $.} 
one can write 
\begin{eqnarray*}
S_{1}(x^{}_{\min}, x^{\max}_1) &=&  \dfrac{\text{erf}[x_{\min}+1]-\text{erf}[x_{\min}-1]}{2}-\dfrac{\text{erf}[x^{\max}_1+1]-\text{erf}[x^{\max}_1-1]}{2}
,\\  S_{\eta}(x^{}_{\min}, x^{\max}_\eta,\epsilon)  &=& \dfrac{\text{erf}[\dfrac{x_{\min}+\eta_{}}{\epsilon}]-\text{erf}[\dfrac{x_{\min}-\eta_{}}{\epsilon}]}{2\eta_{} /\epsilon}- \dfrac{\text{erf}[\dfrac{x^{\max}_\eta+\eta_{}}{\epsilon}]-\text{erf}[\dfrac{x^{\max}_\eta-\eta_{}}{\epsilon}]}{2\eta_{} /\epsilon}
. \end{eqnarray*}
With allowance for the general distribution function $S_{}(x,\eta,\epsilon)$ from 
(\ref{eq:4-S-def-main}), one has
\begin{eqnarray*}
S_{1}(x^{}_{\min}, x^{\max}_1) &=&  S(x_{\min},1,1)-S(x^{\max}_1,1,1)
,\\ S_{\eta}(x^{}_{\min}, x^{\max}_\eta,\epsilon) &=& S_{}(x^{}_{\min},\eta_{},\epsilon) -S_{}(x^{\max}_\eta,\eta_{},\epsilon)
.\end{eqnarray*}
\begin{figure}[h]
\vspace*{-20pt}\hspace*{-30pt}
\includegraphics[scale=0.5]{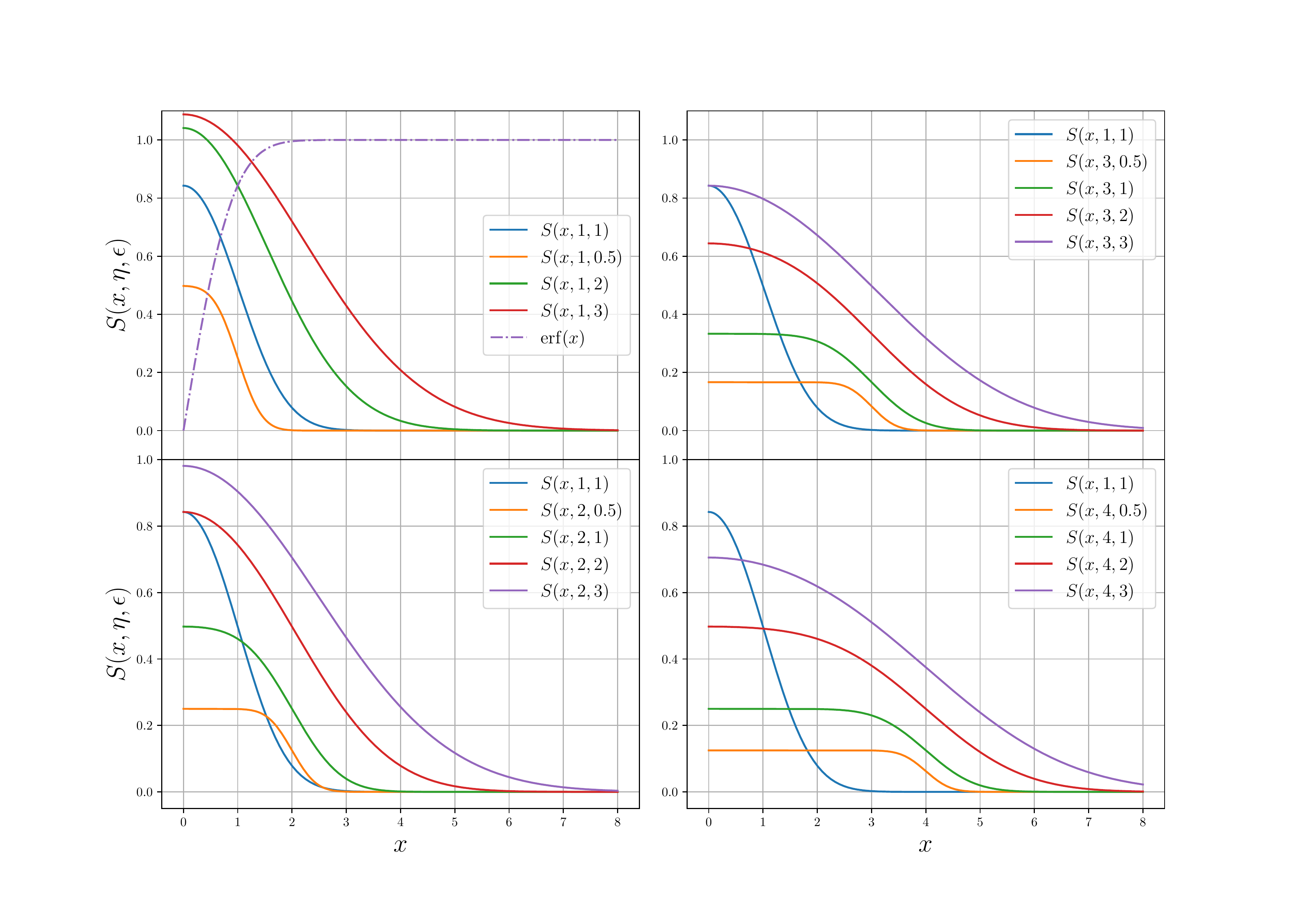}\vspace*{-30pt}
\caption{\small   Dependence of the error function erf$(x)$ 
and function (\ref{eq:4-S-def-main}) on $x$  for $\eta_{}=\{ 1;\ 2;\ 3;\ 4\}$ and $\epsilon = \{0.5; \ 1; \ 2; \ 3\}$.  
Moreover, $x^{\max}_\eta(\eta)=\{ 3.5; \ 4.5; \  5.5; \ 6.5 \}$.} 
\label{fig:4DM-Erf+}
\end{figure}%
Figure~\ref{fig:4DM-Erf+} shows function (\ref{eq:4-S-def-main}) as a
function of  $x$ for $\eta=1;2;3;4$ and $\epsilon = 0.5; 1; 2; 3$.
Since $x^{\max}_1\equiv x^{\max}_{\eta=1}\simeq 2.5+1$ and $x^{\max}_\eta=2.5 + \eta_{}$, 
the maximal values of $x^{\max}_\eta$ as a function of $\eta$ are as follows: 
$\{ 3.5;\ 4.5; \ 5.5; \ 6.5 \}$.
\par
It can be seen that for $x\le 1$ the function (blue curve) $S(x_{\min}\le 1,1,1)\ge 0.5$,
while for $x \simeq 3.0 \le x^{\max}_1$ the function $S_{}(x^{}_{\max}>3,1,1)\le 0.001$.
Therefore, due to the monotonic decrease of this function, one can neglect
its contribution when $x> x^{\max}_1=3.5$, i.e.,  one can accept that
$S_{}(x>x^{\max}_1=3.5,1,1)\simeq 0$, and the following approximation is always valid
$$S_{1}(x^{}_{\min}, x^{\max}_1) = S_{}(x^{}_{\min},1,1) .$$
One can see from Fig.~\ref{fig:4DM-Erf+} that the function $S_{}(x,\eta_{}, \epsilon)$ for $\eta,\epsilon \ne 1$
does not vanish as fast as $S_{}(x>3.5,1,1)$.
Therefore, when $\eta > 1$, we will further use the expression
\begin{eqnarray}
S_{\eta}(x^{}_{\min}, x^{\max}_\eta,\epsilon) &=&   S_{}(x_{\min},\eta_{},\epsilon)-S_{}({2.5} + \eta_{},\eta_{},\epsilon) \equiv \Delta S_{}(x_{\min},\eta_{},\epsilon) 
,\end{eqnarray}
which becomes $S_{}(x_{\min},1,1)$ for $\eta, \epsilon=1$.
 \par
In this approximation formula (\ref{eq:3DM-DiffEventRates-Coh+Inc-SHM+str+Fp=Fn-Qci}) for the differential
event rates has the form
\begin{eqnarray} 
\label{eq:4DM-DiffEventRates-Coh+Inc-SHM+str+Fp=Fn-Qci}
\frac{ d R(T_A)_\text{c/i} }{ d T_A}&=& R_0(A,\chi)   Q_{\rm c/i}(A)\Phi(T_A)_{\rm c/i} C_{\rm c/i}(x_{\min},\eta_{},\epsilon,T_A), \text{~~~where~~~}
\\  \nonumber C_{\rm c/i}(x_{\min},\eta_{},\epsilon,T_A)&=&\Big\{\gamma^{}_{1} S_{}(x_{\min},1,1)
\Theta(T_{A^{}/A^{*},1}^{\max}-T_A)
+\gamma^{}_{\eta} \Delta S_{}(x_{\min},\eta_{},\epsilon)\Theta(T_{A^{}/A^{*},\eta}^{\max}-T_A) \Big\}
. \nonumber  \end{eqnarray} 
The expressions for the maximal nuclear recoil energy used in 
(\ref{eq:4DM-DiffEventRates-Coh+Inc-SHM+str+Fp=Fn-Qci})  
are given below in formula (\ref{eq:4DM-T_A-general}).
Similarly, for the total event rates from (\ref{eq:3DM-EventRates-Coh+Inc-SHM+str-Fp=Fn+}) one has
\begin{eqnarray}
\label{eq:4DM-EventRates-Coh+Inc-SHM+str-Fp=Fn+}
\dfrac{R(T_A^{\min})_\text{c/i}}{R_0(A,\chi) } &=&  \gamma^{}_{1}S(x^{}_{\min},1,1)
Q_{\rm c/i}(A) \int_{T_A^{\min}}^{T_{A^{(*)},1}^{\max}}dT_A \Phi(T_A)_{\rm c/i} \Theta(T_{A^{(*)},1}^{\max}-T_A)
\\&+& \nonumber 
\gamma^{}_{\eta} \Delta S_{}(x_{\min},\eta_{},\epsilon)
Q_{\rm c/i}(A) \int_{T_A^{\min}}^{T_{A^{(*)},\eta}^{\max}}\!\! dT_A  \Phi(T_A)_{\rm c/i}
\Theta(T_{A^{(*)},\eta}^{\max}-T_A)
. \end{eqnarray}
This event rate depends on the detection threshold  $T_A^{\min}$ 
through the relation $x^{}_{\min} = \Big[\dfrac{T_A^{\min}}{T_A^{0}}\Big]^{1/2}$.
According to (\ref{eq:3DM-xmin-xmax}),
\footnote{Note that $ \dfrac{m_\chi c^2}{2} \Big(\dfrac{v_0}{c}\Big)^2 
= \dfrac12 \Big[\dfrac{m_\chi}{1 \text{~GeV}}\Big] \Big[\dfrac{220\text{~km/sec}}{300 \cdot 10^3\text{~km/sec}} \Big]^2 10^6 \text{keV}\simeq  0.269 \Big[\dfrac{r m_A}{1 \text{~GeV}}\Big]  \text{keV}$.
\label{footnote:T0chi}} 
one has 
\begin{eqnarray*}
T_A^{0} &\equiv& \mu_A\dfrac{m_\chi  v_0^2}{2} 
 =  \dfrac{4 r}{(1+r)^2} \dfrac{m_\chi c^2}{2} \Big(\dfrac{v_0}{c}\Big)^2
 \simeq   \dfrac{4 r^2 0.269 }{(1+r)^2} \Big[\dfrac{m_A}{1 \text{~GeV}}\Big]  \text{keV}
.\end{eqnarray*}
Therefore, the dimensionless minimal velocity of the DM particle (relative to the nucleus at rest) is
\begin{eqnarray}\label{eq:4DM-x-min}
x^{}_{\min} &\simeq& \sqrt{\dfrac{\Big[ \dfrac{T_A^{\min}}{1\, \text{keV}}\Big] }{\dfrac{1.076 r^2 }{(1+r)^2} \Big[\dfrac{m_A}{1 \text{~GeV}}\Big] } } \simeq \dfrac{(1+r)}{1.037 r }\sqrt{
\dfrac{\Big[ \dfrac{T_A^{\min}}{1\, \text{keV}}\Big] }{\Big[\dfrac{m_A}{1 \text{~GeV}}\Big] } } 
.\end{eqnarray} 
\begin{figure}[h!]\centering \vspace*{-30pt}
\includegraphics[width=0.6\linewidth]{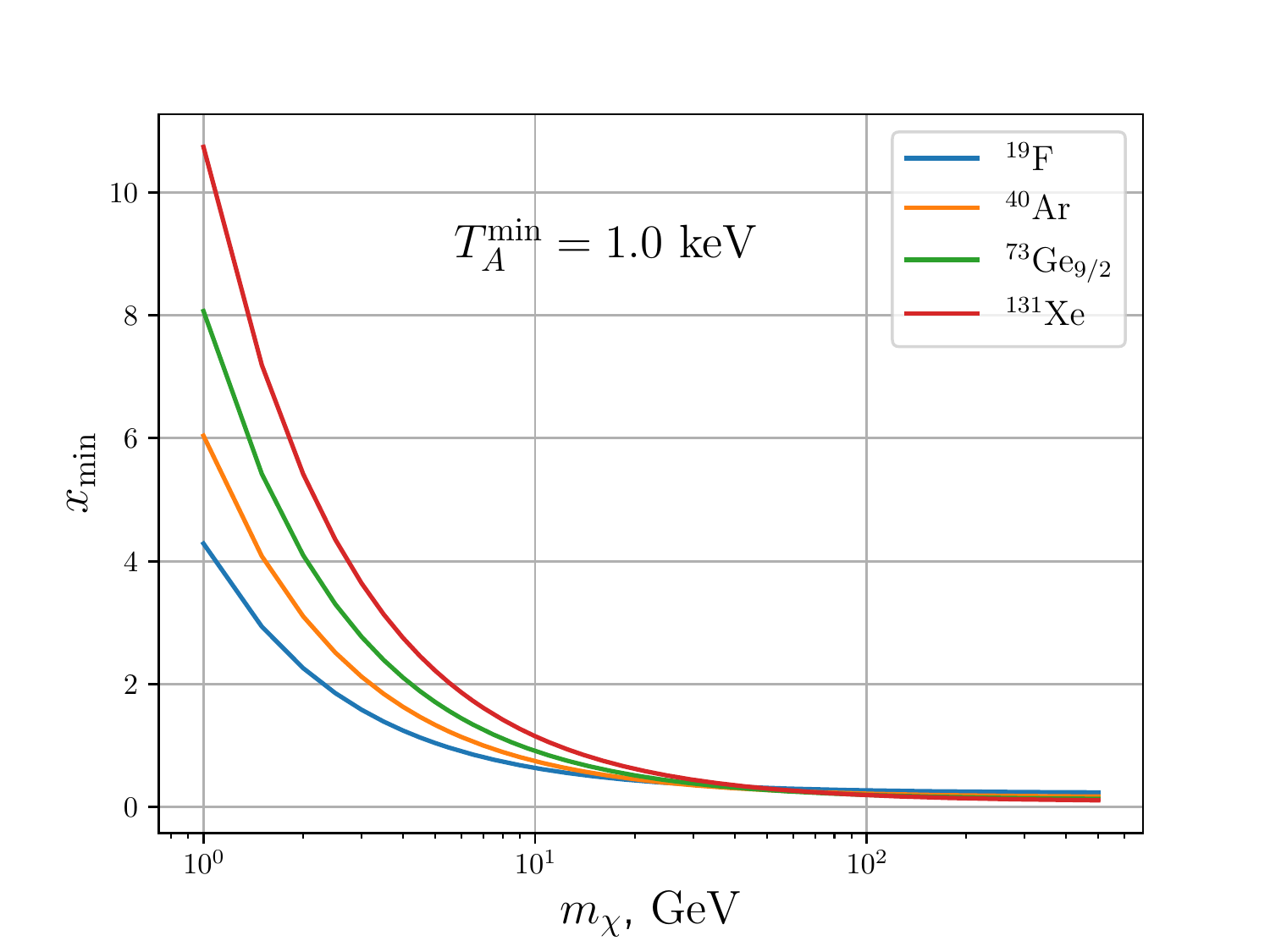} 
 \vspace*{-10pt}
 \caption{\small Dependence of $x_{\min}$ on $m_A$ and $m_\chi$ at the threshold $T^{\min}_A=1\, $keV.
  For fixed $m_A$ and $m_\chi$, the value of $x_{\min}$ is proportional to the square root of $T^{\min}_A$ according to (\ref{eq:4DM-x-min}).}   \label{fig:4DM-Xmin-vs-T_A-mchi}
 \end{figure}
Comparing Fig.~\ref{fig:4DM-Xmin-vs-T_A-mchi} with Fig.~\ref{fig:4DM-Erf+}, 
one can conclude that for very small $m_\chi$ and sufficiently large $m_A$
the function $S_{}(x,\eta,\epsilon)$ from
 (\ref{eq:4-S-def-main}) vanishes (because of the large value of $x=x_{\min}\ge 8$)
 for almost all reasonable values of parameters.
The latter means that both coherent and incoherent differential event rates vanish as well 
according to formulas (\ref{eq:4DM-DiffEventRates-Coh+Inc-SHM+str+Fp=Fn-Qci}).
\par
Expression (\ref{eq:4DM-EventRates-Coh+Inc-SHM+str-Fp=Fn+})
for the total event rates also depends   on the maximal dimensionless velocities 
of the DM distributions, $x^{\max}_1\simeq 3.5$  and $x^{\max}_\eta= 2.5 + \eta_{}$ 
(i.e., on the flow velocity $\eta_{}$), via relations (\ref{eq:3DM-TAmax+TAmax}) and (\ref{eq:3DM-Inelastic-TAmax}).
\par
Therefore,  to calculate the event counting rates with formulas
(\ref{eq:4DM-DiffEventRates-Coh+Inc-SHM+str+Fp=Fn-Qci}) and
(\ref{eq:4DM-EventRates-Coh+Inc-SHM+str-Fp=Fn+}),  one has 
to define  the following values:
$$ m_\chi, \ n_\chi =\rho/m_\chi, \ v_0^2,  \quad \gamma^{}_{1}, \ \gamma^{}_{\eta}, \ \eta_{}, \ \epsilon,  
\text{~~as well as~~}   m_A, N_A , T_A^{\min}, \Delta\varepsilon_{mn}, $$
and use the common  parameters 
$G_{\text{F}}=1.166 \times 10^{-5}\mbox{GeV}^{-2}$ and 
$m = 0.938  \mbox{~GeV} \simeq 1 \mbox{~GeV}.$
\par
To estimate the $T_A$-dependent coherent and incoherent
form-factor functions (\ref{eq:3DM-DiffEventRates-T_A-formfactors}),
the Helm nuclear form factor parameterization \cite{PhysRev.104.1466} is used
$$F(\bm{q})  = 3 \dfrac{j_1(qR_0)}{qR_0}e^{-\dfrac{(s q)^2}{2}},  
\quad  j_1(x) = \dfrac{\sin x}{x^2} - \dfrac{\cos x}{x},  \ s = 0.9 \text{~fm}, \  R_0 = 1.14 A^{1/3} \text{~fm}
.$$
According to the definition 
(\ref{eq:42chiA-CrossSection-ScalarProducts-ChiEta-All-WeakGeneralCurrent-definition})
of the weak current scalar product 
\cite{Bednyakov:2021pgs}, 
the effective coupling constants of the form
$\alpha_f = \chi_V h^f_V,\   \beta_f = \chi_V h^f_A, \  \gamma_f = \chi_A  h^f_V 
,\  \delta_f = \chi_A h^f_A$ are free parameters.
In the SM, neutrinos have $\chi_V = - \chi_A = 1$, and nucleons have 
$$ h_V^p=g_V^p =  \frac12 -2 \sin^2\theta_W, \quad h_V^n=g_V^n = - \frac12; 
\quad h_A^p = g_A^p =   \frac{g_A}{2}, \quad h_A^n = g_A^n =  - \frac{g_ A}{2}. 
$$
Then the effective parameters in the SM are the following:
$$\alpha_f = \chi_V h^f_V  = + g_V^f, \quad \delta_f = \chi_A h^f_A = + g_A^f,
\quad \gamma_f = \chi_A  h^f_V = - g_V^f, \quad  \beta_f = \chi_V h^f_A = - g_A^f .$$
With allowance for $\sin^2\theta_W = 0.239$ and $g_A = 1.27$, one has the following 
numerical values:
\begin{eqnarray}\nonumber
 \label{eq:4DM-Results-SM-alpha+delta}
\alpha_p  &=&  + \frac12 -2 \sin^2\theta_W  = - \gamma_p  \simeq 0.02  
 ,\quad \delta_p  =  \frac{g_A}{2} = - \beta_p  \simeq 0.64
; \\  \alpha_n  &=& - \frac12  = - \gamma_n  = - 0.5
,\qquad \qquad\qquad \delta_n  =  - \frac{g_ A}{2} = - \beta_n  \simeq  - 0.64 
. \end{eqnarray}
These parameters determine the generalized weak nuclear charges from 
(\ref{eq:3DM-DiffEventRates-Qc-Qi})
\begin{eqnarray}\nonumber
 \label{eq:4DM-Q_c-and-Q_i-def}
Q_c(A)&=& \big[\sum^{}_{f=p,n}\alpha_f A_f \big]^2+\big[\sum^{}_{f=p,n}\delta_f{\Delta A_f}\big]^2
=\big[\alpha_p A_p+\alpha_n A_n \big]^2+ \big[\delta_p {\Delta A_p}+\delta_n {\Delta A_n} \big]^2
,\\Q_i(A)&=& \sum_{f=p,n} A^f \big[\alpha^2_f + 3 \delta^2_f \big] =A^p \big[\alpha^2_p+ 3 \delta^2_p \big] + A^n \big[\alpha^2_n + 3 \delta^2_n \big]
.\end{eqnarray}
In calculation of  the cross sections and their ratios in the rest frame of the nucleus, 
one varies {\em only}\/ the nuclear recoil energy $T_A$ in the interval from 
$T_A^{\min}= \epsilon_A^{\text{threshold}}$ to the maximal $T_A^{\max}$,  
which is determined by the DM distribution from condition (\ref{eq:3DM-Inelastic-TAmax}).
With allowance for (\ref{eq:3DM-gi=gc+Fp=Fn})--(\ref{eq:3DM-DiffEventRates-T_A-formfactors}) and "cosmic factors"\/
\begin{eqnarray} \label{eq:4DM-Cosmic-CIhC-CCoh}
C_{\rm i/c}(x_{\min},\eta_{},\epsilon,T_A)\!=\!
\gamma^{}_1 S_{}(x_{\min},1,1) \Theta(T_{A^*/A,1}^{\max}-T_A)  + \gamma^{}_{\eta} \Delta S_{}(x_{\min},\eta_{},\epsilon)\Theta(T_{A^*/A,\eta}^{\max}-T_A)
, \quad 
\end{eqnarray}
ratio (\ref{eq:3DM-Diff-EventRates-Ratio}) between the differential incoherent and coherent rates becomes
\begin{eqnarray}\label{eq:4DM-DiffEventRates-Ratios}
R_{\rm I}(A,T_A)&=& {\dfrac{d R_\text{i} }{d T_A}}/{\dfrac{d R_\text{c} }{d T_A}}
=R_{\rm I}(A) R_A(T_A) R_C(x_{\min},\eta_{},\epsilon,T_A), \quad\text{where}
\\  \nonumber &&R_{\rm I}(A)=\dfrac{A Q_{\rm i}(A)}{Q_{\rm c}(A)},\quad 
R_A(T_A) = \dfrac{1- |F_{}(T_A)|^2}{A|F_{}(T_A)|^2} \quad\text{and}
\\\label{eq:4DM-DiffEventRates-Ratios-R_C} &&R_C(x_{\min},\eta_{},\epsilon,T_A)=\dfrac{C_{\rm i}(x_{\min},\eta_{},\epsilon,T_A)}{C_{\rm c}(x_{\min},\eta_{},\epsilon,T_A)}
.\end{eqnarray}
With (\ref{eq:3DM-gi=gc+Fp=Fn})--(\ref{eq:3DM-DiffEventRates-T_A-formfactors}),  
the ratio of the incoherent total event rate to the coherent one  (\ref{eq:3DM-EventRates-Ratio-SHM+str-Fp=Fn+})
takes the form
\begin{eqnarray} \label{eq:4DM-EventRates-Ratios}
\dfrac{R_\text{i} }{R_\text{c} }(T_A^{\min})&=&\dfrac{Q_i(A)}{Q_c(A)}\Theta(T_A-T_A^{\min}) 
\times \\&\times&   \dfrac{\gamma^{}_{1}S(x^{}_{\min},1,1)\int_{T_{A}^{\min}}^{T_{A^{*},1}^{\max}} dT_A \Phi_{\rm i}(T_A)  +\gamma^{}_{\eta} \Delta S_{}(x_{\min},\eta_{},\epsilon)\int_{T_A^{\min}}^{T_{A^*,\eta}^{\max}} dT_A  \Phi_{\rm i}(T_A)}
{\gamma^{}_{1} S(x^{}_{\min},1,1)\int_{T_A^{\min}}^{T_{A,1}^{\max}} dT_A \Phi_{\rm c}(T_A) + \gamma^{}_{\eta} \Delta S_{}(x_{\min},\eta_{},\epsilon)\int_{T_A^{\min}}^{T_{A,\eta}^{\max}} dT_A  \Phi_{\rm c}(T_A)}
\nonumber
.\end{eqnarray} 
The expression $R_A(T_A)$ from (\ref{eq:4DM-DiffEventRates-Ratios}) does not depend on the interaction, 
it completely defines the dependence of $R_{\rm I}(A,T_A)$ from (\ref{eq:4DM-DiffEventRates-Ratios}) on $T_A$. 
However, in contrast to the {\em nonrelativistic}\/ differential 
incoherent-to-coherent cross section ratio \cite{Bednyakov:2022dmc}, which has the form
\begin{eqnarray}\label{eq:4DM-Results-Inc2CohRatio-A} 
\dfrac{\dfrac{d\sigma^{\text{total}}_\text{inc}}{d T_A}(\chi A\to \chi A^{*})}{\dfrac{d\sigma^{\text{total}}_\text{coh}}{d T_A}(\chi A\to \chi A^{})}= R_I(A) R_A(T_A) , \qquad
 \end{eqnarray}
the ratio of event rates from (\ref{eq:4DM-DiffEventRates-Ratios}) has the meaning only   
for the recoil energies in the interval
\begin{equation}\label{eq:4DM-working-T_A-interval}
T_A^{\min} \le T_A\le T_{A^{*},\eta}^{\max}(r,\Delta\varepsilon_{mn})
.\end{equation} 
If  $T_A >T_{A^{*},\eta}^{\max}(r,\Delta\varepsilon_{mn})$, 
both differential and total inelastic event rates vanish (the inelastic process is not possible).
If $T_A < T_A^{\min}$, no recoil energy can be measured.
Taking into account that with $r$ from (\ref{eq:3DM-Inelastic-TAmax}) 
the "elastic"\/ nuclear recoil energy has the form
$$T_{A,\eta}^{\max}(r,0) \equiv T_{A^{*},\eta}^{\max}(r,\Delta\varepsilon_{mn}=0) =
 \dfrac{4 T_{\eta}^{\max}   r}{(r+1)^2} ,$$
the right side of condition (\ref{eq:4DM-working-T_A-interval})  can be expressed as follows:
\begin{eqnarray} \label{eq:4DM-T_A-general}
T_{A^{*},\eta}^{\max}(r,\Delta\varepsilon_{mn})  &=&
T_{A,\eta}^{\max}(r,0)\dfrac{1-\omega/2+\sqrt{1- \omega } }{2} \le T_{A,\eta}^{\max}(r,0),
\\\text{where~~} \omega&\equiv& \dfrac{( r  +1) \Delta\varepsilon_{mn}}{ T_{\eta}^{\max}} 
, \text{~~and~according to footnote \ref{footnote:T0chi}}, \nonumber
\\  T_{\eta}^{\max} &=& ( x^{\max}_\eta)^2 T^{0}_\chi   
 \simeq  0.27  \big(2.5+\eta_{}\big)^2 \Big[\dfrac{r m_A}{1 \text{~GeV}}\Big]  \text{keV}, 
 \text{~~and~~}  \eta\ge 1
. \nonumber \end{eqnarray}
One can see that the maximal recoil energy in the elastic collision is 
always larger than the same value in the inelastic collision.
Formula (\ref{eq:4DM-T_A-general}) has the meaning, i.e., 
the value $T_{A^{*},\eta}^{\max}(r,\Delta\varepsilon_{mn})$ is a real (non-negative) number, 
and only under condition (\ref{eq:3DM-Inelastic-condition}) is written as
\footnote{If $\omega> 1$, the incoherent process with these $m_A$, $m_\chi$, $\eta_{}$ and $\Delta\varepsilon_{mn}$ is impossible.} 
\begin{eqnarray}\label{eq:4DM-Inelastic-condition}
\omega&=& (r  +1) \dfrac{\Delta\varepsilon_{mn}}{ T^{\max}_{\eta}} \le 1
\text{~~~or~~~} 0.27  \big(2.5+\eta_{}\big)^2 \ge  \Big(1+\dfrac{1}{r}\Big)
\dfrac{\Big[\dfrac{\Delta\varepsilon_{mn}}{1 \text{~keV}}\Big] }{\Big[\dfrac{ m_A}{1 \text{~GeV}}\Big] } 
.\end{eqnarray}
The case $\omega= 1$ means that {\em only}\/ the maximal (for given DM distribution) 
energy of the DM particle, $T_{\eta}^{\max}$, is enough to excite the target nucleus 
at the level with energy $\Delta\varepsilon_{mn}$, 
provided the mass ratio of the DM particle and the nucleus is $r$.
Moreover, this maximal energy is directly connected with the nuclear excitation energy, 
$T_{\eta}^{\max}=(r+1)\Delta\varepsilon_{mn}$.
In other words,  if the nucleus is "so lucky"\/ to have the excitation level just with the energy 
$\Delta\varepsilon_{mn}=T_{\eta}^{\max}/( r +1)$, 
the excitation of the nucleus is possible exactly to this "unique level"\/.
In this case, the kinetic energy of the nuclear recoil, which inevitably accompanies the excitation
 of this {\em particular}\/ nuclear  level is, according to (\ref{eq:4DM-T_A-general}),
\begin{eqnarray}
\label{eq:4DM-TAmax-at-omega=1}
T_{A^{*},\eta}^{,\omega=1}(r) &=& \dfrac{T_{A,\eta}^{\max}(r,0)}{4}=
 \Big[\dfrac{r}{r+1}\Big]^2   0.27  \big(2.5+\eta_{}\big)^2 \Big[\dfrac{m_A}{1 \text{~GeV}}\Big]  \text{keV}
.\end{eqnarray}
From formula (\ref{eq:4DM-TAmax-at-omega=1}) 
one has the "threshold"\/ value of $r$ (and $m_\chi$), at which this "unique excitation" takes place, 
in the following form:
\begin{eqnarray}
\label{eq:4DM-mchimin-at-omega=1}
\dfrac{1}{r} =\sqrt{\dfrac{\Big[\dfrac{m_A}{1 \text{~GeV}}\Big]}{\Big[\dfrac{T_{A^{*},\eta_{}}^{\omega=1}}{1\text{~keV}}\Big]} } \sqrt{0.27}\big(2.5+\eta_{}\big)-1
 \text{~~or~~}
m_\chi^{\min}(T_A^*) = \dfrac{m_A}{\dfrac{\sqrt{ 0.27 }(2.5+\eta_{})}{\sqrt{\Big[\dfrac{T_{A^{*},\eta}^{\omega=1}}{1 \text{~keV}}\Big]  \Big[\dfrac{m_A}{1 \text{~GeV}}\Big]^{-1}}} -1} 
.\quad\end{eqnarray}
Vanishing of the denominator in (\ref{eq:4DM-mchimin-at-omega=1})
means that the excitation to the nuclear level  with $\Delta\varepsilon_{mn}= T^{\max}_{\eta} \dfrac{1}{ r +1}$ accompanied by the recoil energy of the excited nucleus $T_{A^{*},\eta_{} }^{\omega=1}$ from (\ref{eq:4DM-TAmax-at-omega=1}) is possible only for $m_\chi \simeq \infty$. 
Clearly, this process does not occur for any $m_\chi$.
\par
\begin{figure}[h!] 
\centering \vspace*{-35pt}
\includegraphics[width=0.7\linewidth]{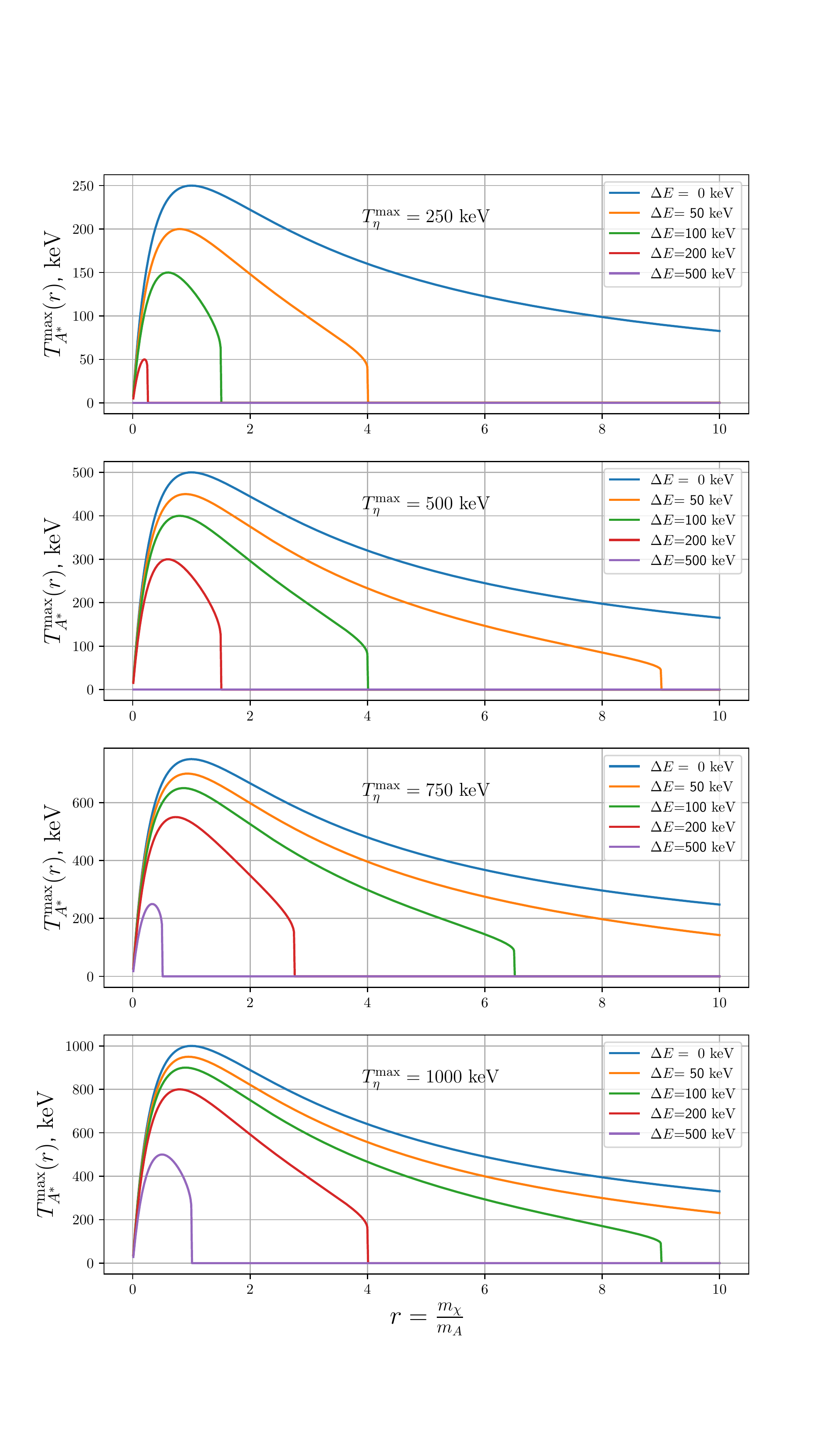}
\vspace*{-50pt}
\caption{\small Theoretical dependence of the maximal recoil energy of the excited nucleus, 
$T_{A^{*},\eta}^{\max}$, (expression (\ref{eq:4DM-T_A-general})) on the mass ratio of the DM particle and the nucleus, $r$,  for four "trial"\/ maximal energies of the DM particles, $T^{\max}_\eta$,  
and five excitation energies  $\Delta E$.}
\label{fig:4DM-4-TEAmax-vs-r}
\end{figure}

Figure~\ref{fig:4DM-4-TEAmax-vs-r} shows the behavior of the maximal recoil energy of 
the excited nucleus, $T_{A^{*},\eta}^{\max}$, as a function of the mass ratio of the DM particle and the nucleus, $r$, for some typical maximal energies of the DM particles, $T^{\max}_\eta$,  and some nuclear excitation energies $\Delta E$.
One can see that the maximal recoil energy of excited nuclei, $T_{A^{*},\eta}^{\max}$, is achieved
when the masses of the $\chi$ particle and the nucleus coincide ($r=1$).
As the ratio $r$ increases, the maximal recoil energy of the excited nucleus rather quickly decreases. 
The sharp drop of some curves down to zero occurs when the increasing
$\omega = \dfrac{( r +1) \Delta\varepsilon_{mn}}{ T_{\eta}^{\max}}$ from
(\ref{eq:4DM-Inelastic-condition})  reaches one.
For example, in the top plot, this happens at $T^{\max}_\eta=250$~keV,
$\Delta E= 50$~keV and $r=4$.
There are noticeable regions of the parameters $r$ and $T^{\max}_\eta$ (for example, at sufficiently high excitation energies of nuclear levels, $\Delta E\ge 200$~keV), where $\omega >1$.
In these regions, $T_{A^{*},\eta}^{\max}$ is set equal to zero,
which reflects the impossibility of the nuclear excitation.
\par
\begin{figure}[h!] 
\centering \vspace*{-50pt}
\includegraphics[width=0.75\linewidth]{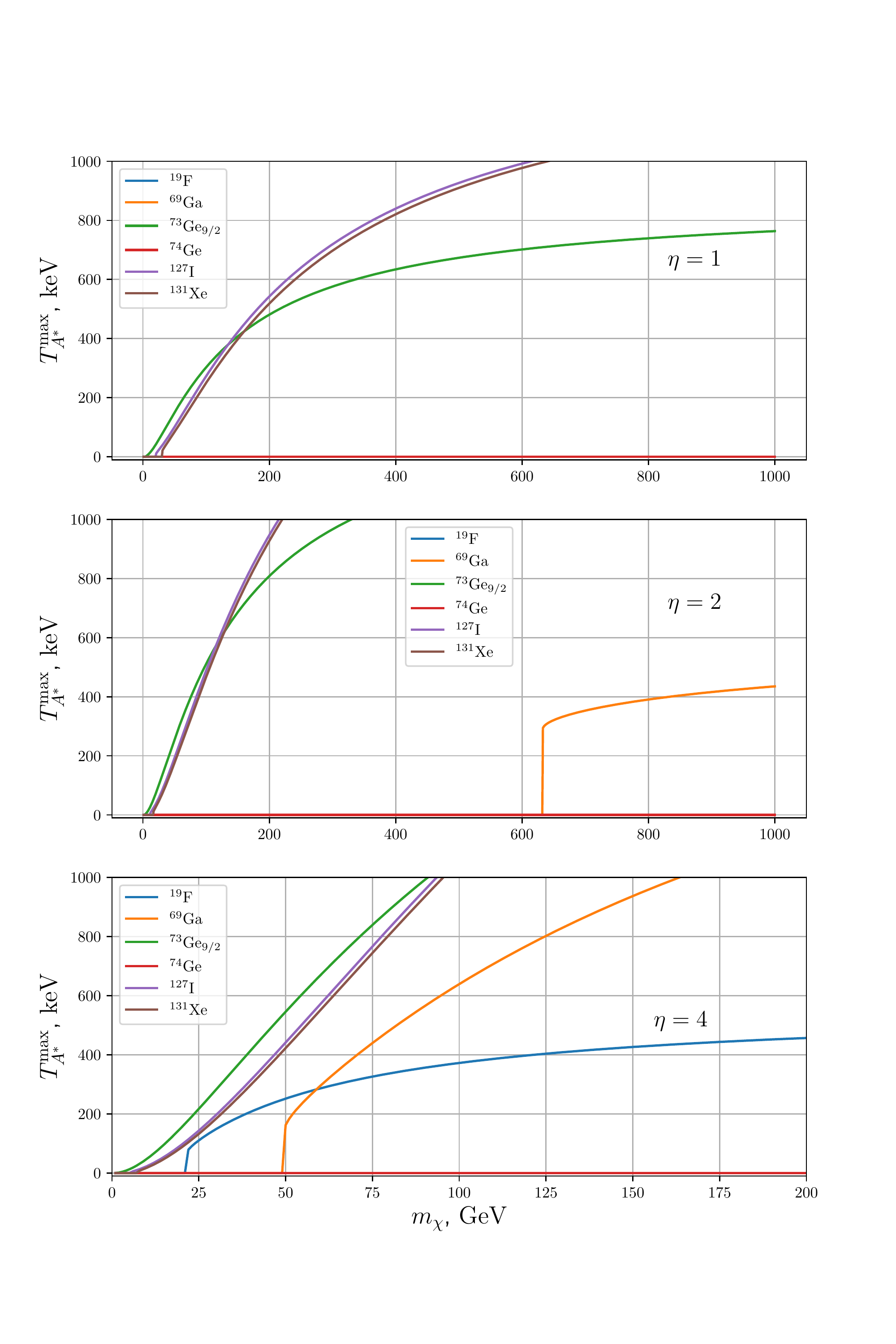} 
\vspace*{-50pt}
\caption{\small Dependence of the maximal recoil energy of the excited nuclei, 
$T_{A^{*},\eta}^{\max}(r,\Delta\varepsilon_{mn})$, on $m_\chi$ for typical targets and the DM-flow velocities $\eta=1;2;4$. 
The value $\eta=1$ corresponds to the Standard Halo Model. }\label{fig:4DM-TAEmaxes}
\end{figure} 
In Fig.~\ref{fig:4DM-TAEmaxes}, the maximal recoil energy of the excited nucleus  (\ref{eq:4DM-T_A-general})
is given as a function of the mass of the 
$\chi$ particle for some dimensionless velocities of the DM flow $\eta$ and some typical target nuclei 
(with specific values of their masses and excitation energies, 
given in Table~\ref{tab:50chiA-Ratios-RIad-vs-A} below).
One can see that in the case of the Standard Halo Model ($\eta=1$)
the inelastic excitation of the quite heavy nuclei $^{127}$I and $^{131}$Xe as well as $^{73}$Ge
(due to their low excitation energies) is allowed for any $m_\chi$.
The recoil energies of these (excited) nuclei reach the MeV scale.
For lighter nuclei (fluorine, gallium and $^{74}$Ge), despite the difference in their energy excitation, 
the DM particles from the SHM distribution are unable to cause the inelastic process.
With $\eta=2$, the situation changes insignificantly.
The recoil energies for heavy nuclei only increase, 
and the inelastic channel in gallium opens at sufficiently large $m_\chi$.
Finally, in the case of a strong DM flow ($\eta=4$), 
the energy of DM particles is quite enough to excite all the nuclei
with noticeable recoil energies, which increase rapidly with $m_\chi$.
A "small exception"\/ is fluorine, where the recoil energy increase is not so "rapid"\/
due to the small mass of the fluorine nucleus, 
and $^{74}$Ge, for which the inelastic channel opens with 
large DM masses ($m_\chi\ge 200$ GeV) due to the 
very high excitation energy (596 keV).
\par
"Conversion"\/ of relation (\ref{eq:4DM-T_A-general}) gives 
$T_{A^*}$-dependence of the minimal DM particle mass, $m_\chi^{\min}$,
starting from which the excitation of the nucleus becomes possible together with 
the recoil energy $T_{A^*}$ of the nucleus.
This dependence has the form 
\begin{eqnarray}\label{eq:4DM-r-via-T_A}
m^{\min}_\chi (T_{A^*}, A,\eta) &=&   \dfrac{m_A T_A }{ \sqrt{2 m_A T_A   ( x^{\max}_\eta)^2 v^2_0 }- (T_A+\Delta\varepsilon_{mn} )}
=\\&=& \dfrac{m_A }{2(2.5+\eta_{})\sqrt{0.27 } \sqrt{\Big[\dfrac{m_A}{1 \text{~GeV}}\Big]
\Big[\dfrac{T_{A^{*}}}{1 \text{~keV}}\Big]^{-1} }-1 - \dfrac{\Delta\varepsilon_{mn}}{T_{A^{*}}}}
.\nonumber \end{eqnarray}
\begin{figure}[h!] 
\vspace*{-20pt} \centering
\includegraphics[width=0.95\linewidth]{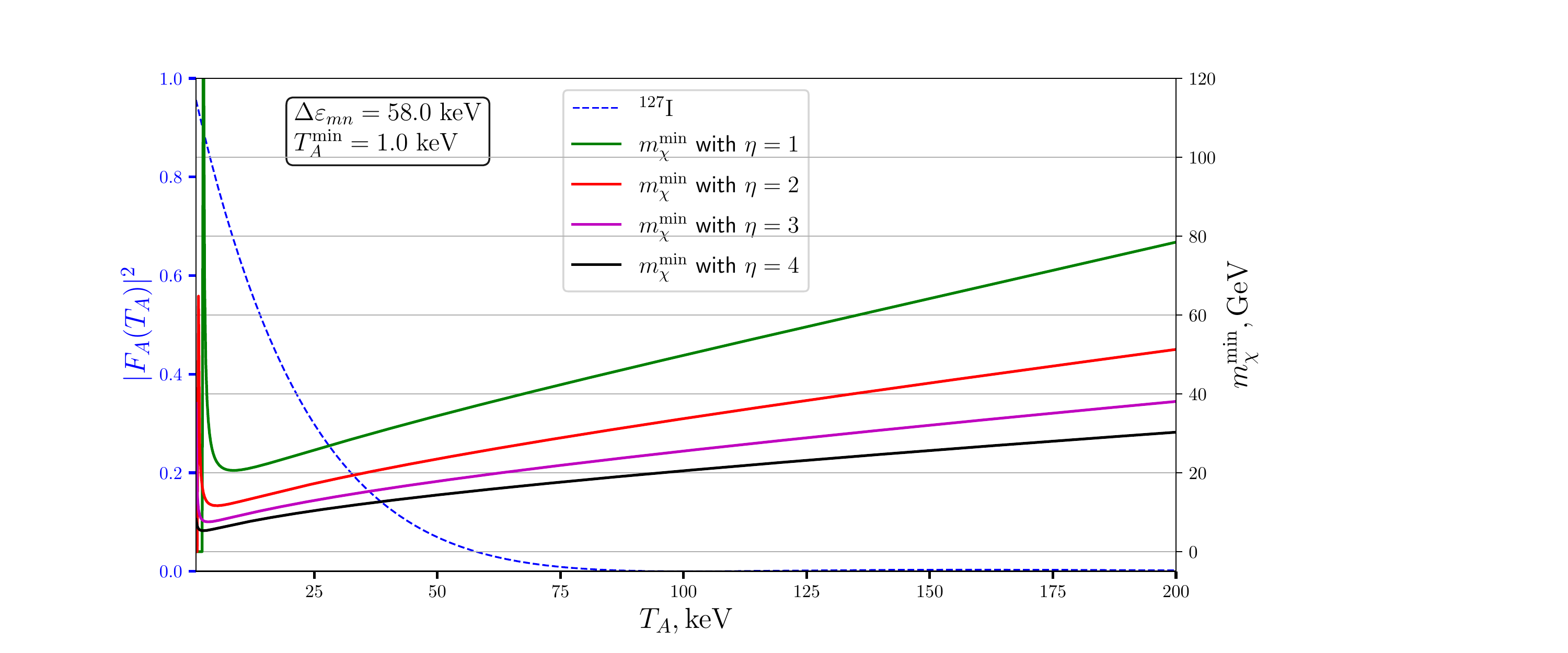}\vspace*{-10pt}
\includegraphics[width=0.95\linewidth]{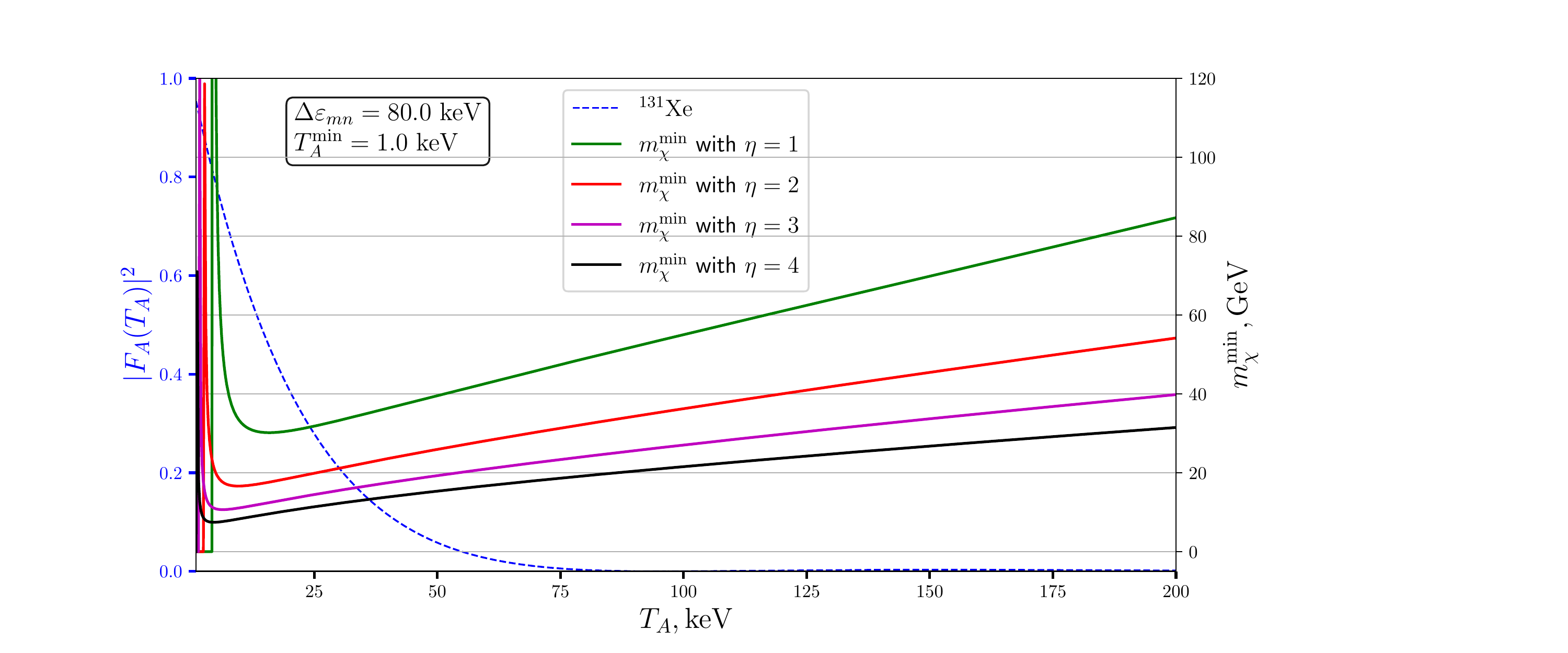}\vspace*{-10pt}
\includegraphics[width=0.95\linewidth]{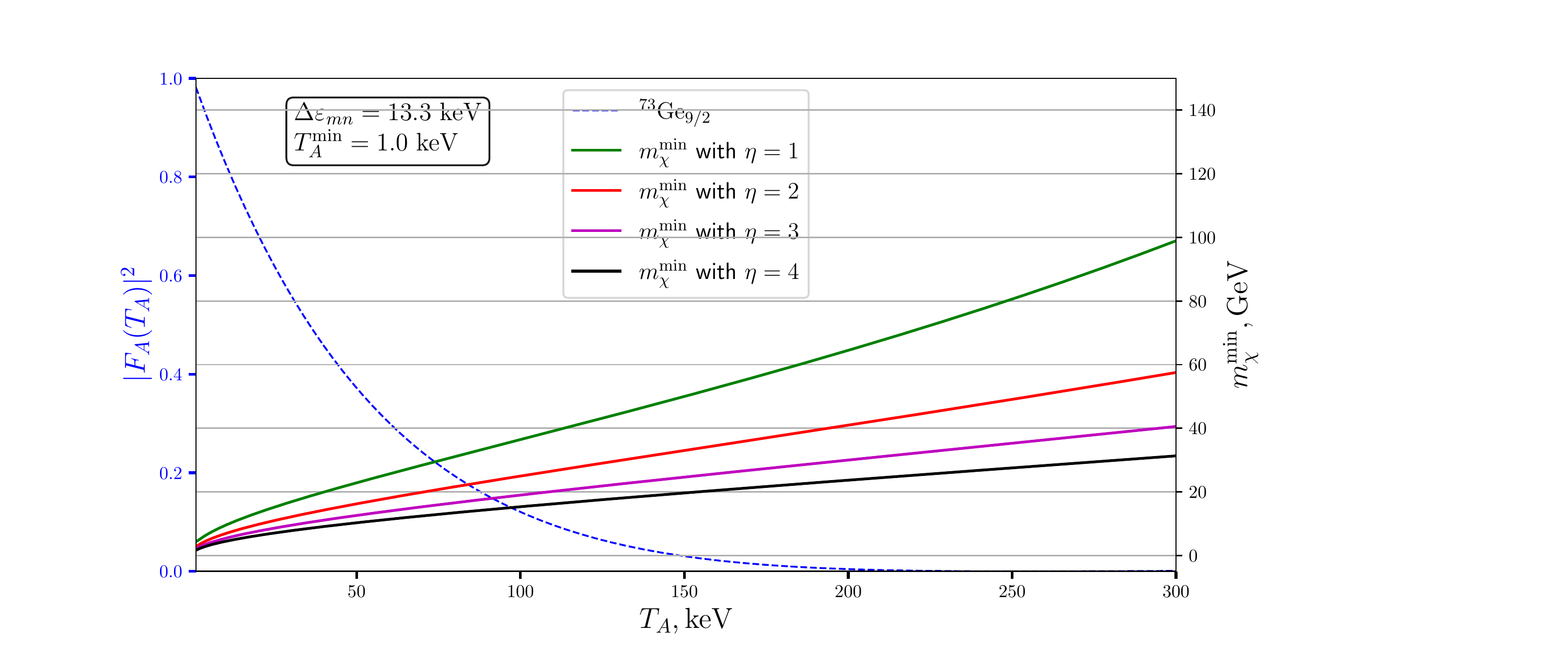}\vspace*{-10pt}
\caption{\small The minimal masses $m_\chi^{\min}$ are given (right y-axis)  
as functions of the recoil energy, $T_{A^{*}}$, of the (excited) nuclei $^{131}$Xe, $^{127}$I, and $^{73}$Ge
for the DM distributions with the maximal flow velocities $\eta_{}=1$ (green curve), 
$\eta_{}=2$ (red) $\eta_{}=3$ (purple) and $\eta_{}=4$ (dark curve).
The left vertical y-axis illustrates the behavior of the square form factor of  the nucleus.}
\label{fig:4DM-F2+mchi_min-vs-T_A-eta-IXeGe}
\end{figure}
In Fig.~\ref{fig:4DM-F2+mchi_min-vs-T_A-eta-IXeGe} and 
Fig.~\ref{fig:4DM-F2+mchi_min-vs-T_A-eta-F19-Ga69} the
 $m^{\min}_\chi$-function (\ref{eq:4DM-r-via-T_A}) 
is shown for  the DM distributions with maximal velocities  
$\eta_{}=1$ (green curve), $\eta_{}=2$ (red curve), $\eta_{}=3$ (purple curve) 
and $\eta_{}=4$ (dark curve). 
\begin{figure}[h!] 
\vspace*{-10pt}
\includegraphics[width=0.95\linewidth]{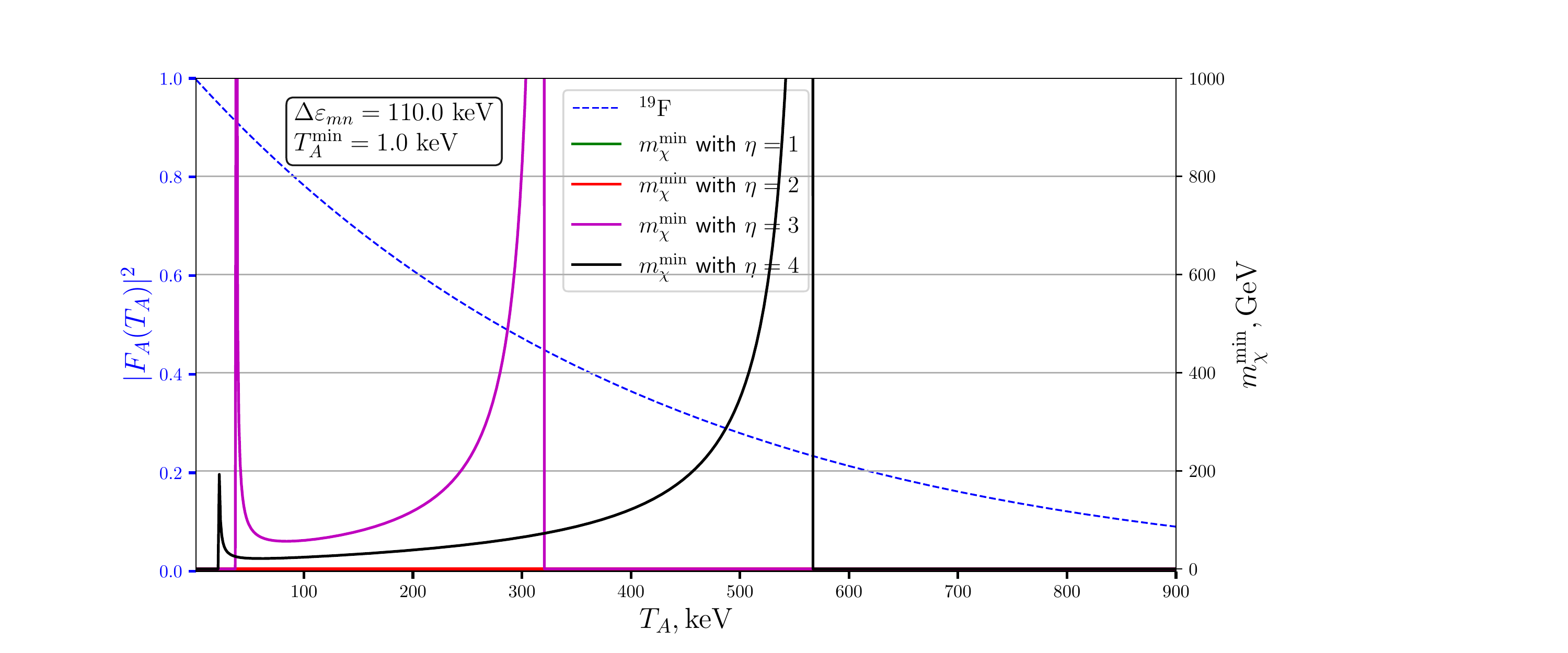}\vspace*{-10pt}
\includegraphics[width=0.95\linewidth]{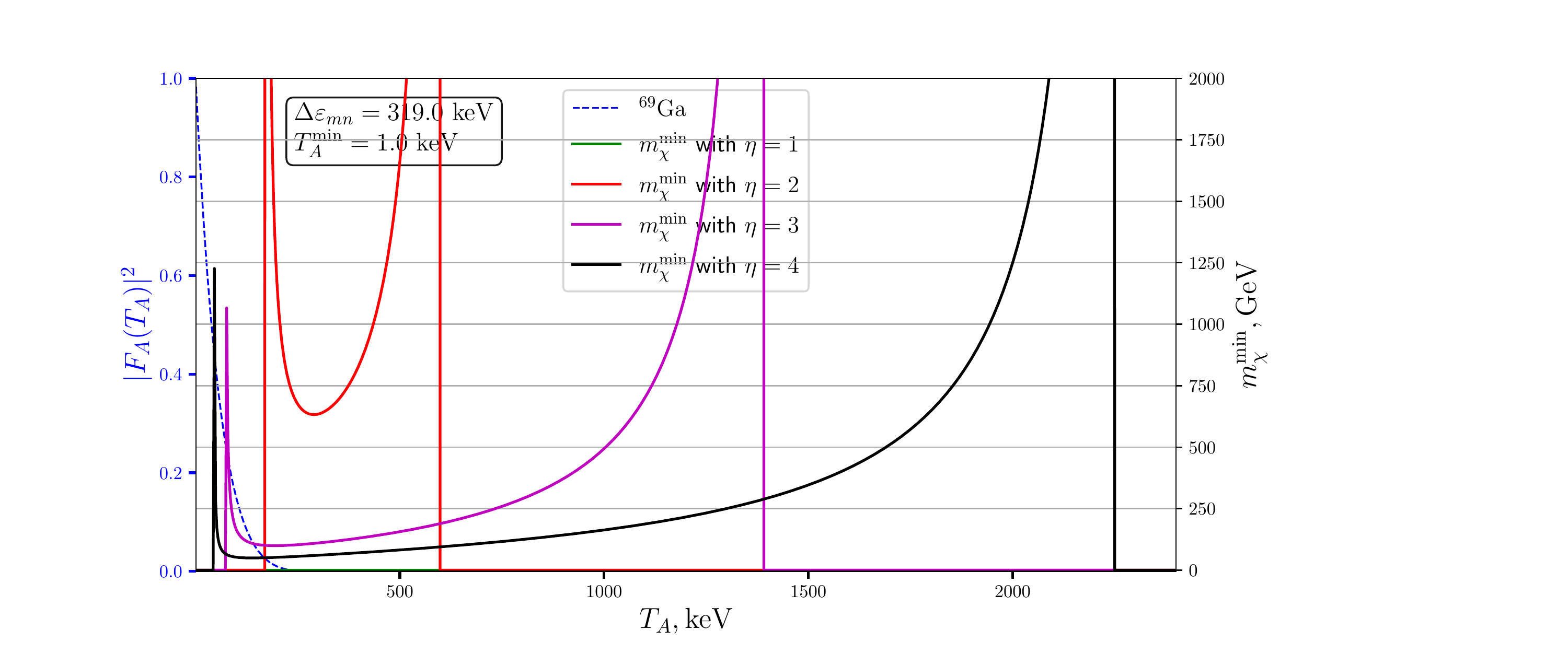}\vspace*{-10pt}
\caption{\small Same as in Fig.~\ref{fig:4DM-F2+mchi_min-vs-T_A-eta-IXeGe}, but for the $^{19}$F and $^{69}$Ga nuclei.}
\label{fig:4DM-F2+mchi_min-vs-T_A-eta-F19-Ga69}
\end{figure} 
From Fig.~\ref{fig:4DM-F2+mchi_min-vs-T_A-eta-IXeGe} one can seen that if $\eta=2$ (red curve), 
the recoil energy $T_A\simeq 50~$keV in $^{131}$Xe can be caused only by the DM particle with 
$m_\chi \ge 25$ GeV/$c^2$.
For the SHM-model ($\eta=1$, green curve),  the same nuclear recoil  $T_A\simeq 50~$keV can 
only be produced by the DM particle, with $m_\chi  \ge 40$ GeV/$c^2$.
For nuclei in Fig.~\ref{fig:4DM-F2+mchi_min-vs-T_A-eta-IXeGe}
(with "a small exception" of small $T_A$), 
the inelastic channel of interaction can be open for almost any value of $m_\chi$.
\par
In Fig.~\ref{fig:4DM-F2+mchi_min-vs-T_A-eta-F19-Ga69}, similar functions are shown for the light nucleus
$^{19}$F  ($\Delta\varepsilon_{mn}=110\,$keV) and  the $^{69}$Ga nucleus ($\Delta\varepsilon_{mn}=319\,$keV).
Comparing with Fig.~\ref{fig:4DM-F2+mchi_min-vs-T_A-eta-IXeGe}, 
one can notice some difference in the behavior of the curves
\footnote{It is due to a smaller value of the ratio $\dfrac{m_A}{\Delta\varepsilon_{mn}}$ for $^{19}$F 
and $^{69}$Ga than for $^{131}$Xe and $^{73}$Ge, respectively, despite the relatively small difference in their excitation energies.}. 
For both nuclei the green curve ($\eta=1$) "lies"\/ completely at $m_\chi=0 $ (it is not visible in the plots).
This means that the DM particle with the SHM distribution is unable to excite the $^{19}$F and $^{69}$Ga 
 nuclei (together with production of the recoil energy, $T_A$, within the intervals indicated in the plots).
With increase of the maximal DM velocity  (purple and black curves), the excitation of 
 $^{19}$F becomes possible with accompanying recoil energy
$T_A$ in the interval $25 \div 575\,$keV ($40 \div 320\,$keV) when $\eta=4(3)$, respectively.
Note that to achieve the recoil energy $T_{A^*}$ at the edges of these intervals, 
an infinite initial kinetic energy  is required, which is equivalent to $m_\chi \simeq \infty$.
This is seen from the sharp pole peaks in the curves
and forbidden regions for $T_A$ outside these peaks both at small and  large $T_A$.
These (in this case two) poles correspond to the zero denominator in 
formula (\ref{eq:4DM-r-via-T_A}).
In Fig.~\ref{fig:4DM-F2+mchi_min-vs-T_A-eta-IXeGe},  
the similar pole-like behavior can be noticed for $^{129}$I and $^{131}$Xe
in the region of very small $T_A$. 
\par
The excitation of $^{69}$Ga (to the lower level)  can occur already at $\eta=2$, and is possible only in
some limited intervals of the recoil energy.
However, unlike $^{19}$F, these allowed $T_A$ intervals are located almost completely 
at (very) large $T_A$ (see Fig.~\ref{fig:4DM-F2+mchi_min-vs-T_A-eta-F19-Ga69}), where the coherent contribution is already vanished, because the square of the form factor (blue dotted line)
has long been zero.
The range of allowed recoil energy, $T_{A^*}$, which is inevitably connected 
with the excitation of the $^{19}$F and $^{69}$Ga nuclei  
(see Fig.~\ref{fig:4DM-F2+mchi_min-vs-T_A-eta-F19-Ga69}), is limited, 
and its width depends on the maximal velocity of the DM particle.
For example, if with the $^{19}$F target  the 110-keV $\gamma$-ray is registered together 
with $T_{A^*}\simeq 370\,$keV, 
one concludes that the DM particle which caused this $\gamma$-quantum 
has $m_\chi \ge$ 100 GeV$/c^2$ and belongs to the DM flow with $\eta\simeq 4$.
Similar conclusions can be made for the $^{69}$Ga nucleus.
\par
\begin{figure}[h!] \centering 
\vspace*{-10pt}\hspace*{-30pt}
\includegraphics[width=\linewidth]{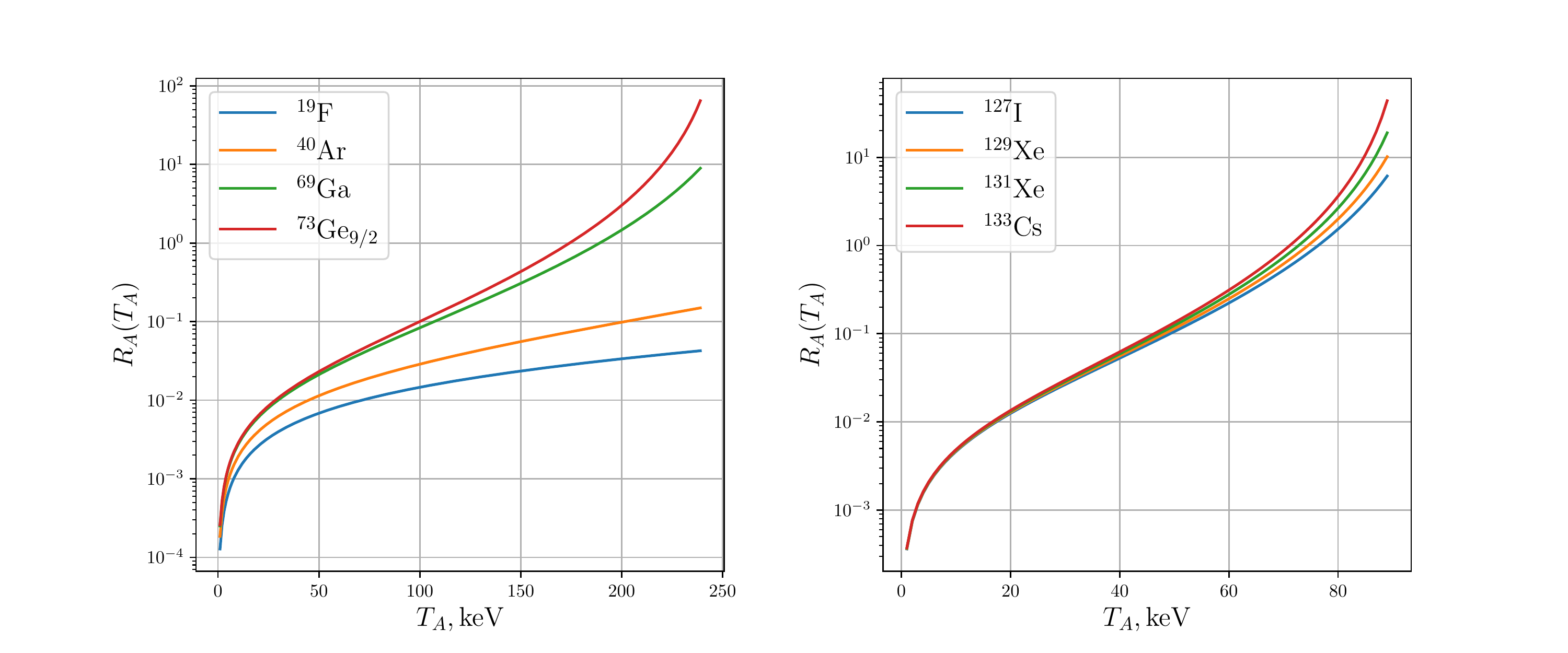}\vspace*{-10pt}
\caption{\small "Nuclear"\/ ratios $R_{A}(T_A)$ from (\ref{eq:4DM-DiffEventRates-Ratios}) as functions of 
recoil energy of the nucleus $T_A^{}$.} \label{fig:4DM-Pure-R_A}
\end{figure}

In Fig.~\ref{fig:4DM-Pure-R_A},  the nuclear factor $R_{A}(T_A)$ from formula 
(\ref{eq:4DM-DiffEventRates-Ratios}) is shown  as a function of $T_A$ for some typical target nuclei.
Figure~\ref{fig:4DM-Pure-R_A} illustrates the effect of "pure nuclear incoherence"\/, 
which is completely specified by the squared nuclear form factor behavior with increasing $T_A$.
The expression $R_{A}(T_A)$ from (\ref{eq:4DM-DiffEventRates-Ratios})
has a pole when $|F(T_A)|^2\simeq 0$.
This condition was used to define the maximal recoil energy $T_A^{\max}$ in the figures (x-axis).
The nuclear factor $R_{A}(T_A)$ allows one to estimate the prospects  of the
target nucleus (with specific $m_A$ and $\Delta\varepsilon_{mn}$) 
for registration of the inelastic excitation by the DM particle.
\begin{table}[h!] \begin{center}
{\small \begin{tabular}{|r|cccccc|c|c|}  \hline
$^A$Nucleus$(Z,N)$& $Z_+$ &$ Z_-$& $\Delta Z$ & $N_+$  &$N_-$ &$\Delta N$ & $\Delta\epsilon$, keV& $R_{\rm SM}(A)$ \\ \hline
${}^{4}\text{He}^{0+}(2,2)$ &1 & 1 & 0 & 1 & 1 & 0   & 20210$^{0+}$,$21010^{0-}$,21840$^{2-}$ &  23.44 \\ 
${}^{6}\text{Li}^{1+}(3,3)$ &2 & 1 & 1 & 2 & 1 & 1   &  2186$^{3+}$,3563$^{0+}$,4310$^{2+}$ &  23.44 \\
${}^{12}\text{C}^{0+}(6,6)$ &3 & 3 & 0 & 3 & 3 & 0   & 4438$^{2+}$,7654$^{0+}$,9641$^{3-}$  &  23.44 \\ 
${}^{19}\text{F}^{\frac{1}{2}+}(9,10)$ &5 &4 &1 & 5 & 5 & 0 & {\bf 110}$^{\frac{1}{2}-}$, 197$^{\frac{5}{2}+}$,1346$^{\frac{5}{2}-}$,1459$^{\frac{3}{2}-}$   &20.69
\\${}^{23}\text{Na}^{\frac{3}{2}+}(11,12)$ &6&5&1 & 7 & 5 & 2 & {\bf 440}$^{\frac{5}{2}+}$,2076$^{\frac{7}{2}+}$,2391$^{\frac{1}{2}+}$  &21.18
\\${}^{35}\text{Cl}^{\frac{3}{2}+}(17,18)$ & 9 & 8 & 1 & 10 & 8 & 2  &{\bf 1220}$^{\frac{1}{2}+}$,1763$^{\frac{5}{2}+}$,2693$^{\frac{3}{2}+}$ &21.06 
\\${}^{40}\text{Ar}^{0+}(18,22)$ & 9 & 9 & 0 & 11 & 11 & 0  &{\bf 1461}$^{2+}$,2121$^{0+}$,2524$^{2+}$
 &19.22\\  ${}^{48}\text{TI}^{0+}(24,24)$ & 12 & 12 & 0 & 12 & 12 & 0  &{\bf 984}$^{2+}$,2296$^{4+}$,2421$^{2+}$&19.84 \\    
${}^{69}\text{Ga}^{\frac{3}{2}-}(31,38)$ & 16 & 15 & 1 & 20 & 18 & 2  & {\bf 319}$^{\frac{1}{2}-}$,574$^{\frac{5}{2}-}$, 872$^{\frac{3}{2}-}$ ,1029$^{\frac{1}{2}-}$ &19.14
\\ ${}^{73}\text{Ge}^{\frac{9}{2}+}(32,41)$ & 16 & 16 & 0 & 25 & 16 & 9  &{\bf 13.3}$^{\frac{5}{2}+}$,67$^{\frac{1}{2}-}$, 69$^{\frac{7}{2}+}$,354$^{\frac{5}{2}-}$ &17.00\\ 
${}^{74}\text{Ge}^{0+}(32,42)$ & 16 & 16 & 0 & 21 & 21 & 0  &{\bf 596}$^{2+}$,1204$^{2+}$, 1483$^{0+}$ &18.00 \\   
${}^{127}\text{I}^{5/2+}(53,74)$&27&26&1&39&35&4  &
{\bf 58}$^{\frac{7}{2}+}$,203$^{\frac{3}{2}+}$,375$^{\frac12+}$, 418$^{\frac{5}{2}+}$  &17.01 \\ 
 ${}^{129}\text{Xe}^{\frac{1}{2}+}(54,75)$&27&27&0&38&37&1 & {\bf 40},236,318$^{\frac{3}{2}+}$,322$^{\frac{5}{2}+}$,412$^{\frac{1}{2}+}$ &17.13 \\ 
${}^{131}\text{Xe}^{\frac{3}{2}+}(54,77)$&27&27&0&40&37&3  & {\bf 80}$^{\frac{1}{2}+}$,164$^{\frac{11}{2}+}$, 365$^{\frac{5}{2}+}$, 405$^{\frac{3}{2}+}$ & 16.72 \\ 
${}^{133}\text{Cs}^{\frac{7}{2}+}(55,78)$&28&27&1&46&32&6  &{\bf 81},161$^{\frac{5}{2}+}$,384$^{\frac{3}{2}+}$ 437$^{\frac12+}$,633$^{\frac{11}{2}+}$  & 16.72
\\   ${}^{207}\text{Pb}^{\frac{1}{2}-}(82,125)$&41&41&0&63&62&1 &{\bf 570}$^{\frac{5}{2}-}$, 898$^{\frac{3}{2}-}$1633$^{\frac{13}{2}+}$, 2340$^{\frac{7}{2}-}$ &15.86\\ \hline
\end{tabular}}\end{center}
\caption{\small Characteristics of some target nuclei used. Here $Z=A_p,\ N=A_n$ and $\Delta A_p \equiv \Delta Z = Z_+ - Z_-$, $ \Delta A_n \equiv \Delta N= N_+ - N_-$.
The total nuclear spin is calculated by the formula $\frac{\Delta Z}{2}+\frac{\Delta N}{2}$.
The indices $\pm$ denote the direction of the nucleon spin with respect to the quantization axis.
Values for $ R_{\rm SM}(A)\equiv R_I(A)$ from formula (\ref{eq:4DM-Results-Inc2CohRatio-WeakCharge}) are
obtained with the SM coupling constants $ \alpha_p \simeq 0.02, \ \alpha_n = - 0.5, \ \delta_p = - \delta_n \simeq 0.64$ from (\ref{eq:4DM-Results-SM-alpha+delta}).
}   \label{tab:50chiA-Ratios-RIad-vs-A}
\end{table} 

\subsection*{\em Event rates with the Standard Model weak couplings}
\begin{figure}[h!] \centering 
\includegraphics[width=0.7\linewidth]{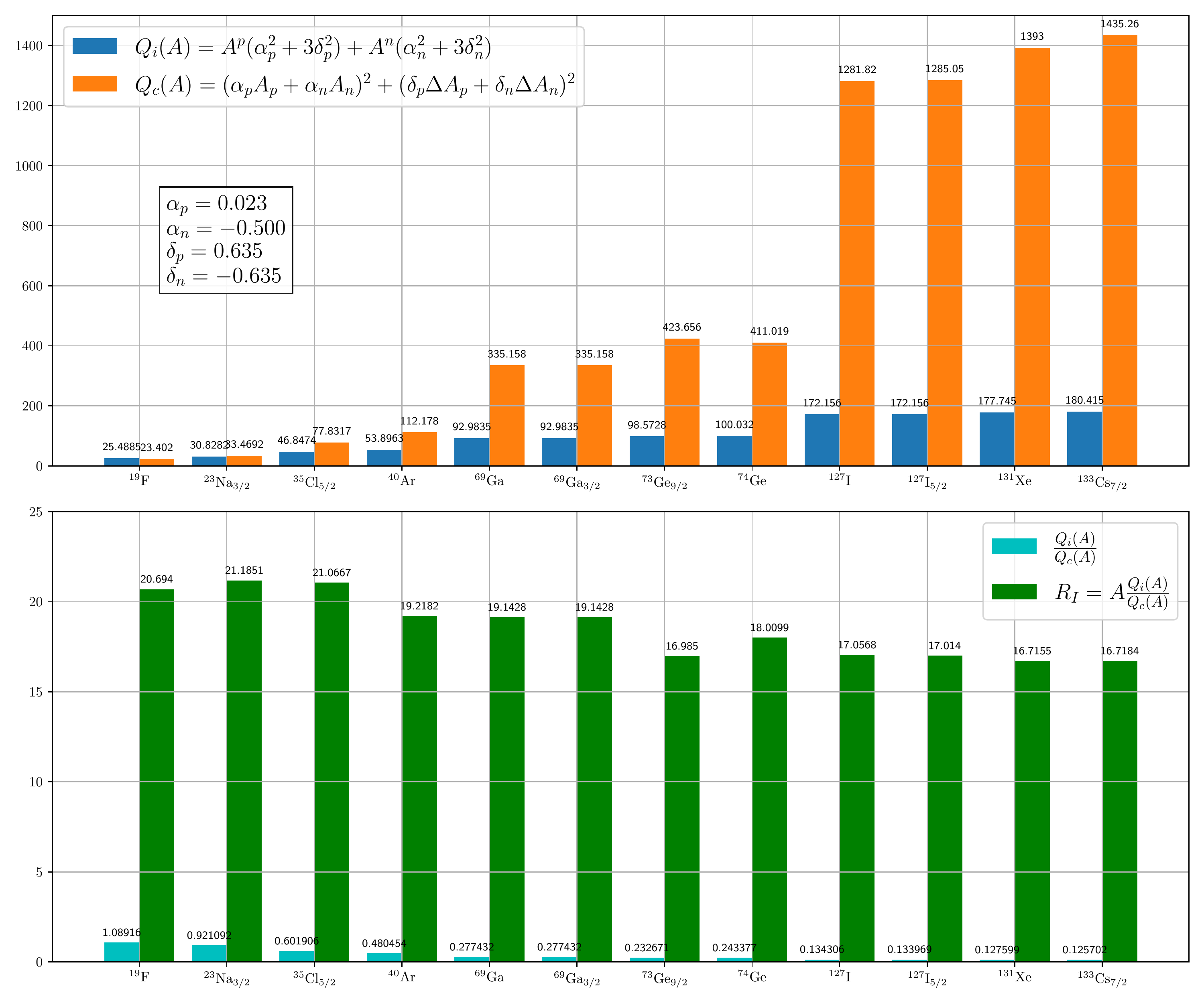}
\vspace*{-10pt}
\caption{\small Nuclear charge factors from formulas (\ref{eq:4DM-Q_c-and-Q_i-def}) and
(\ref{eq:4DM-Results-Inc2CohRatio-WeakCharge})
with SM parameters from (\ref{eq:4DM-Results-SM-alpha+delta}).}
\label{fig:4DM-Table-with-Qs-in-SM}
\end{figure}

The incoherent-to-coherent rate ratio dependence on the nature of the $\chi$-nucleon interaction 
is completely determined by the first "charge-nuclear"\/ factor in 
(\ref{eq:4DM-DiffEventRates-Ratios}), written as follows:
\begin{eqnarray}
\label{eq:4DM-Results-Inc2CohRatio-WeakCharge} 
R_{\rm I}(A) = A\dfrac{Q_{\rm i}(A)}{Q_{\rm c}(A)} =
\dfrac{(A_p+A_n)[ A^p (\alpha^2_p+ 3 \delta^2_p) + A^n (\alpha^2_n + 3 \delta^2_n )]}
{(\alpha_p A_p+\alpha_n A_n)^2+ (\delta_p {\Delta A_p}+\delta_n {\Delta A_n} )^2}
 . \end{eqnarray}
In the nonrelativistic approximation, it is determined only by the effective coupling constants $\alpha_{p,n}$ 
and $\delta_{p,n}$ from (\ref{eq:42chiA-CrossSection-ScalarProducts-ChiEta-All-Weak-Nonrel}),
which are "weighted"\/ with the proton--neutron structure parameters of the nucleus.
 This factor $R_{\rm I}(A)$ can either enhance or weaken the  "pure"\/ effect of the nuclear structure 
  shown in Fig.~\ref{fig:4DM-Pure-R_A}.
The SM-values of this factor for a number of target nuclei are given 
in Table~\ref{tab:50chiA-Ratios-RIad-vs-A} and Fig.~\ref{fig:4DM-Table-with-Qs-in-SM}.
The "charge-nuclear"\/ factors $R_{\rm SM}(A)$ from
(\ref{eq:4DM-Results-Inc2CohRatio-WeakCharge})
give more than an order of magnitude increase in the "incoherence effect"\/.
Therefore, the dominance of the inelastic contribution can begin at twice as low recoil energies, i.e., 
the incoherent nuclear excitation may be accompanied by lower recoil energy $T_A$.
\par
\begin{figure}[h!] \centering 
\vspace*{-60pt}\hspace*{-20pt}
\includegraphics[width=0.72\linewidth]{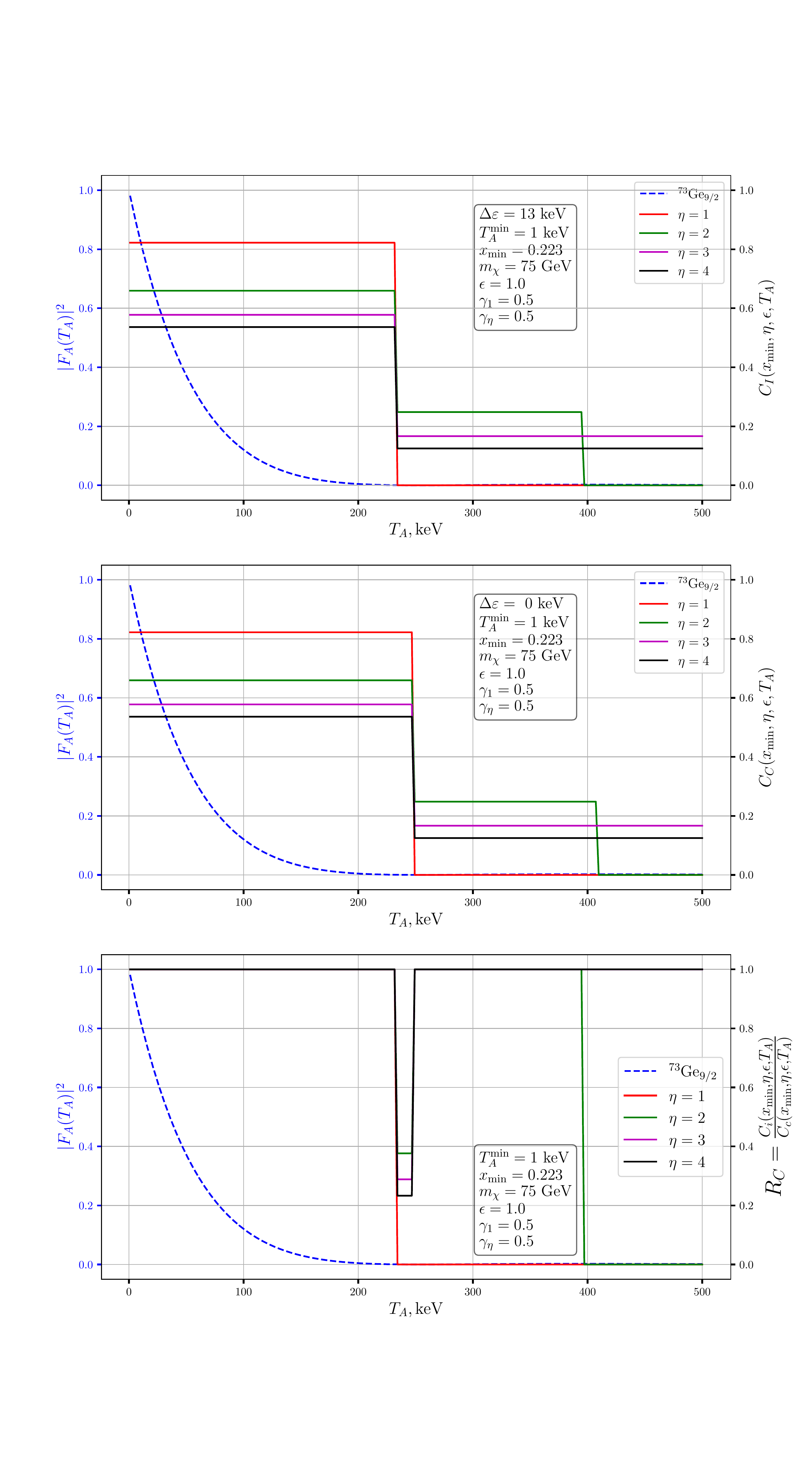}
\vspace*{-50pt}
\caption{\small  For $^{73}$Ge as an example, the coefficient
$C_{\rm i}(x_{\min},\eta_{},\epsilon,T_A)$ (upper plot),
$C_{\rm c}(x_{\min},\eta_{},\epsilon,T_A)$ (middle plot)
from (\ref{eq:4DM-DiffEventRates-Coh+Inc-SHM+str+Fp=Fn-Qci}),
and their ratios $R_C(x_{\min},\eta_{},\epsilon,T_A)$ (lower plot) from (\ref{eq:4DM-DiffEventRates-Ratios-R_C}) 
are given as functions of  the nuclear recoil energy $T_A$ for $\eta=1\div 4$, $T_A^{\min}=1\,$keV,
  $m_\chi = 75\,$GeV$/c^2$ and $\epsilon=1, \gamma_1=\gamma_\eta=0.5$.
The dashed line shows the squared $^{73}$Ge form factor. 
 } \label{fig:4DM-Ge73-Cosmos-vs-TA-3figs}
\end{figure}

 
The third, "cosmic"\/ factor in formula (\ref{eq:4DM-DiffEventRates-Ratios}), i.e., the 
ratio $R_C(x_{\min},\eta_{},\epsilon,T_A)$ from (\ref{eq:4DM-DiffEventRates-Ratios-R_C}), 
is always equal to one if only one DM velocity distribution (for example, SHM) is allowed.
If one takes into account, as presented in formula (\ref{eq:4DM-DiffEventRates-Ratios-R_C}),
the two DM distributions simultaneously (SHM and $\eta>1$), 
the relation $R_C(x_{\min},\eta_{},\epsilon,T_A)$ 
is equal to one or  zero (see lower plot in Fig.~\ref{fig:4DM-Ge73-Cosmos-vs-TA-3figs})
in almost entire range of parameters.
The zero value occurs when increasing $T_A$ becomes greater than $T_{A^*,\eta}^{\max}$
but still smaller than $T_{A,\eta}^{\max}$. 
In this case, the numerator of this ratio vanishes (upper plot in Fig.~\ref{fig:4DM-Ge73-Cosmos-vs-TA-3figs})
along with the incoherent contribution.
There are also cases, where the first term in the numerator of $R_C(x_{\min},\eta_{},\epsilon,T_A)$ 
corresponding to smaller velocity $\eta$ vanishes, for example,  
a little earlier than it happens with the first term in the denominator.
This situation can be seen from the comparison of the top and middle plots 
in Fig.~\ref{fig:4DM-Ge73-Cosmos-vs-TA-3figs}.
As a result, a characteristic "dip"\/ occurs in $R_C(x_{\min},\eta_{},\epsilon,T_A)$,
shown in the bottom plot in Fig.~\ref{fig:4DM-Ge73-Cosmos-vs-TA-3figs}.
The width of this dip is determined by $\Delta\epsilon_{mn}$.
As $m_\chi$ increases, this dip shifts to higher values of $T_A$, where the coherent contribution almost completely disappears.
  \par
\begin{figure}[h!]
\centering 
\vspace*{-50pt}\hspace*{-20pt}
\includegraphics[width=0.9\linewidth]{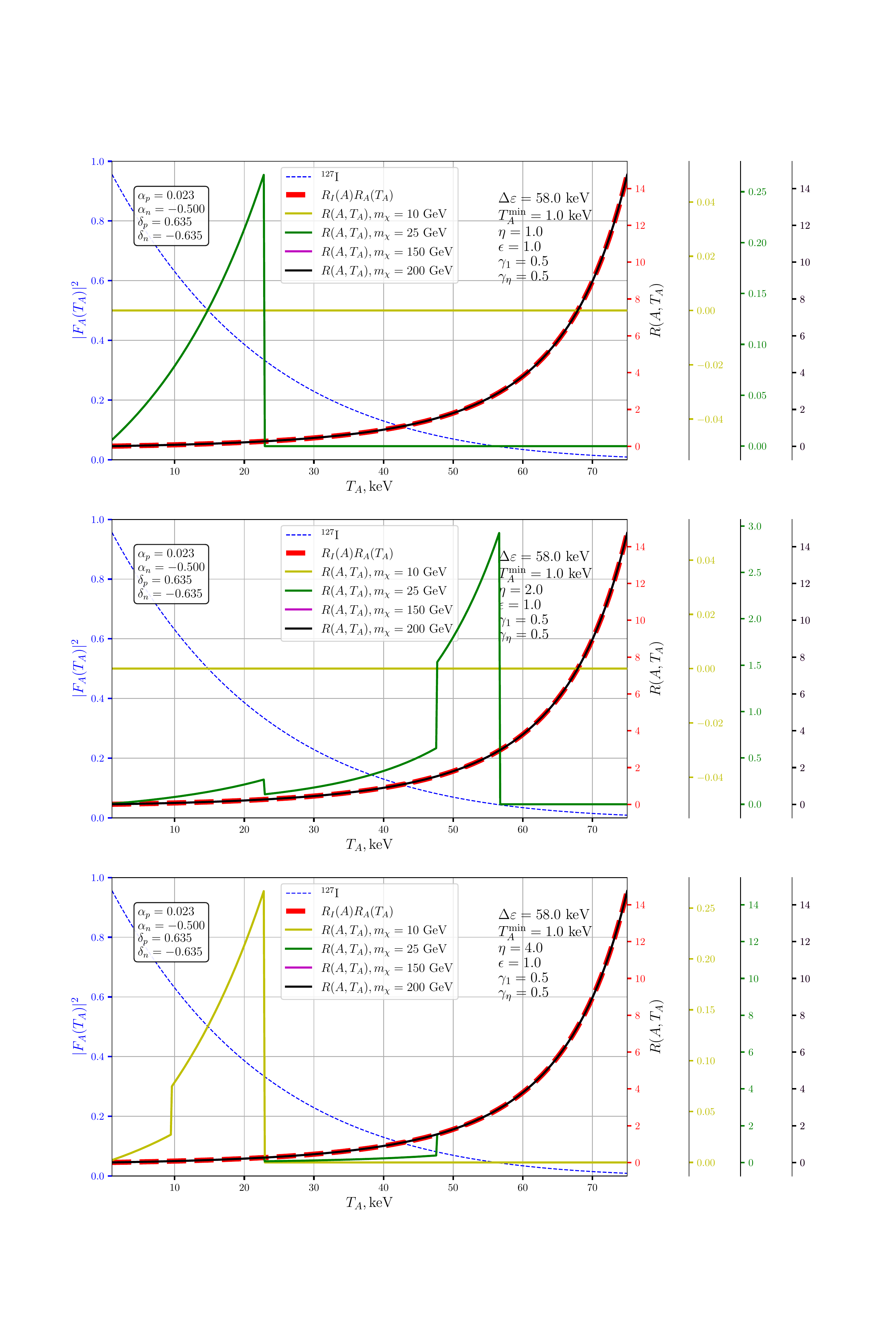}
\vspace*{-70pt}
\caption{\small The role of the  two-DM-component factor $R_C(x_{\min},\eta_{},\epsilon,T_A)$ from 
(\ref{eq:4DM-DiffEventRates-Ratios-R_C})  is shown for $^{127}$I.
 It manifests itself in differences of the solid curves $R_{}(A,T_A)$
 marked with different colors for different $m_\chi$, 
 from the "pure"\/ product $R_{A}(T_A)R_{\rm SM}(A)$ (red bold dashed line)
 only for small $m_\chi\le 25~$GeV$/c^2$.
 The colors of the right-side y-axes correspond to these masses.
The blue thin dotted line shows the square of the $^{127}$I form factor $|F_A(T_A)|^2$ (left y-axis). }
\label{fig:4DM-I127-RC-role-1}
\end{figure}

 
The influence of $R_C(x_{\min},\eta_{},\epsilon,T_A)$ from (\ref{eq:4DM-DiffEventRates-Ratios-R_C})
on the total incoherent-to-coherent rate ratio (\ref{eq:4DM-DiffEventRates-Ratios})  for $^{127}$I
is shown in Fig.~\ref{fig:4DM-I127-RC-role-1}. 
The influence is noticeable only for sufficiently small values of $m_\chi = 10\div 35$ GeV$/c^2$.
For example, if $\eta=2$ (middle plot), 
the incoherent process in $^{127}$I is impossible for $m_\chi = 10\,$ GeV$/c^2$.
The corresponding (yellow) curve vanishes for all $T_A$.
The green curve for $m_\chi =25\,$ GeV$/c^2$ (and $\eta=2$) is strongly 
"suppressed"\/ by the factor $R_C(x_{\min},\eta_{},\epsilon,T_A)$.
It has a characteristic dip for middle $T_A$ and vanishes earlier with increasing  $T_A$    
than the red dashed curve of the "pure"\/ product  $R_{A}(T_A)R_{\rm SM}(A)$.
 All these "features"\/ point to the insignificance 
  of the inelastic contribution {\em in the case of small}\/ $m_\chi$.
\begin{figure}[h!] 
\centering
\vspace*{-60pt}\hspace*{-25pt}
\includegraphics[width=0.7\linewidth]{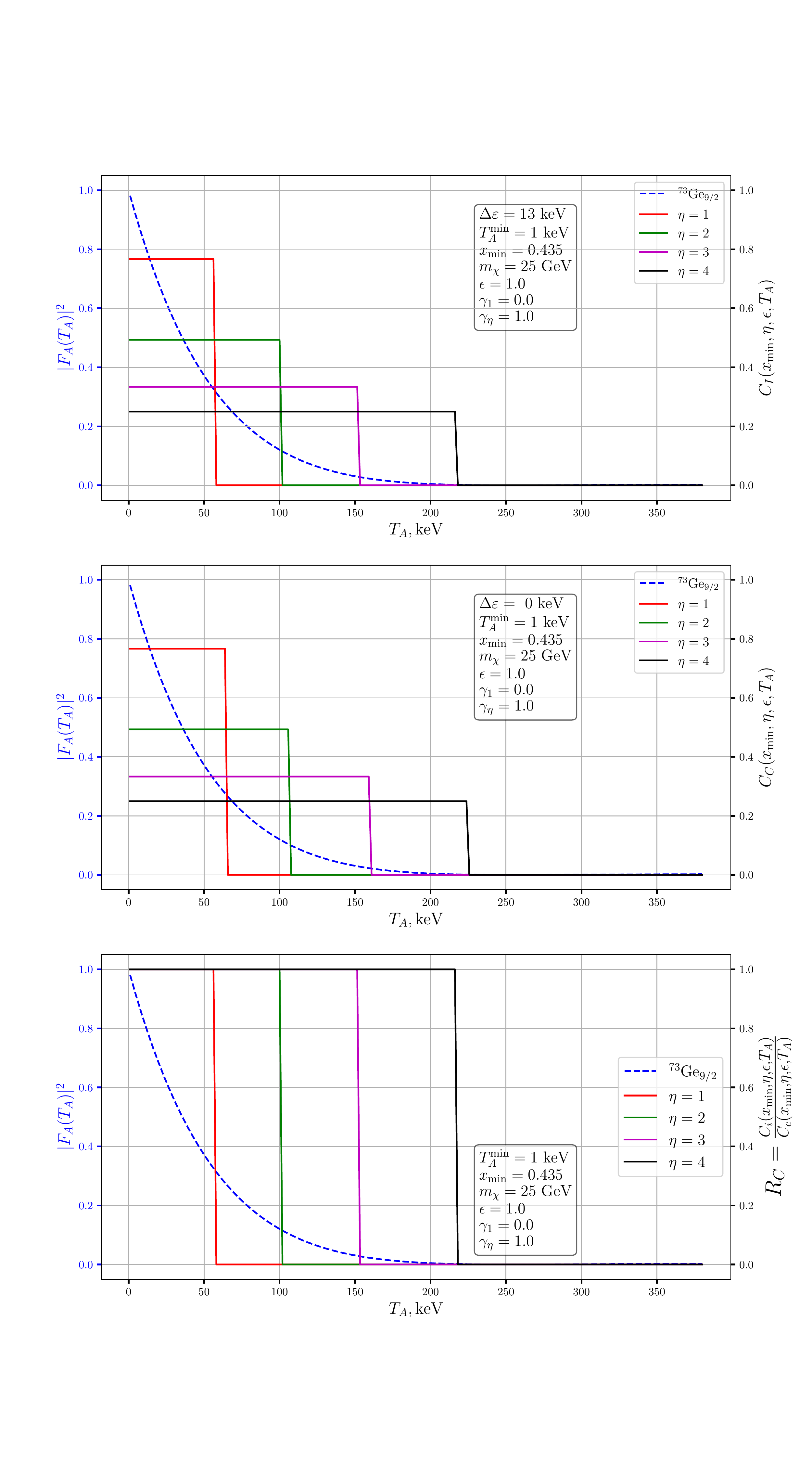}
\vspace*{-60pt}
\caption{\small The "cosmic factors"\/ and $R_C(x_{\min},\eta_{},\epsilon,T_A)$
as functions of $T_A$ are shown in $^{73}$Ge for only one DM distribution with different parameters $\eta$.
The blue dotted line shows  $|F_A(T_A)|^2$.} \label{fig:4DM-C_i+C_c+R_C-1DM}
\end{figure}
 \par 
In general, the role of the factor $R_C(x_{\min},\eta_{},\epsilon,T_A)$,  which is usually equal to one (see
Fig.~\ref{fig:4DM-Ge73-Cosmos-vs-TA-3figs} and \ref{fig:4DM-I127-RC-role-1}),
 disappears as $m_\chi$ and $\eta$ increase
  \footnote{The cases, where it is not true (too small $m_\chi$ and/or too large $\Delta\epsilon_{mn}$), require more careful analysis.}.  
This  means that  the kinetic energy of the incident DM particle becomes sufficient 
 to excite the nucleus and to produce simultaneously large enough nuclear recoil, $T_A$.
 Therefore, the inelastic event rate
  \footnote{In the nonrelativistic SM-weak interaction approximation.}
 appears as a permanent component of the DM interaction,
 in contrast to the elastic one, which quickly disappears with increasing $T_A$.
  \par
The "peculiar"\/ $T_A$-dependence of the rate ratios
(\ref{eq:4DM-DiffEventRates-Ratios}) shown in Fig.~\ref{fig:4DM-I127-RC-role-1} illustrate
in a different way (than in Fig.~\ref{fig:4DM-Pure-R_A}) 
 the interplay between the elastic and inelastic event rates
(or their "transition dynamics"\/ with increasing $T_A$).
The interplay is given for the specific nuclei ($m_A$ and $\Delta\epsilon_{mn}$) 
and the DM masses as a function of the DM kinetic energy $T^{\max}_\eta$.
\par
In other words, one can argue that the ratio between differential incoherent and coherent
event rates  (\ref{eq:4DM-DiffEventRates-Ratios}), $R_I(A, T_A)$, 
coincides with the ratio between the differential incoherent and coherent  $\chi A$ scattering cross sections
(\ref{eq:4DM-Results-Inc2CohRatio-A}) provided conditions (\ref{eq:4DM-working-T_A-interval}) 
 and (\ref{eq:4DM-T_A-general}) are satisfied
\begin{equation*}
T_A^{\min} \le T_A\le T_{A^{*},\eta}^{\max}(r,\Delta\varepsilon_{mn})  
. \end{equation*} 
For fixed $\eta$, $m_A$, $\Delta\varepsilon_{mn}$ and $T_A$ 
it restricts the allowed DM masses $m_\chi$ (see Fig.~\ref{fig:4DM-F2+mchi_min-vs-T_A-eta-IXeGe} and 
Fig.~\ref{fig:4DM-F2+mchi_min-vs-T_A-eta-F19-Ga69}).%
\smallskip\par
The case of two different DM distributions (\ref{eq:3DM-both-speeds-in-Earth-system}), 
despite some versatility, looks quite exotic with existence of just these two DM distributions, 
and looks quite complicated for numerical study with the "pecular"\/ curves 
shown in Fig.~\ref{fig:4DM-I127-RC-role-1}.
\par 
Therefore, in what follows, we consider only one DM distribution 
with a set of maximal dimensionless velocities $\eta=1,2,4$ and weight $\gamma_\eta=1$.
Then the "cosmic factors"\/ (\ref{eq:4DM-Cosmic-CIhC-CCoh}) become identical
in size and differ only in the "length of $T_A$"\/
$$
C_{\rm c/i}(x_{\min},\eta_{},\epsilon,T_A)=C_{}(x_{\min},\eta_{},\epsilon)
\Theta(T_{A^{}/A^{*},\eta}^{\max}-T_A),
 $$
since always $T_{A^{},\eta}^{\max} \ge T_{A^{*},\eta}^{\max}$
(see Fig.~\ref{fig:4DM-C_i+C_c+R_C-1DM}). 
Here one has the notation
\begin{eqnarray} \label{eq:4DM-C1DM}
C_{}(x_{\min},\eta_{},\epsilon) &\equiv& S_{}(x_{\min},\eta_{},\epsilon)-S_{}({2.5} + \eta_{},\eta_{},\epsilon)
.\end{eqnarray}
Then the  total (coherent plus incoherent) differential event rate
(\ref{eq:4DM-DiffEventRates-Coh+Inc-SHM+str+Fp=Fn-Qci}) takes the form
\begin{eqnarray} \nonumber \label{eq:4DM-DiffEventRate-Total-1DM}
\frac{ d R(T_A)_\text{} }{R_0^w  d T_A}&=&C_{}(x_{\min},\eta_{},\epsilon)\dfrac{ m_A  }{A m_\chi }
[ Q_{\rm c}(A)\Phi_{\rm c} (T_A)\Theta(T_{A,\eta}^{\max}\!-\!T_A) 
+ Q_{\rm i}(A)\Phi_{\rm i}(T_A) \Theta(T_{A^{*},\eta}^{\max}\!-\!T_A) ] \qquad
\\&=&
C_{}(x_{\min},\eta_{},\epsilon)\dfrac{ m_A  }{A m_\chi } Q_{\rm c}(A)\Phi_{\rm c} (T_A) \Theta(T_{A,\eta}^{\max}\!-\!T_A) [1+ {R(A,T_A,T_A^{\min})}] , \qquad  
 \\ &&  
\label{eq:4DM-R_0}  
\quad \text{where} \quad R_0^w \equiv  \dfrac{ G^2_F  }{4 \pi}\dfrac{  \rho_\chi  N_{\rm Av} }{ v_0 } 
 \cong  10^{3} \dfrac{\rm events}{\rm~ kg~keV~year}. 
\end{eqnarray}
Relations (\ref{eq:4DM-DiffEventRates-Ratios}) and (\ref{eq:4DM-EventRates-Ratios}) 
are also simplified (because $T_{A^{},\eta}^{\max} \ge T_{A^{*},\eta}^{\max}$) as follows:
\begin{eqnarray}\label{eq:4DM-DiffEventRateRatios-1DM}
R_{\rm }(A,T_A,T_A^{\min})&=&{\dfrac{d R_\text{i} }{d T_A}}/{\dfrac{d R_\text{c} }{d T_A}}=
\dfrac{Q_{\rm i}(A)}{Q_{\rm c}(A)}\dfrac{1- |F_{}(T_A)|^2}{|F_{}(T_A)|^2}
\Theta(T_A-T_A^{\min})\Theta(T_{A^{*},\eta}^{\max}-T_A) 
, \qquad
\\R_{\rm }(A,T_A^{\min})&\equiv&\dfrac{R_\text{i} }{R_\text{c} }(T_A^{\min})=\dfrac{Q_i(A)}{Q_c(A)}\Theta(T_A-T_A^{\min})  \dfrac{\int_{T_A^{\min}}^{T_{A^*,\eta}^{\max}} dT_A  [1- |F_{}(T_A)|^2]}{\int_{T_A^{\min}}^{T_{A,\eta}^{\max}} dT_A  |F_{}(T_A)|^2}
\nonumber.\end{eqnarray} 
\begin{figure}[h!] 
\centering \vspace*{-90pt}\hspace{30pt}
\includegraphics[width=0.78\linewidth]{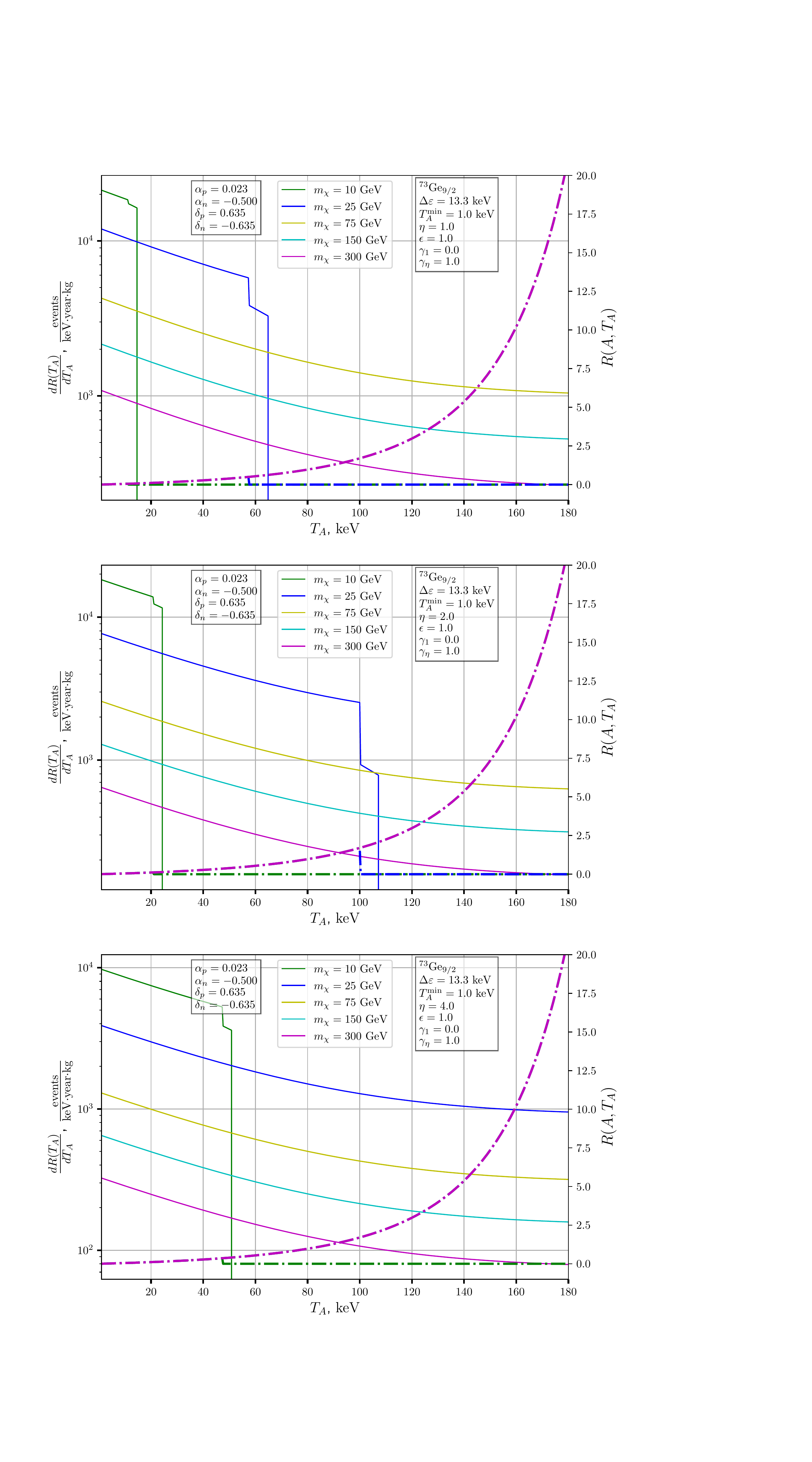}
\vspace*{-60pt}
\caption{\small For the $^{73}$Ge nucleus, the dash-dotted line shows the 
ratio $R_{\rm }(A,T_A,T_A^{\min}=1\,$keV) from (\ref{eq:4DM-DiffEventRateRatios-1DM}) 
as a function of $T_A$ (right y-axis) for three velocities $\eta$ (plots from top to bottom)
and five  $m_\chi$  (marked in different colors).
The solid thin line (left y-axis) shows the 
differential event rate from (\ref{eq:4DM-DiffEventRate-Total-1DM}), which
corresponds to $m_\chi$ (of the same color).
The $\chi$-nucleon interaction is defined by the non-relativistic SM coupling constants 
from (\ref{eq:4DM-Results-SM-alpha+delta}).}
\label{fig:4DM-1DM-Rates-and-Ratios-vs-TA-eta-mchi-Ge73}
\end{figure}
\begin{figure}[h!] 
\centering
\vspace*{-70pt}\hspace{30pt}
\includegraphics[width=0.78\linewidth]{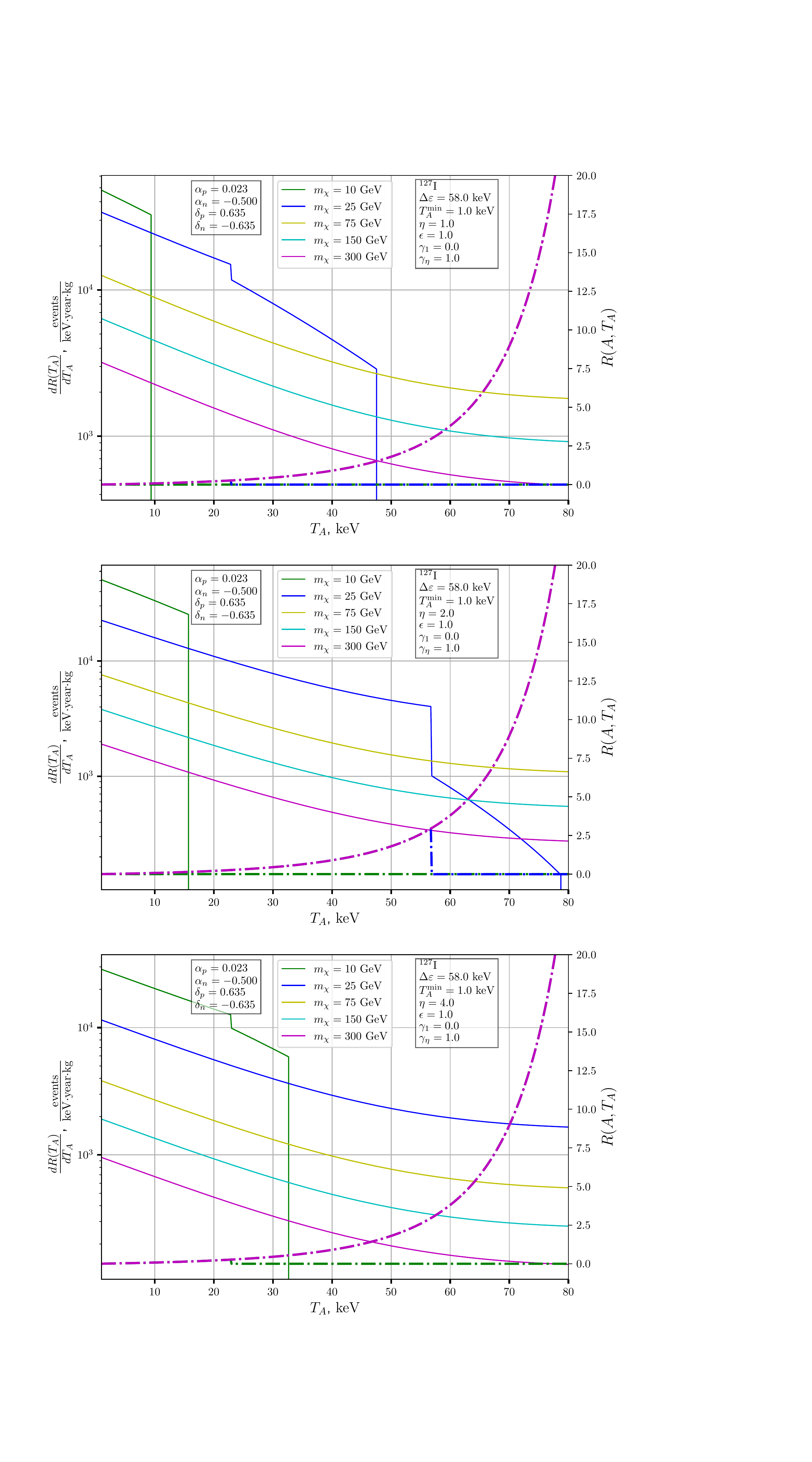}
\vspace*{-80pt}
\caption{\small Same as in Fig.~\ref{fig:4DM-1DM-Rates-and-Ratios-vs-TA-eta-mchi-Ge73}, but for the $^{127}$I nucleus.
} \label{fig:4DM-1DM-Rates-and-Ratios-vs-TA-eta-mchi-I127}
\end{figure} 
The total, integrated over $T_A$, event rate as a function of the energy threshold $T_A^{\min}$, 
as well as of $m_A$, $\Delta\varepsilon_{mn}$ together with  "cosmic parameters"\/, 
$m_\chi, \eta, \gamma_\eta, \epsilon$, takes the form
\begin{eqnarray}\label{eq:4DM-1DME-Total-EventRate}
R_{\text{i+c}}(T_A^{\min})=
C_{}(x_{\min},\eta_{},\epsilon) \Big[
Q_i(A)  \int_{T_A^{\min}}^{T_{A^*,\eta}^{\max}} dT_A  \Phi_{\rm i}(T_A)
 +Q_c(A) \int_{T_A^{\min}}^{T_{A,\eta}^{\max}} dT_A  \Phi_{\rm c}(T_A)\Big]
.\qquad 
\end{eqnarray} 
Expressions (\ref{eq:4DM-DiffEventRate-Total-1DM}), (\ref{eq:4DM-DiffEventRateRatios-1DM})
and (\ref{eq:4DM-1DME-Total-EventRate}) are {\em the main formulas}\/ for further consideration.
\begin{figure}[h!] 
\centering
\vspace*{-70pt}\hspace{30pt}
\includegraphics[width=0.8\linewidth]{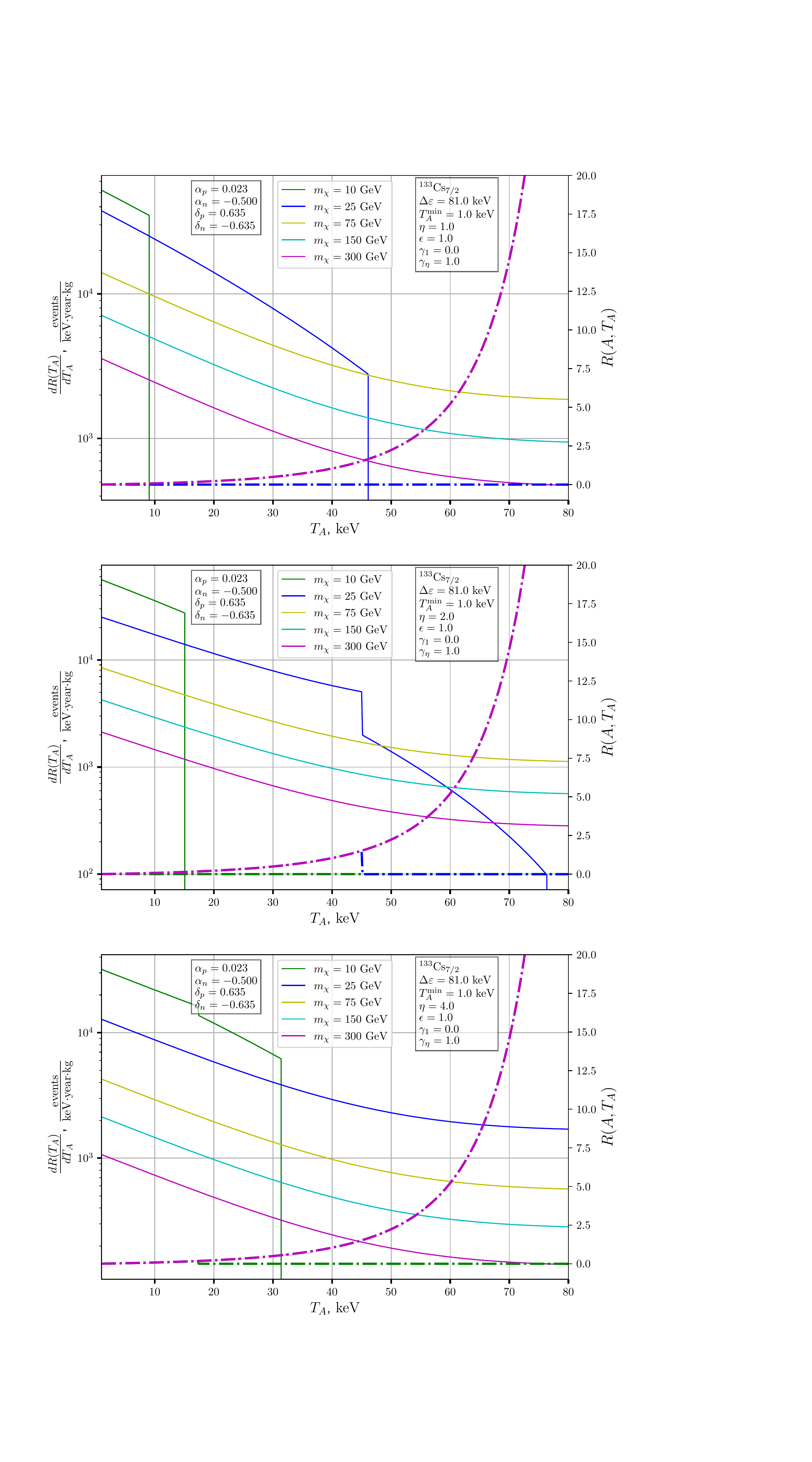}
\vspace*{-50pt}
\caption{\small Same as in Fig.~\ref{fig:4DM-1DM-Rates-and-Ratios-vs-TA-eta-mchi-Ge73}, but for 
the $^{133}$Cs nucleus.
} \label{fig:4DM-1DM-Rates-and-Ratios-vs-TA-eta-mchi-Cs133}
\end{figure}
\par
In Fig.~\ref{fig:4DM-1DM-Rates-and-Ratios-vs-TA-eta-mchi-Ge73}--%
\ref{fig:4DM-1DM-Rates-and-Ratios-vs-TA-eta-mchi-Cs133}, 
the dash-dotted line shows the ratio $R_{\rm }(A,T_A,T_A^{\min})$ given by
(\ref{eq:4DM-DiffEventRateRatios-1DM}) as a function of $T_A$ for some values of $\eta$ and $m_\chi$ 
(highlighted by color) when the $\chi A$ interaction is defined by the SM coupling constants 
in (\ref{eq:4DM-Results-SM-alpha+delta}).
The thin solid lines illustrate the $T_A$-behavior of the (total) differential 
event rates $\dfrac{d R(T_A)_\text{} }{d T_A}$ from (\ref{eq:4DM-DiffEventRate-Total-1DM})
corresponding to the $R_{\rm }(A,T_A,T_A^{\min})$-curves (shown in the same color).
\par 
\begin{figure}[h!] \centering \vspace*{-50pt} 
\includegraphics[width=0.75\linewidth]{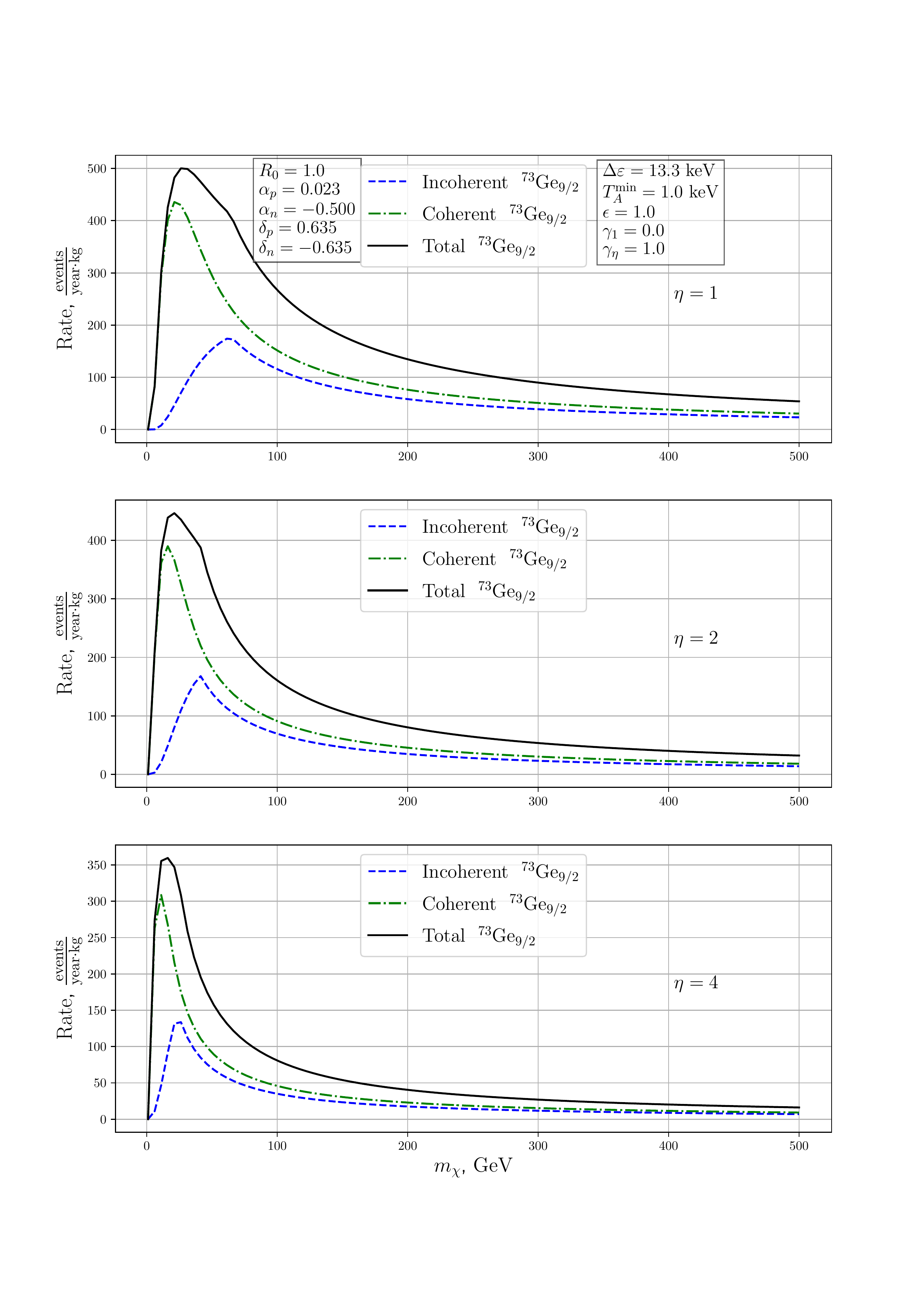}
\vspace*{-50pt}
\caption{\small  Total, integrated over $T_A$, event rate (\ref{eq:4DM-1DME-Total-EventRate})
in $^{73}$Ge as a function  of the DM mass $m_\chi$ and other "cosmic parameters", $\eta, \gamma_\eta, \epsilon$. In contrast to the SM for $R_0$ from (\ref{eq:4DM-R_0}) here one has $R_0=1.0$.
The coherent and incoherent contributions to the total rate are also shown.}
\label{fig:4DM-1DM-Rates-Integratedand-Ge73}
\end{figure} 
\begin{figure}[h!]  \centering \vspace*{-60pt} 
 \includegraphics[width=0.7\linewidth]{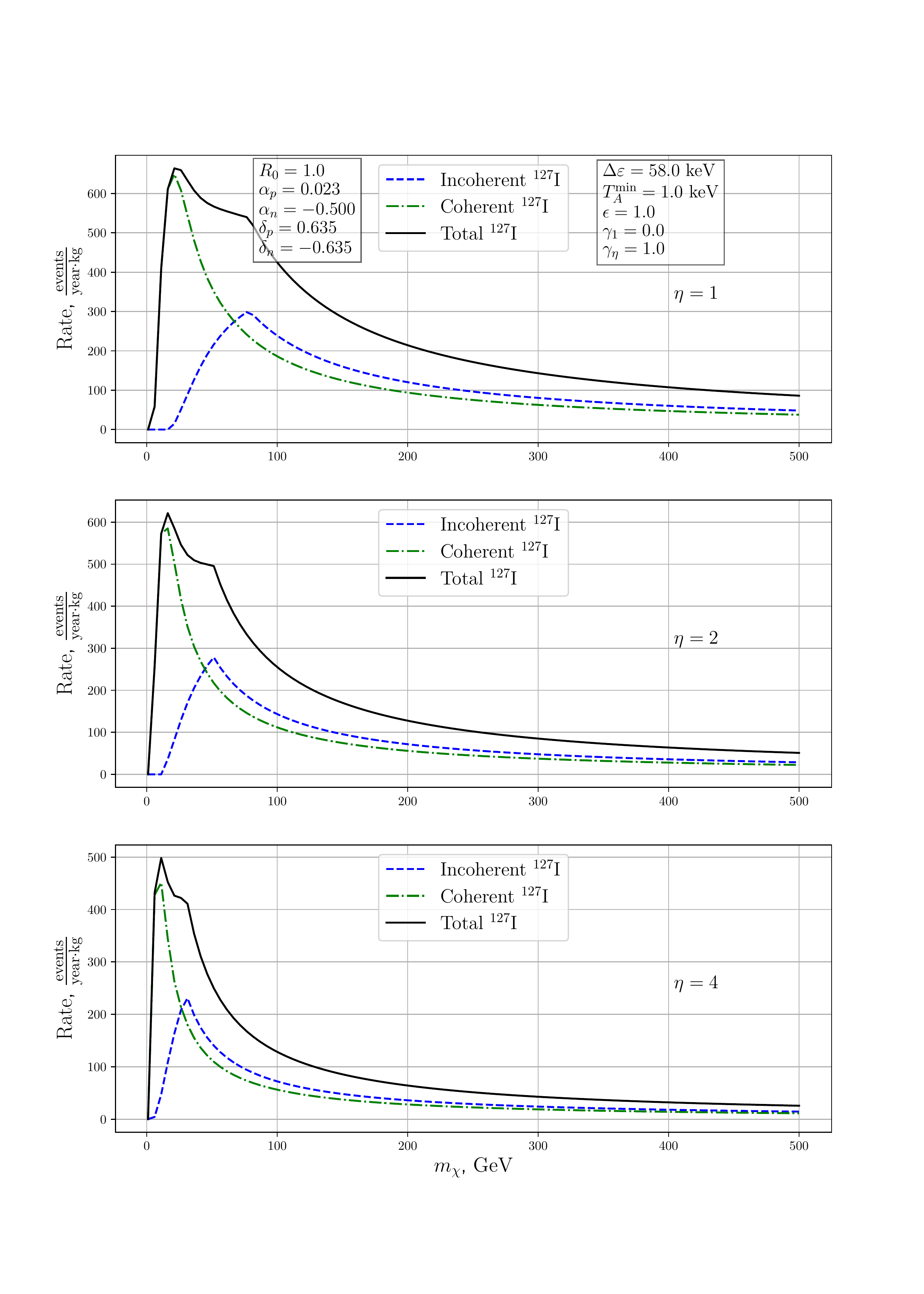}
\vspace*{-50pt}
\caption{\small Same as in Fig.~\ref{fig:4DM-1DM-Rates-Integratedand-Ge73}, but for the $^{127}$I nucleus.} 
\label{fig:4DM-1DM-Rates-Integratedand-I127} 
\end{figure} 
\par
From these plots one can see the following main features.
For small masses $m_\chi$ and small velocities $\eta$,  the inelastic interaction is impossible, and 
the dash-dotted curves  of the inelastic-to-elastic rate ratio "lie at zero".
With $m_\chi$  increasing by an order of magnitude,  
the total rate decreases by about an order of magnitude.
The sharp termination of the rate curves is due to the restrictions on the velocity $\eta$ for fixed $m_\chi$.
The double kink in these curves with increasing $T_A$  
results from the vanishing of the incoherent contribution to the total rate 
and subsequent vanishing of the coherent contribution with the large enough  $T_A$.
Over the entire interval of $T_A$, the total differential rates decrease by no more than one order of magnitude. 
Moreover, with increasing $T_A$,  the purely coherent rate  falls off much faster than 
its sum with the incoherent rate.
\par 
Approximately in the middle of the $T_A$ intervals depicted in Fig.~\ref{fig:4DM-1DM-Rates-and-Ratios-vs-TA-eta-mchi-Ge73}--\ref{fig:4DM-1DM-Rates-and-Ratios-vs-TA-eta-mchi-Cs133}, 
the incoherent rate becomes comparable with the coherent one, 
and as $T_A$ increases, it begins to dominate.
For example, if  the "measured" recoil energy in the $^{133}$Cs nucleus becomes $T_A \ge 60$ keV
(see  Fig.~\ref{fig:4DM-1DM-Rates-and-Ratios-vs-TA-eta-mchi-Cs133}), 
one can conclude that this signal is due to the incoherent interaction. 
In fact, this signal should also be accompanied by a $\gamma$-quantum with the energy of 81 keV.
\par
Furthermore, the total (coherent plus incoherent) expected event rate in $^{73}$Ge and $^{127}$I, 
given by expression (\ref{eq:4DM-1DME-Total-EventRate})
is shown in Fig.~\ref{fig:4DM-1DM-Rates-Integratedand-Ge73}--%
\ref{fig:4DM-1DM-Rates-Integratedand-I127} as a function of $m_\chi$.
It is  seen that the maximal rate is expected, as a rule, for DM masses about or a bit  
smaller than $m_\chi \cong 50~$GeV$/c^2$.
\par
Note that
in Fig.~\ref{fig:4DM-1DM-Rates-and-Ratios-vs-TA-eta-mchi-Ge73}--%
\ref{fig:4DM-1DM-Rates-and-Ratios-vs-TA-eta-mchi-Cs133}
the absolute value of the rate (for illustrative purposes) is determined by the intensity of the weak
lepton--nucleon interaction of the Standard Model, although it is already quite clear
that the interaction of DM particles with matter is noticeably weaker.
Furthermore, since the results of the "classical"\/ direct DM search experiments 
\cite{Smith:1990kw,Vergados:1996hs,Bertone:2004pz,Spooner:2007zh,Bednyakov:2008gv,Saab:2012th,Baudis:2012ig,Cushman:2013zza,Livio:2014gda,Gelmini:2015zpa,Bernabei:2022ath,Bernabei:2020mon,Bednyakov:2022dmc}
are interpreted in terms of {\em spin-independent} and {\em spin-dependent} 
cross sections of the DM particle interaction with nucleons,
let us discuss these two types of $\chi A$ interaction in more detail below.

\begin{figure}[h!] 
\vspace*{-60pt}\hspace*{-20pt}
\includegraphics[width=0.67\linewidth]{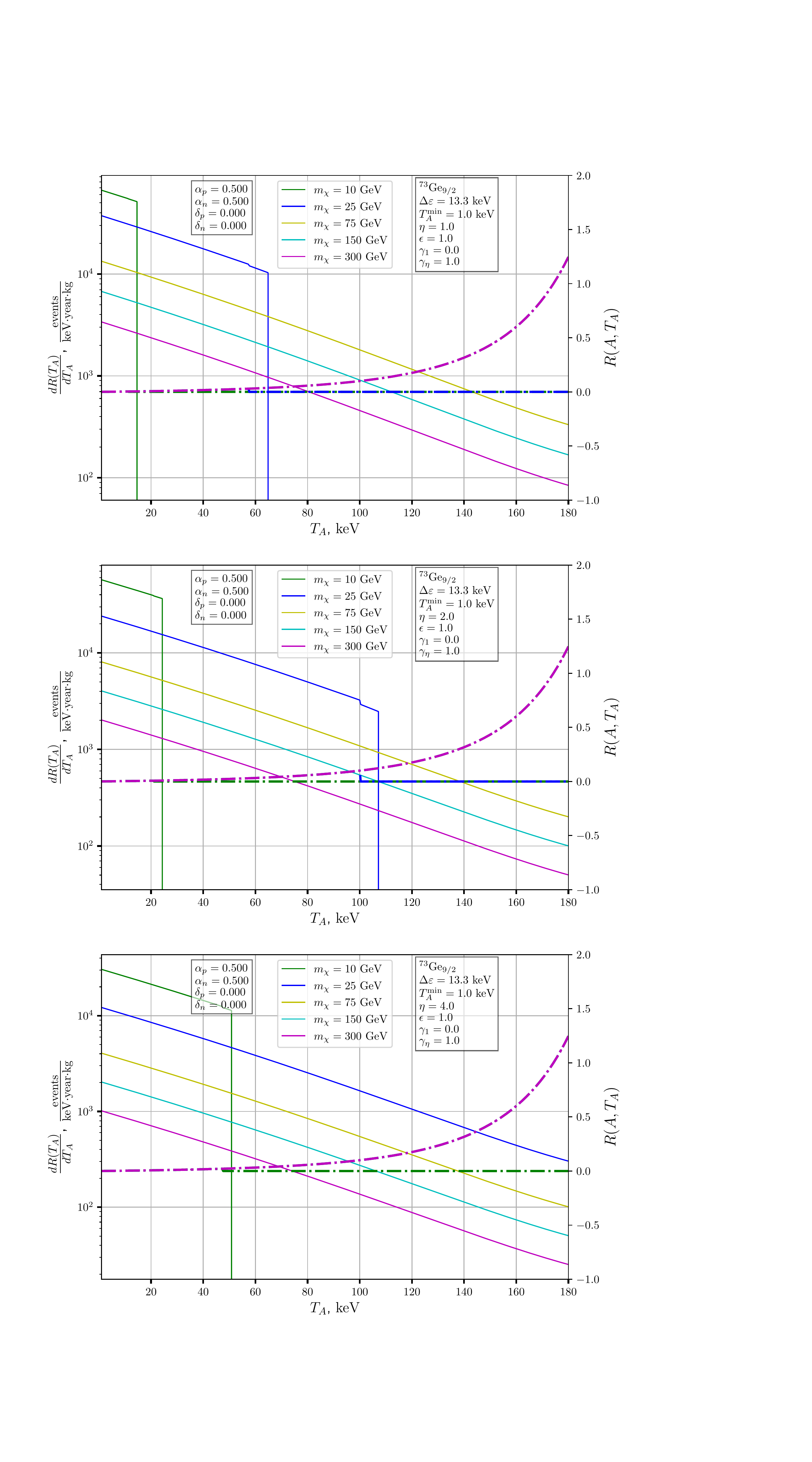}
\hspace*{-80pt}
\includegraphics[width=0.67\linewidth]{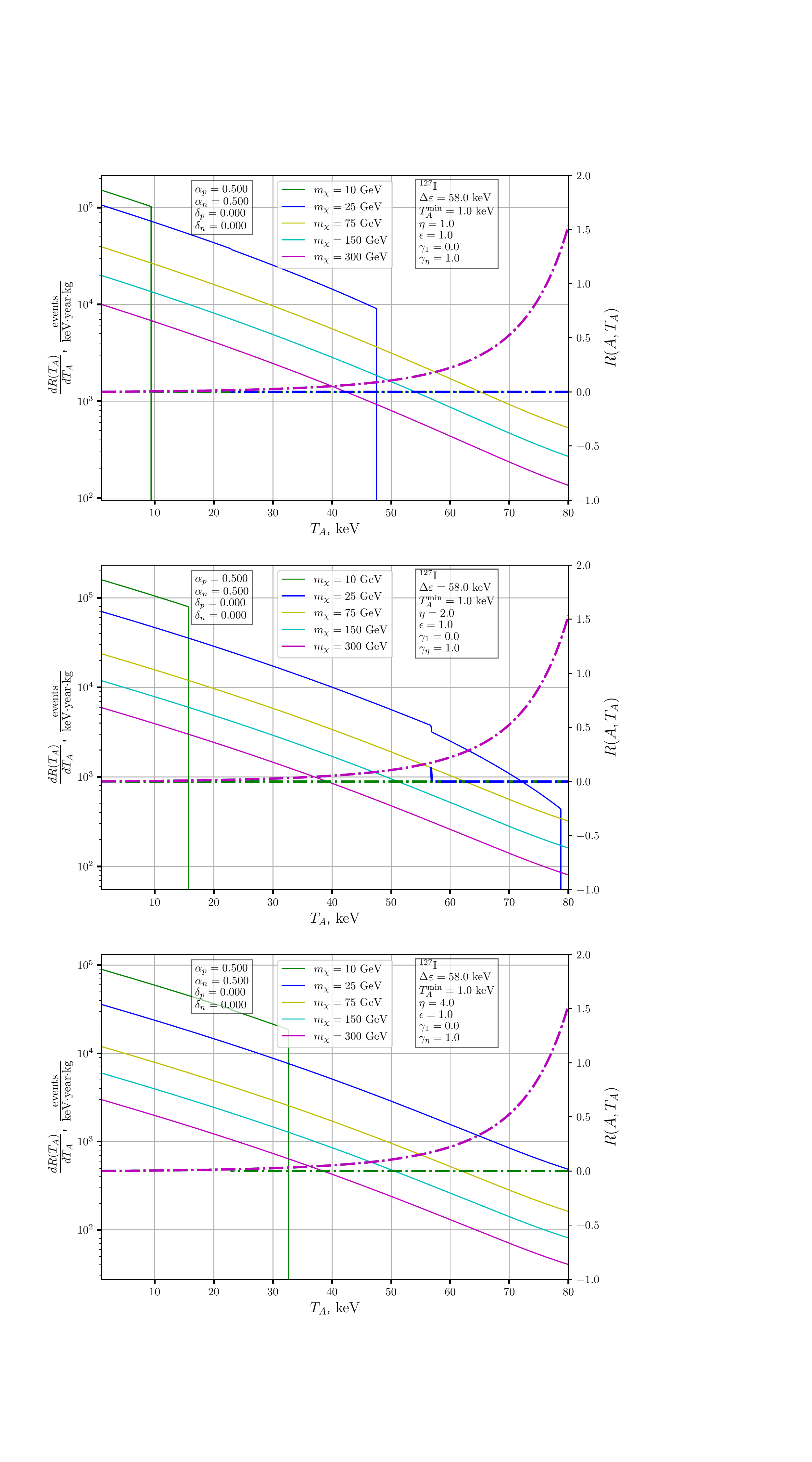}
\vspace*{-60pt}
\caption{\small For $^{73}$Ge and $^{127}$I, the ratio $ R^{\text{iso}}_{\text{Scalar}}(A,T_A)$ 
of the incoherent-to-coherent rate (right y-axis) from (\ref{eq:4DM-Results-Inc2CohRatio-Scalar-Iso}) 
and the differential rate (\ref{eq:4DM-DiffEventRate-Total-1DM}) (left y-axis)
are given as a function of the nuclear recoil energy, $T_A$, in the case of 
{\em isoscalar}\/ scalar $\chi A$ interaction ($\alpha_p=\alpha_n=0.5$).}
\label{fig:4DM-1DM-Rates-and-Ratios-vs-TA-eta-mchi-ScalarIsoScalar}
\end{figure} 

\subsection*{\em Spin-independent event rates}
In the nonrelativistic approximation 
\cite{Bednyakov:2018mjd,Bednyakov:2019dbl,Bednyakov:2021ppn}, 
the {\em scalar}\/ (in terminology of the direct DM  search the
{\em spin-independent}) interaction  has the same form as the purely vector one, 
when in the scalar products 
(\ref{eq:42chiA-CrossSection-ScalarProducts-ChiEta-All-WeakGeneralCurrent-definition})
only  constants $\alpha_p$ and $\alpha_n$ remain nonzero
\cite{Bednyakov:2022dmc}.
Then "Scalar"\/ ratio (\ref{eq:4DM-DiffEventRates-Ratios})  
of the differential incoherent-to-coherent rate can be written as follows:
\begin{eqnarray} \label{eq:4DM-Results-Inc2CohRatio-Scalar}
 R^{}_{\text{Scalar}}(A,T_A) &=&R_A(T_A)A\dfrac{A_p \alpha_p^2+ A_n \alpha_n^2}{(A_p \alpha_p+A_n \alpha_n )^2}  \Theta(T_A - T_A^{\min})  \Theta(T_{A^{*},\eta}^{\max}(r,\Delta\varepsilon_{mn})- T_A ) 
 . \qquad\end{eqnarray}
In the isoscalar case (when $\alpha_p= \alpha_n$), 
the dependence on these constants cancels out completely,  and 
  this formula behaves like the "pure nuclear ratio"\/ $R_A(T_A)$, i.e.,  with $R_I(A)=1$
 \begin{eqnarray} \label{eq:4DM-Results-Inc2CohRatio-Scalar-Iso}
 R^{\text{iso}}_{\text{Scalar}}(A,T_A) =  R_A(T_A) 
 \Theta(T_A - T_A^{\min})  \Theta(T_{A^{*},\eta}^{\max}(r,\Delta\varepsilon_{mn})- T_A ) 
 .\qquad \end{eqnarray}
If one of these parameters is significantly smaller than the other, for example, as in the SM, where $\alpha_p \ll \alpha_n$ (or, for simplicity, $\alpha_p \simeq 0$), one has the relation 
$$ R^{\alpha_p = 0}_{\text{Scalar}}(A,T_A)= R_A(T_A) \dfrac{A_p+A_n}{A_n}=\dfrac{1-F^2_A(T_A) }{A_n F^2_A(T_A)}, $$
which is also independent of $\alpha_n $ and differs from $R_A(T_A)$ only by about a factor of 2.
\par
\begin{figure}[h!] 
\centering \includegraphics[width=0.7\linewidth]{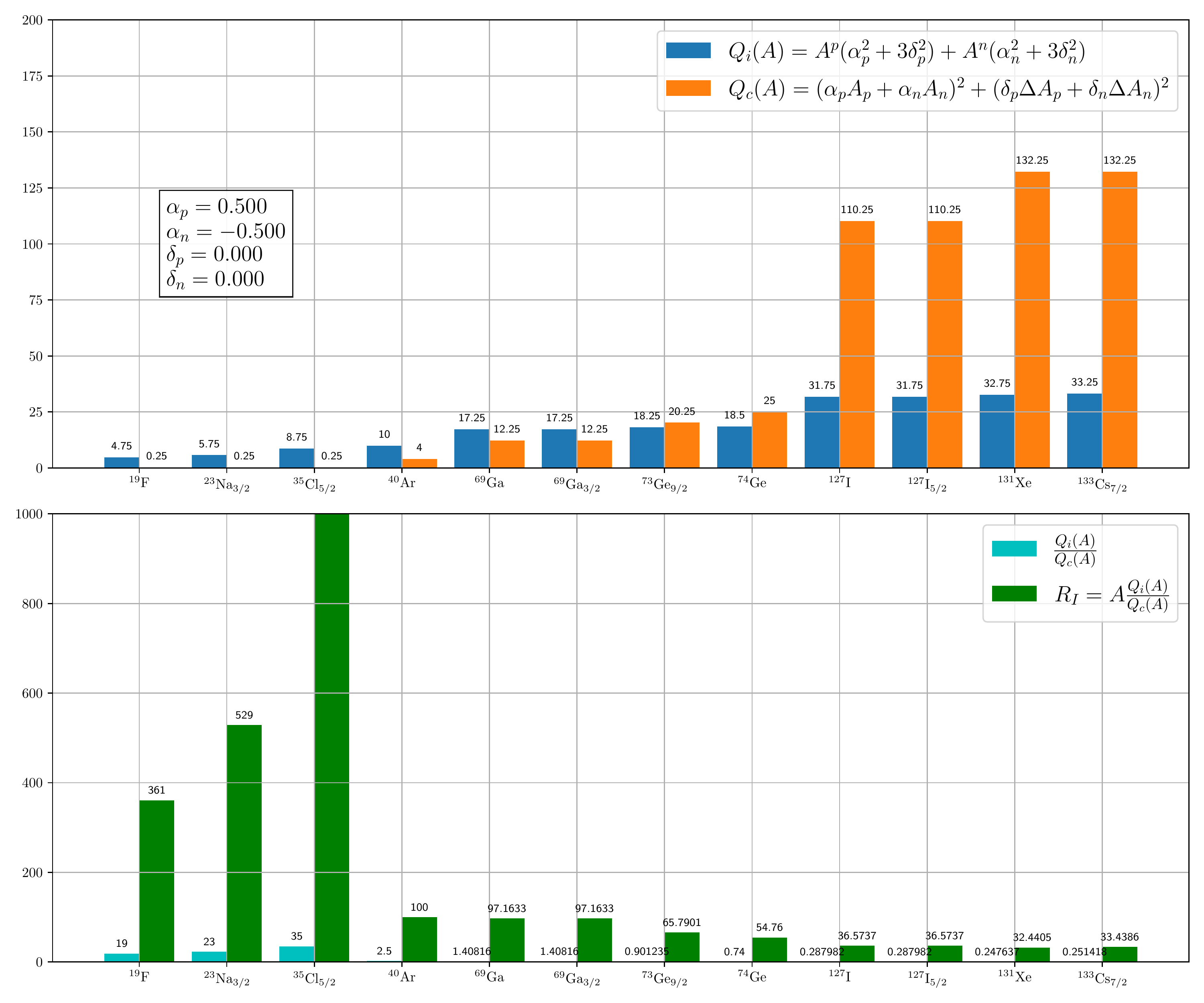}
\vspace*{-10pt}
\caption{\small Nuclear charge factors from formulas (\ref{eq:4DM-Q_c-and-Q_i-def}) and
(\ref{eq:4DM-Results-Inc2CohRatio-WeakCharge}) with parameters $\alpha_p= - \alpha_n=0.5$.} \label{fig:4DM-Table-with-Qs-in-SM-Scalar}
\end{figure} 
In Fig.~\ref{fig:4DM-1DM-Rates-and-Ratios-vs-TA-eta-mchi-ScalarIsoScalar}, 
isoscalar ratios (\ref{eq:4DM-Results-Inc2CohRatio-Scalar-Iso})
are  shown for $^{73}$Ge and $^{127}$I
along with differential total (coherent plus incoherent) rates from (\ref{eq:4DM-DiffEventRate-Total-1DM})
calculated with equal scalar couplings $a_p= a_n=0.5$.
For {\em this} scalar $\chi A$ interaction the inelastic rate does not noticeably contribute
 to the total differential event rate.
The ratio $ R^{\text{iso}}_{\text{Scalar}}(A,T_A)$ is less than one almost everywhere.
\par
In the {\em anti}\/-isoscalar case, where $\alpha_n= - \alpha_p$, as in the SM, one has
$$ R^{\text{SM}}_{\text{Scalar}}(A,T_A)=R_A(T_A)(A_p+A_n)\dfrac{A_p \alpha_p^2+ A_n \alpha_n^2}
{(A_p \alpha_p -A_n \alpha_n )^2}=R_A(T_A)\Big(\dfrac{p+1}{p-1}\Big)^2, 
\text{~~where~~} p =\dfrac{A_n}{A_p}
.$$
\begin{figure}[h!] 
\vspace*{-60pt}\hspace*{-20pt}
\includegraphics[width=0.67\linewidth]{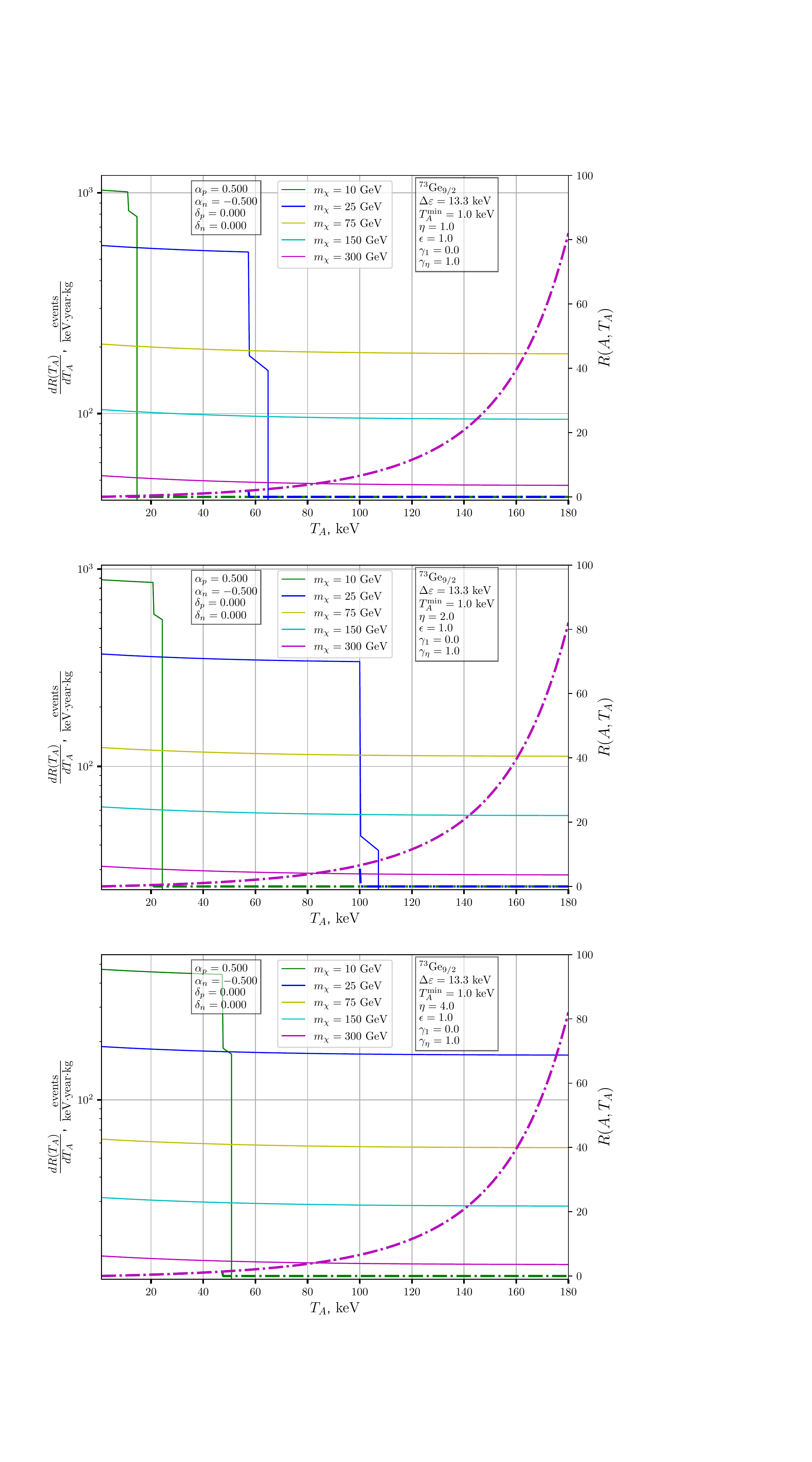}
\hspace*{-80pt}
\includegraphics[width=0.67\linewidth]{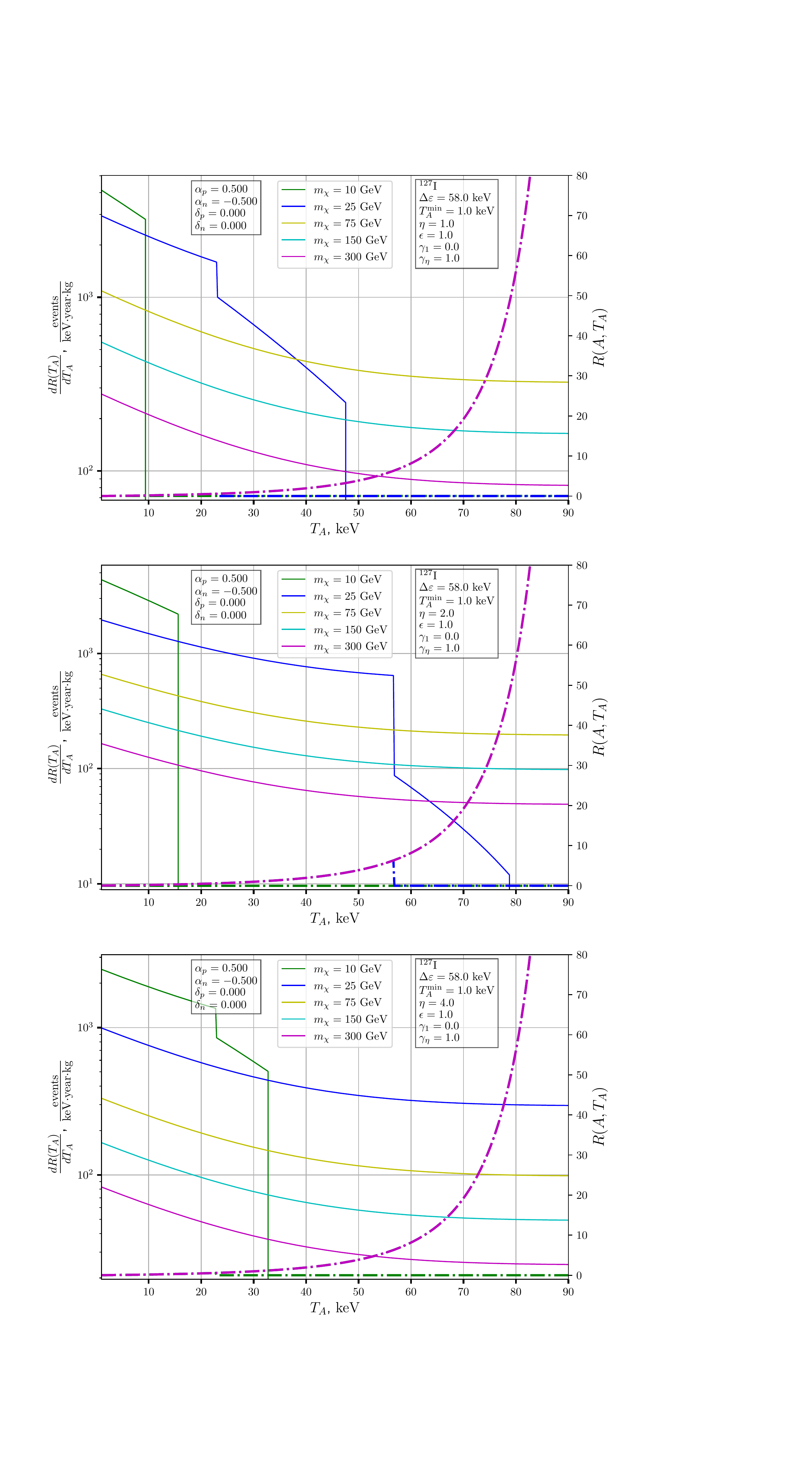}
\vspace*{-70pt}
\caption{The ratio $ R^{\text{SM}}_{\text{Scalar}}(A,T_A)$  of the incoherent-to-coherent rate from
(\ref{eq:4DM-Results-Inc2CohRatio-Scalar-Iso})  (right y-axis) 
and the differential rates (\ref{eq:4DM-DiffEventRate-Total-1DM}) (left y-axis) 
for $^{73}$Ge and $^{127}$I as a function of the nuclear recoil energy $T_A$ in the case of 
{\em anti-isosalar}\/  scalar $\chi A$ interaction ($\alpha_p= -\alpha_n= 0.5$).}
\label{fig:4DM-1DM-Rates-and-Ratios-vs-TA-eta-mchi-SMScalar}
\end{figure} 
\par
\begin{figure}[h!]
\vspace*{-30pt}\hspace{-10pt}
\includegraphics[width=0.55\linewidth]{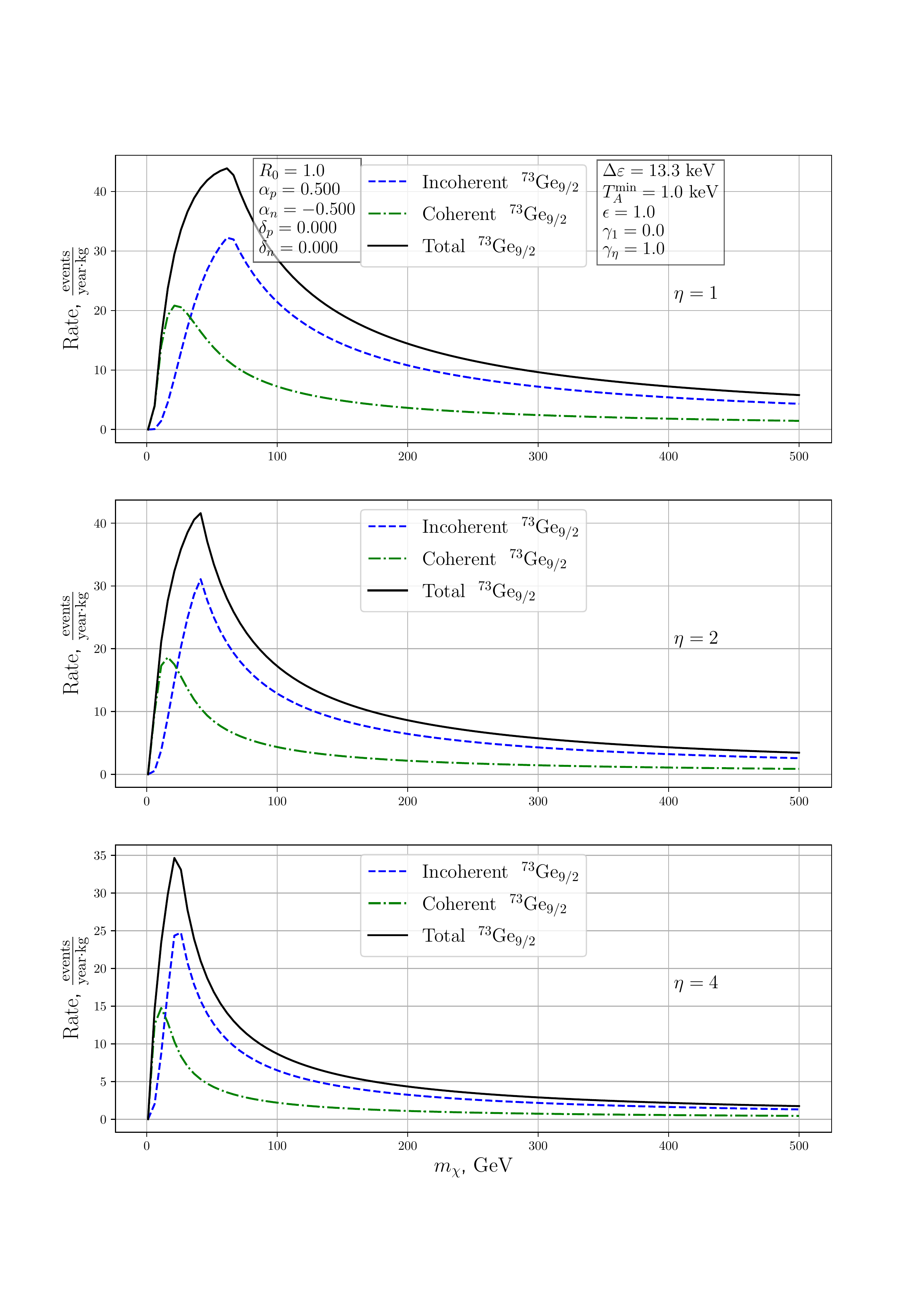}\hspace{-30pt}
\includegraphics[width=0.55\linewidth]{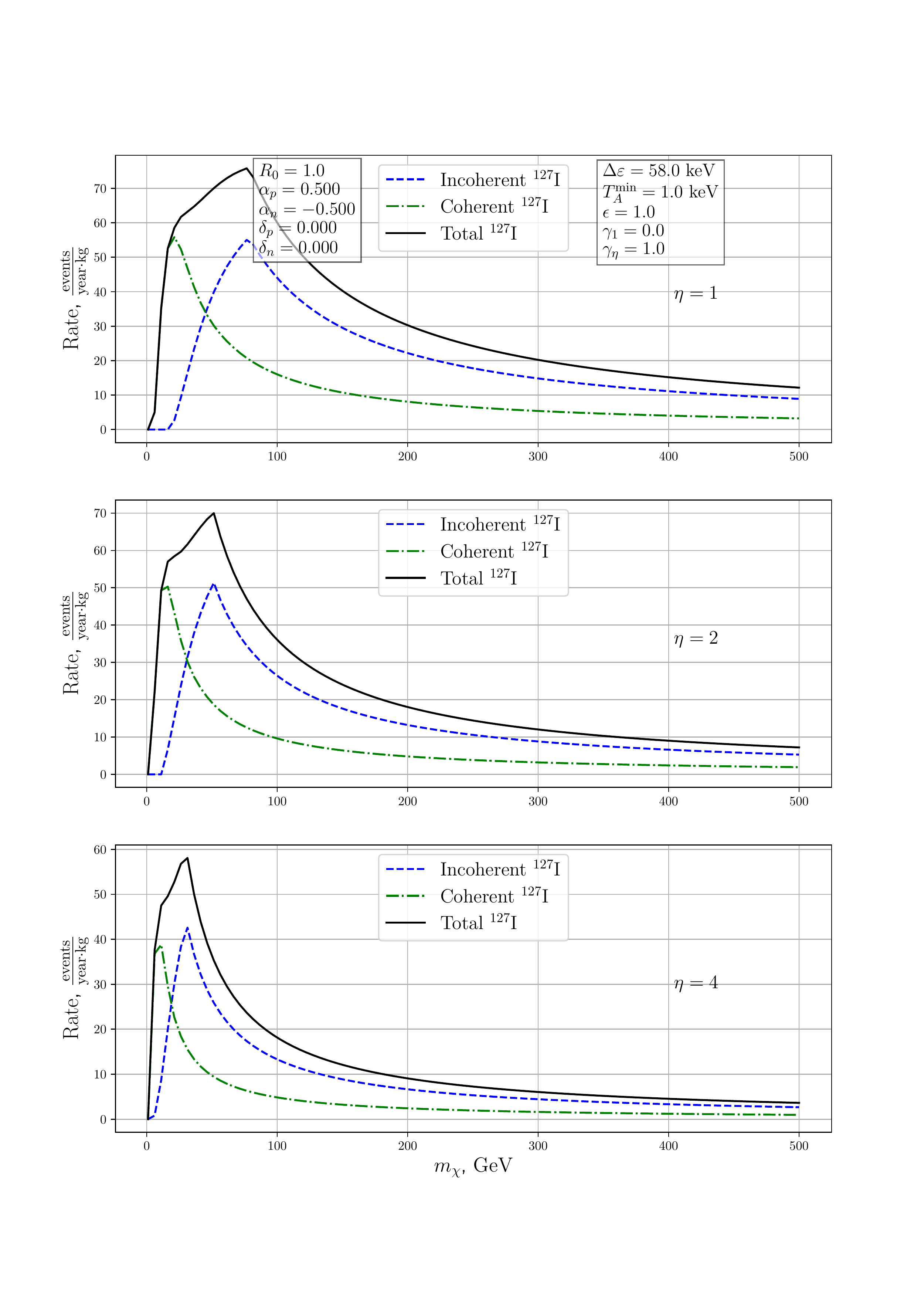}
\vspace*{-40pt}
\caption{\small Integrated event rate in the $^{73}$Ge and $^{127}$I nuclei
as a function of the mass $m_\chi$ and other "cosmic parameters"\/ in the case of
{\em anti-isoscalar}\/  scalar $\chi A$ interaction ($\alpha_p= -\alpha_n= 0.5$).
The coherent and incoherent contributions to the total rate are also shown.}
\label{fig:4DM-1DM-Rates-Integratedand-Ge73-SMScalar}
\end{figure}

From Table~\ref{tab:50chiA-Ratios-RIad-vs-A} and Fig.~\ref{fig:4DM-Table-with-Qs-in-SM-Scalar}
one can see that for the lightest nuclei there is no coherent contribution in this case, 
since the number of protons coincides with the number of neutrons and $p=1$.
For fluorine, sodium, and chlorine, the elastic contribution is strongly suppressed, and the inelastic charge factor is orders of magnitude higher than the elastic one.
The heavier mass of the target nucleus, the greater the value of $p$,
and the difference between $R^{\text{SM}}_{\text{scalar}}(A,T_A)$ and $R_A(T_A)$
presented in Fig.~\ref{fig:4DM-1DM-Rates-and-Ratios-vs-TA-eta-mchi-ScalarIsoScalar} is smaller.
Figure~\ref{fig:4DM-1DM-Rates-and-Ratios-vs-TA-eta-mchi-SMScalar} illustrates 
the anti-isocalar scalar $\chi A$ interaction by means of 
$R^{\text{SM}}_{\text{Scalar}}(A,T_A)$ and differential event rate.
The total, integrated over $T_A$, event rate as a function of $m_\chi$  in approximation  (\ref{eq:4DM-1DME-Total-EventRate}) 
is shown in Fig.~\ref{fig:4DM-1DM-Rates-Integratedand-Ge73-SMScalar}
for the scalar (with $\alpha_n= - \alpha_p$) interaction with (simplification) $R_0=1.0$.
The coherent and incoherent contributions to the total rate are also given in 
Fig.~\ref{fig:4DM-1DM-Rates-Integratedand-Ge73-SMScalar}.
\par
From Fig.~\ref{fig:4DM-1DM-Rates-and-Ratios-vs-TA-eta-mchi-ScalarIsoScalar} and
\ref{fig:4DM-1DM-Rates-and-Ratios-vs-TA-eta-mchi-SMScalar} one can see the
"main trend".
There is a smooth change of the "content"\/ in the measurable event rate as $T_A$ increases, i.e.,  
the elastic contribution to the rate is being smoothly replaced by the inelastic (incoherent) contribution.
The total (elastic+inelastic) expected event rate decreases, but not significantly 
(less than by an order of magnitude).
It can be believed that  the number of expected events 
will not be too small to completely escape the registration, despite these events change their origin.
From Fig.~\ref{fig:4DM-1DM-Rates-Integratedand-Ge73-SMScalar}
one can also see that the inelastic (incoherent) scalar contribution to the total integrated
rate dominates over the elastic (coherent) one with increasing $m_\chi$ ($m_\chi> 30\div 50\, $GeV$/c^2$).
\par 
Therefore, if a device is configured to register only the elastic scattering events, 
it rather quickly begins to see nothing, since if $T_A$ increases, the elastic processes quickly decrease.
At the same time, one has an increase of the expected inelastic events, which the device is not able of see. 
\begin{figure}[h!] 
\vspace*{-60pt}\hspace*{-20pt}
\includegraphics[width=0.65\linewidth]{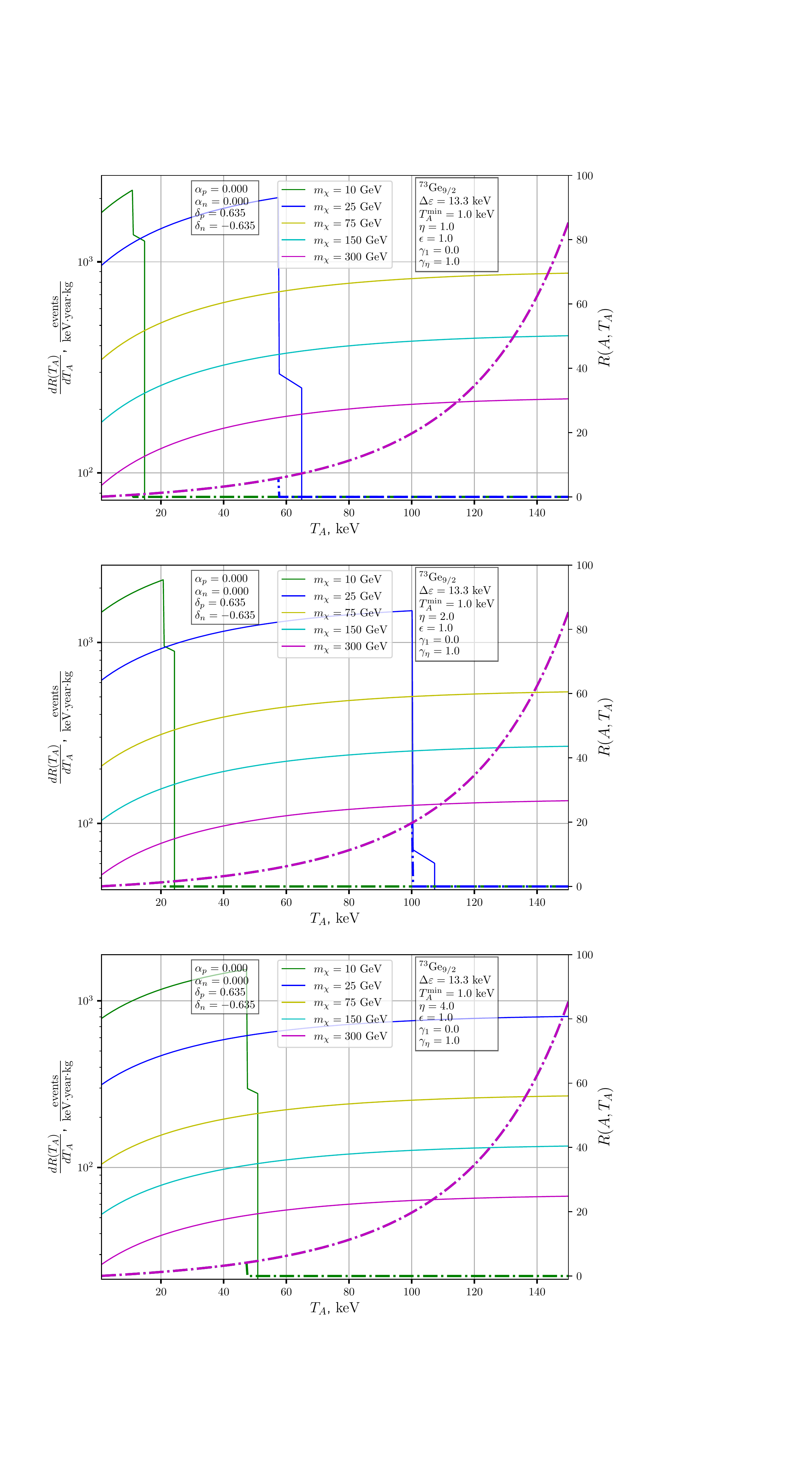}
\hspace*{-80pt}
\includegraphics[width=0.65\linewidth]{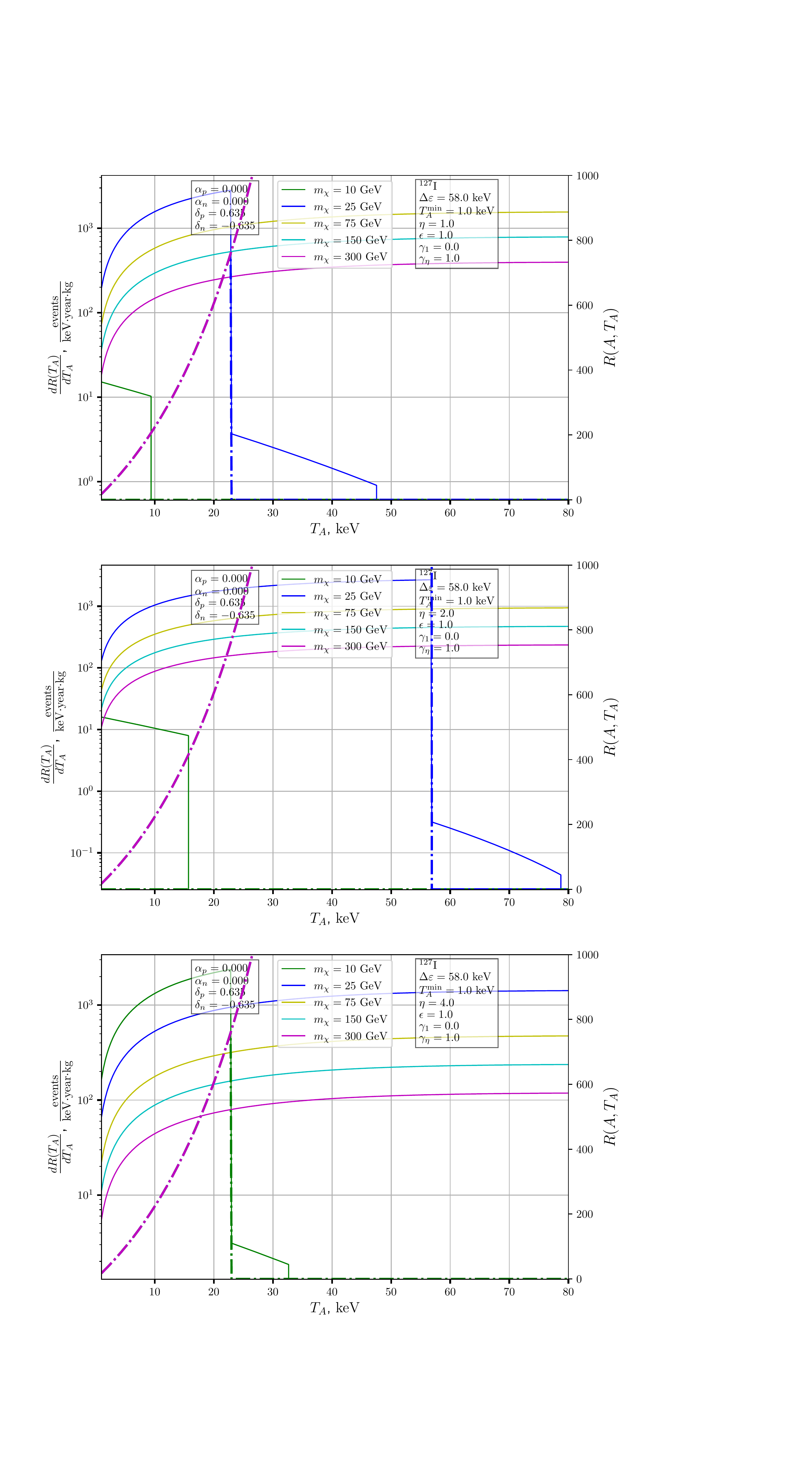}
\vspace*{-60pt}
\caption{The ratio $ R^{\text{SM}}_{\text{Axial}}(A,T_A)$ of the total incoherent-to-coherent rate
from (\ref{eq:4DM-Results-Inc2CohRatio-Axial}) and the differential rates (\ref{eq:4DM-DiffEventRate-Total-1DM}) 
are given for $^{73}$Ge and $^{127}$I 
as functions of the nuclear recoil energy $T_A$ for the 
SM-like axial SM interaction when $\delta_n=-\delta_p $
(the dependence on these parameters survives only in the rates).}
\label{fig:4DM-1DM-Rates-and-Ratios-vs-TA-eta-mchi-Axial-SM}
\end{figure}
The ability of the device to detect interaction of the DM particles with sufficiently large mass also falls sharply.
It turns out that the "wanted" $\chi A$ interaction could be possible,
but no way to detect it.

\subsection*{\em Spin-dependent event rates}
Let us turn to the {\em axial-vector}\/ $\chi A$ interaction, which is called {\em spin-dependent}\/
in the terminology of the direct DM search.
In this case,  the vector coupling constants $\alpha_p=\alpha_n=0$
in formula (\ref{eq:4DM-Results-Inc2CohRatio-WeakCharge}), and 
relation (\ref{eq:4DM-DiffEventRates-Ratios}) becomes
\begin{eqnarray}
 \label{eq:4DM-Results-Inc2CohRatio-Axial}
R^{\text{}}_{\text{Axial}}(A,T_A) = R_A(T_A)
\dfrac{3A(A^p \delta^2_p + A^n \delta^2_n)} {(\delta_p \Delta A_p + \delta_n \Delta A_n)^2}
 \Theta(T_A - T_A^{\min})  \Theta(T_{A^{*},\eta}^{\max}(r,\Delta\varepsilon_{mn})- T_A ) 
 . \quad \end{eqnarray}
One can see that for nuclei with zero spin (more precisely, when $\Delta A_{p/n}=0$)
the coherent contribution to the total cross section (rate as well) from the purely axial-vector interaction
is absent, and this relation loses its meaning.
\begin{figure}[h!]  
 \centering \vspace*{-40pt}\hspace{-25pt}
\includegraphics[width=0.55\linewidth]{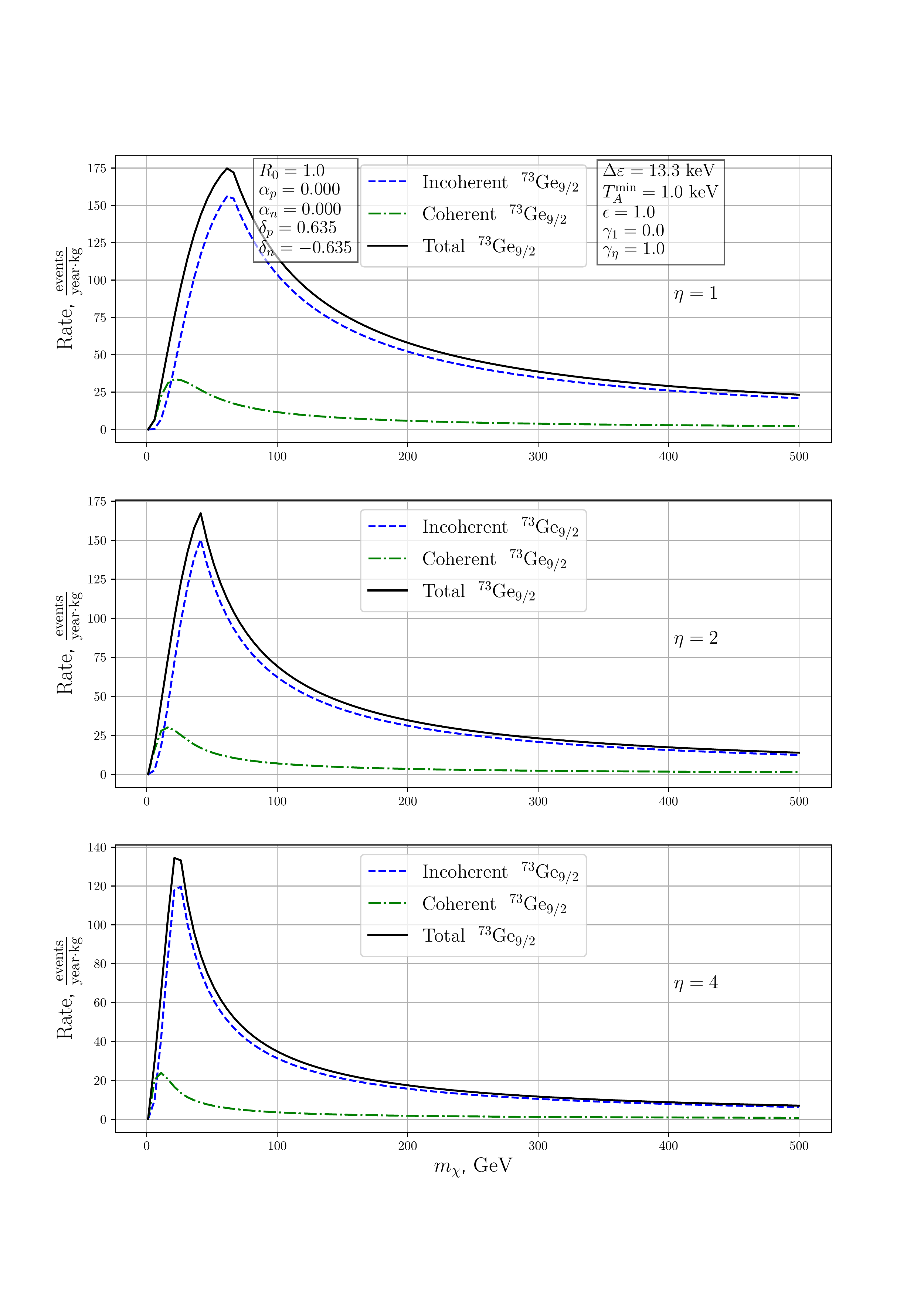}\hspace{-30pt}
\includegraphics[width=0.55\linewidth]{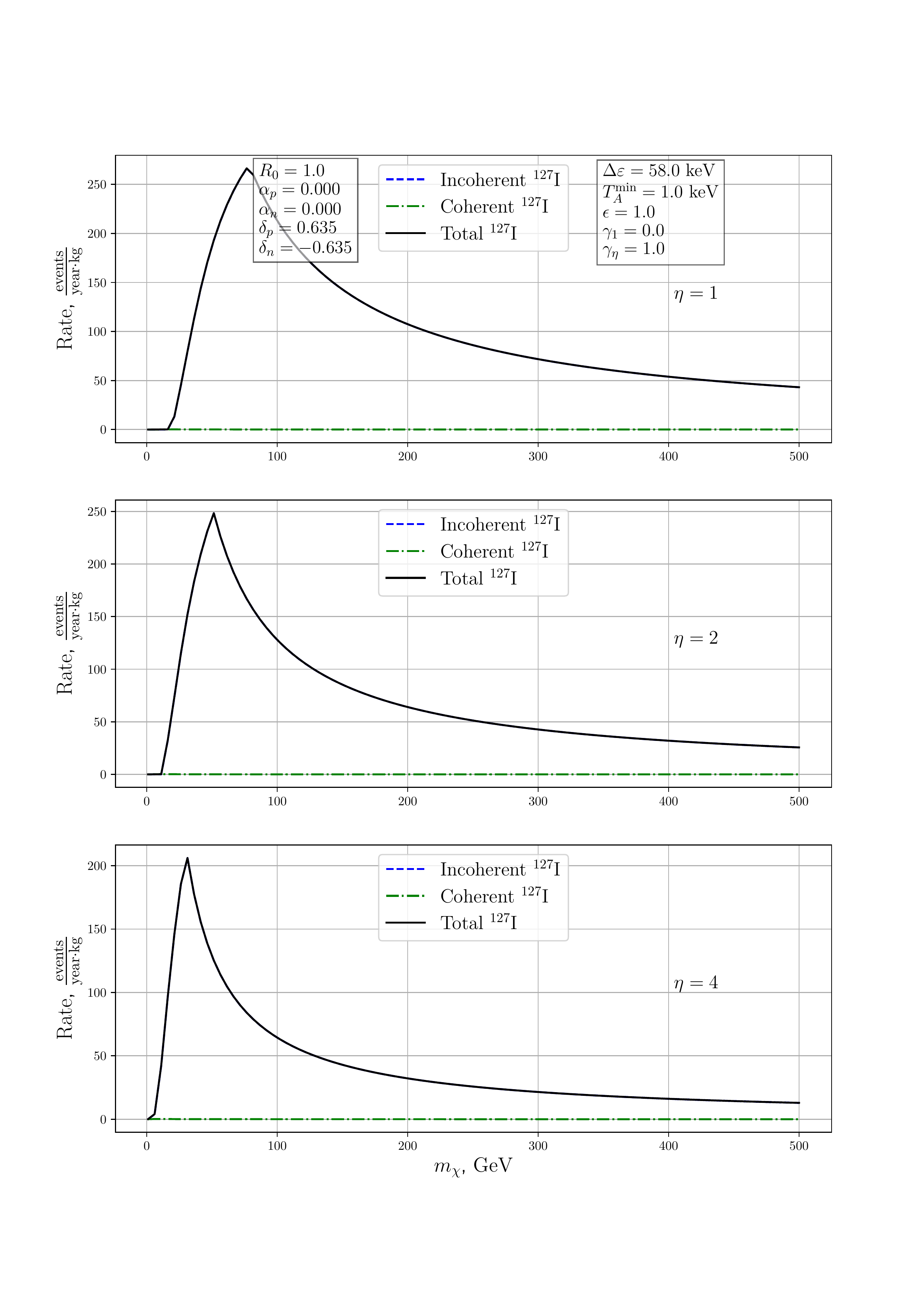}
\vspace*{-30pt}
\caption{\small Integrated total event rate in $^{73}$Ge and $^{127}$I as a function of the mass  $m_\chi$ and other "cosmic parameters"\/ in the case of SM-like {\em  axial-vector}\/ $\chi A$ interaction 
($\delta_n= - \delta_p$). The coherent and incoherent contributions to the total rate are also shown.}
\label{fig:4DM-1DM-Rates-Integratedand-Ge73-AxialSM}
\end{figure}
\par
 Since in the SM the axial effective constants of the proton and neutron are related,  
$\delta_n=-\delta_p$, the formula
$$  R^{\text{SM}}_{\text{Axial}}(A,T_A)= \dfrac{3 R_A(T_A) A^2} {(\Delta A_p - \Delta A_n)^2}
$$
does not depend on the axial coupling constant $\delta_p$.
This ratio is not suppressed by $A$, but, on the contrary, 
it is directly proportional to $A$, provided $\Delta A_p \ne \Delta A_n$
(see Fig.~\ref{fig:4DM-1DM-Rates-and-Ratios-vs-TA-eta-mchi-Axial-SM}).
Otherwise, relation (\ref{eq:4DM-Results-Inc2CohRatio-Axial}) loses its meaning again.
The total (coherent+incoherent), integrated over $T_A$, event rate as a function of $m_\chi$
for $^{73}$Ge and $^{127}$I with approximation (\ref{eq:4DM-1DME-Total-EventRate}) 
is shown in Fig.~\ref{fig:4DM-1DM-Rates-Integratedand-Ge73-AxialSM}
for SM-like axial-vector interaction ($\delta_n= - \delta_p$), where $R_0=1.0$ is taken for simplicity.
One can see from 
Fig.~\ref{fig:4DM-1DM-Rates-and-Ratios-vs-TA-eta-mchi-Axial-SM}
and Fig.~\ref{fig:4DM-1DM-Rates-Integratedand-Ge73-AxialSM}
that the coherent contribution to the total event rate is either very small 
or completely absent in this case of $\chi A$ interaction.
\begin{figure}[h!] 
\vspace*{-60pt}\hspace*{-20pt}
\includegraphics[width=0.67\linewidth]{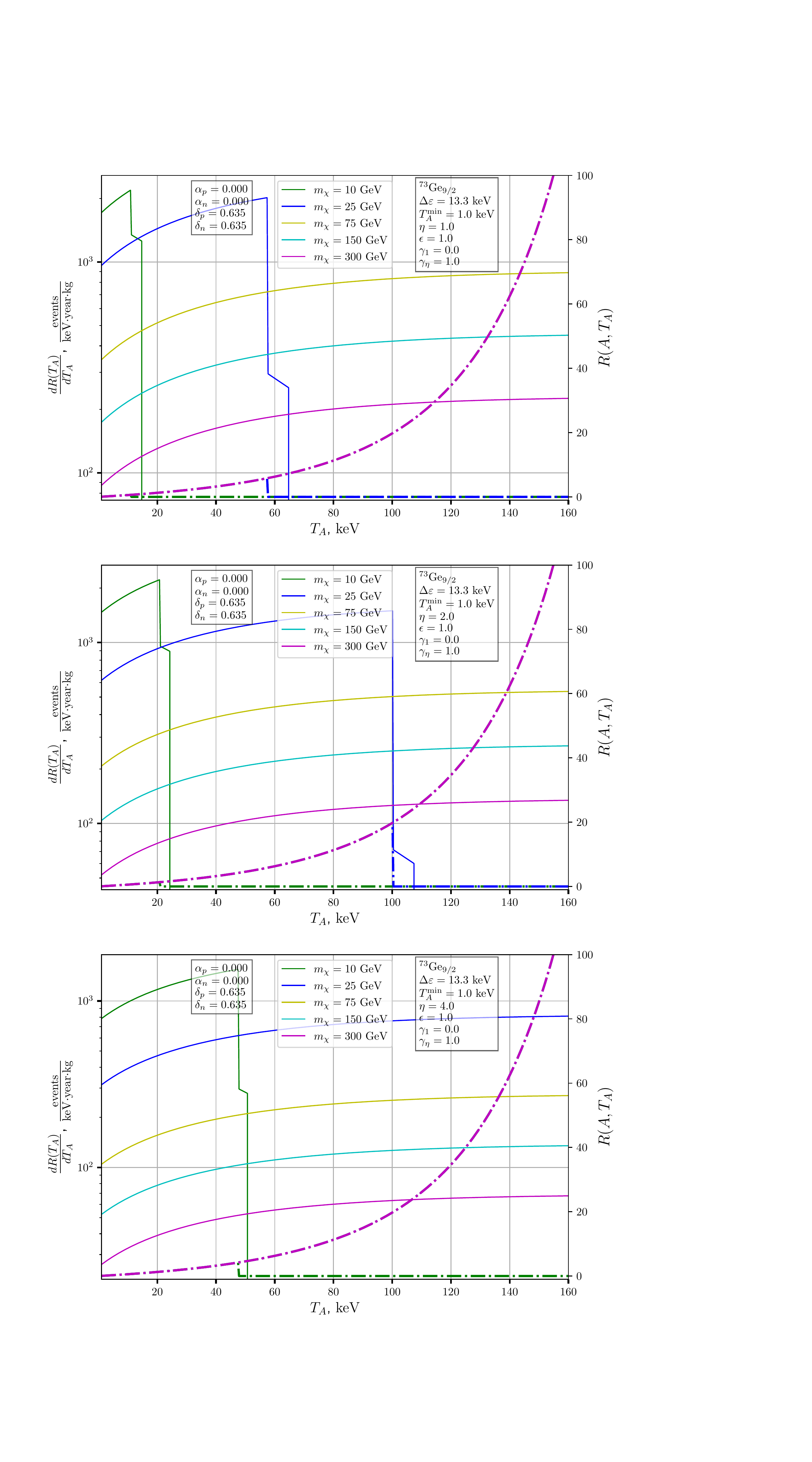}
\hspace*{-80pt}
\includegraphics[width=0.67\linewidth]{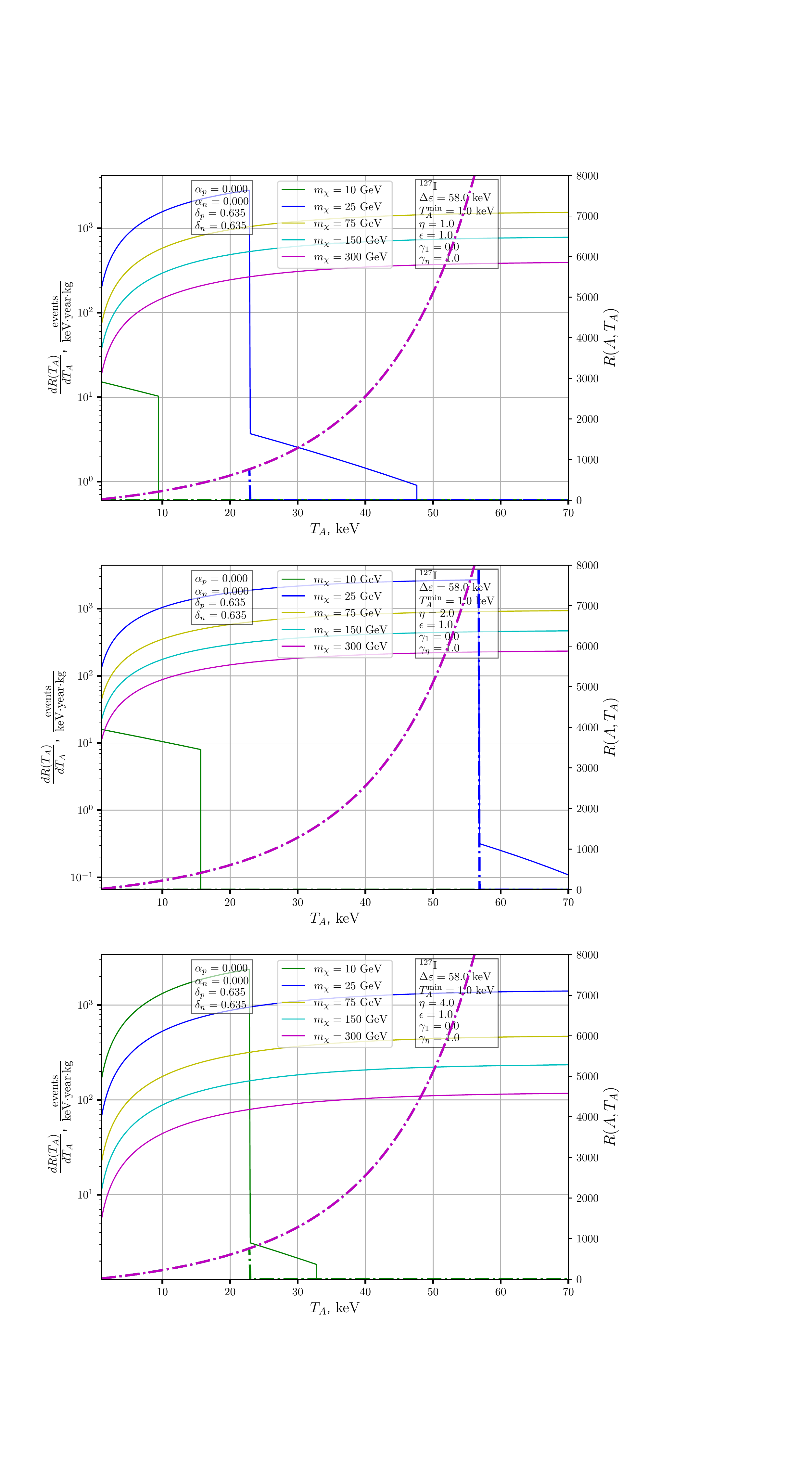}
\vspace*{-80pt}
\caption{The ratio $ R^{\text{spin}}_{\text{axial}}(A,T_A)$ from 
(\ref{eq:4DM-Results-Inc2CohRatio-Axial}) 
(right y-axis) and differential rate (\ref{eq:4DM-DiffEventRate-Total-1DM}) (left y-axis)
are presented for $^{73}$Ge and $^{127}$I 
as a function of the nuclear recoil energy $T_A$ in the case of {\em isoscalar}\/ 
axial-vector interaction, when $\delta_n=\delta_p$, (dependence on these parameters is preserved only in the rates).}
\label{fig:4DM-1DM-Rates-and-Ratios-vs-TA-eta-mchi-Axial-Iso}
\end{figure} 
\par
\begin{figure}[h!] \centering \vspace*{-30pt}\hspace{-25pt} 
\includegraphics[width=0.55\linewidth]{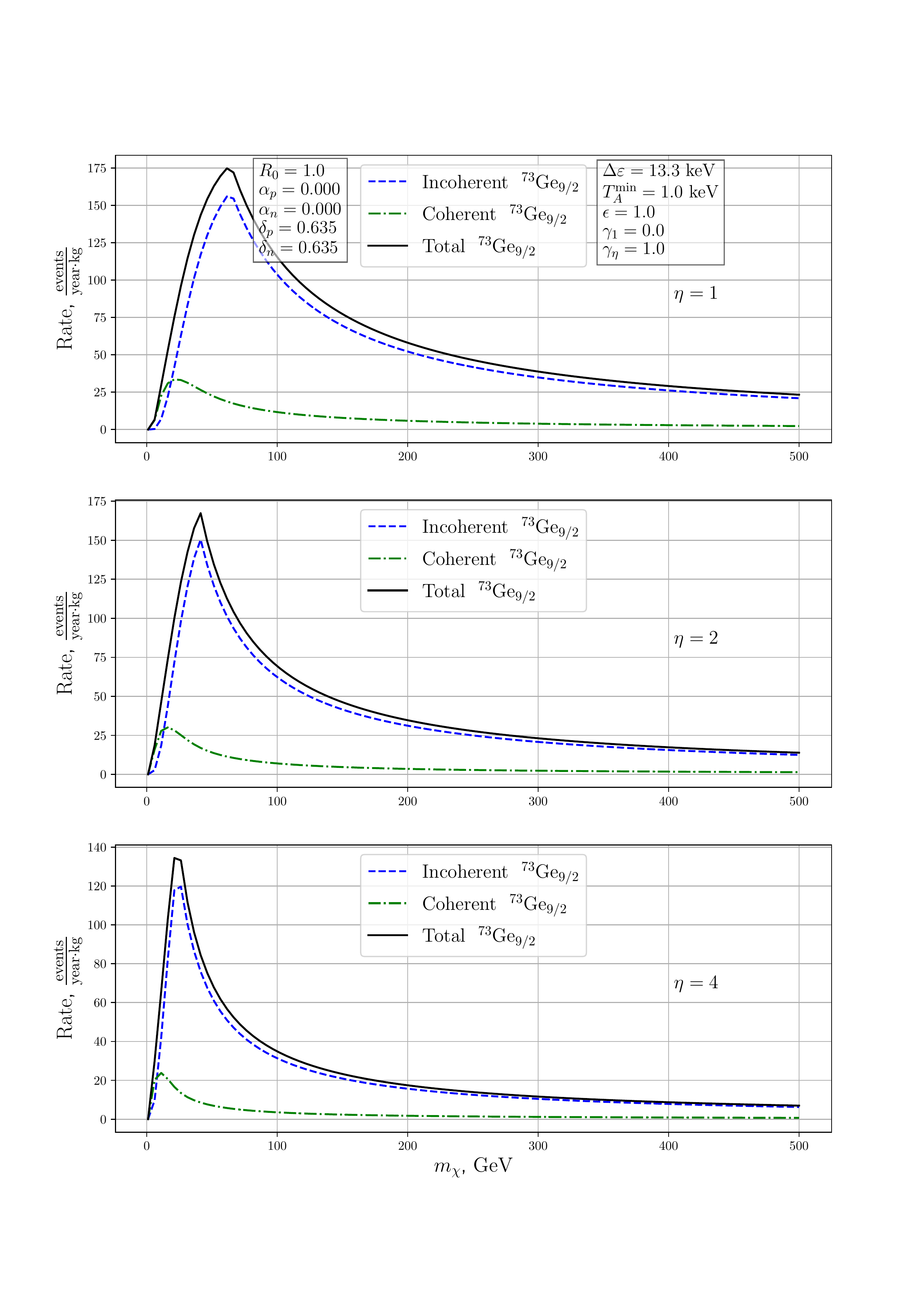} \hspace{-35pt}
\includegraphics[width=0.55\linewidth]{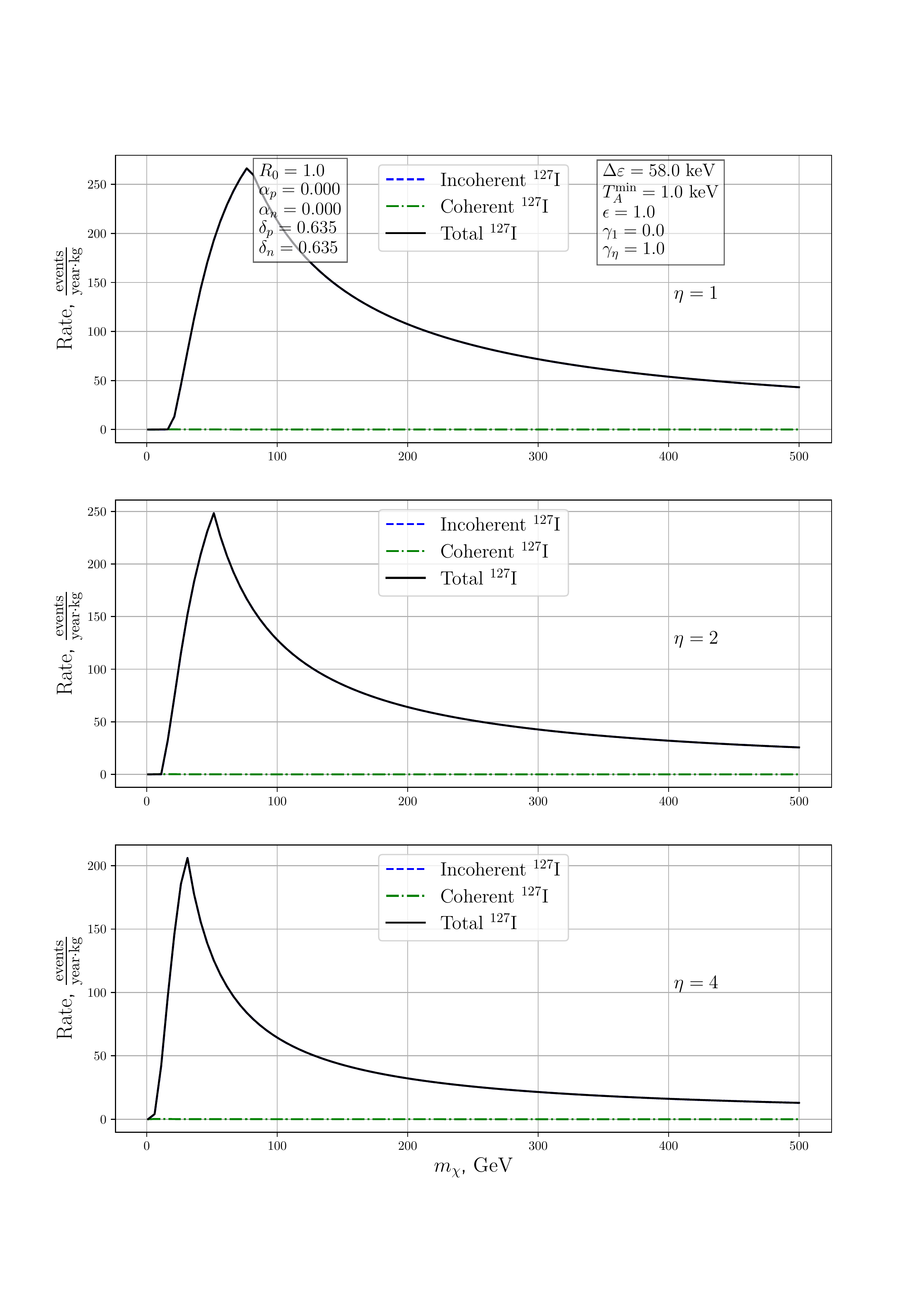}
\vspace*{-40pt}
\caption{\small Counting rate integrated  over $T_A$  in $^{73}$Ge and $^{127}$I
as a function of the mass $m_\chi$ and other
"cosmic parameters"\/ in the case of {\em isoscalar}\/  ($\delta_n=\delta_p$) axial 
(spin-dependent) $\chi A$ interaction.
The coherent and incoherent contributions to the total count rate are also shown.}
\label{fig:4DM-1DM-Rates-Integratedand-Ge73-AxialIso}
\end{figure}

When $\delta_n=\delta_p$, relation (\ref{eq:4DM-Results-Inc2CohRatio-Axial})  also
loses any dependence on the axial couplings and takes the form
\begin{eqnarray*}
R^{\text{spin}}_{\text{Axial}}(A,T_A)&=&  \dfrac{3 R_A(T_A)A^2} {( \Delta A_p + \Delta A_n)^2}
, \end{eqnarray*}
where $\Delta A \equiv \Delta A_p +\Delta A_n$ plays the role of the total spin of nucleus $A$.
This situation can be considered as  traditional spin-dependent interaction of (dark matter) $\chi$ particles with the target nuclei, since the coherent cross section is proportional to the square of the nuclear spin.
In Fig.~\ref{fig:4DM-1DM-Rates-and-Ratios-vs-TA-eta-mchi-Axial-Iso},  the set of plots
relevant to this kind of $\chi A$ interaction is given.
The total (coherent+incoherent), integrated over $T_A$, event rate as a function of $m_\chi$
for $^{73}$Ge and $^{127}$I nuclei (see (\ref{eq:4DM-1DME-Total-EventRate})) 
is shown in Fig.~\ref{fig:4DM-1DM-Rates-Integratedand-Ge73-AxialIso}
for the {\em iso-scalar}\/ ($\delta_n=\delta_p$) axial-vector $\chi A$ interaction
with $R_0=1.0$ taken for simplicity.
\par
From Fig.~\ref{fig:4DM-1DM-Rates-and-Ratios-vs-TA-eta-mchi-Axial-SM} and 
Fig.~\ref{fig:4DM-1DM-Rates-and-Ratios-vs-TA-eta-mchi-Axial-Iso}
the "general trend"\/ in the balance between the coherence and incoherence becomes "obvious".
Indeed, if  $\chi$ particles interact with nucleons via axial-vector current  {\em only}, 
it is not possible to detect these particles by means of their coherent (elastic) interaction
with zero-spin target nuclei, simply because the interaction  does not exist.  
The only way to discover $\chi$ particles is registration of their  inelastic (incoherent) interaction 
with nuclei.
Moreover, even if the target consists of nuclei with a nonzero spin,
the coherent (elastic, proportional to the square of the nuclear spin) contribution
to the expected total rate  is "not visible"\/ above the dominant "background" of the incoherent
(inelastic, proportional to the atomic mass $A$)
contribution almost in the  entire range of recoil energy $T_A$.
\par
Thus, in the case of axial-vector $\chi A$ interaction {\em only}, the elastic scattering channel, 
which is traditionally used with nonzero spin target nuclei in the spin-dependent DM search, is doomed to see nothing. 
This is due to the fact that all "signal-forming"\/ $\chi A$ interactions occur via the inelastic channel, to which traditional DM detectors are, as a rule,  insensitive.
The detector usually aimed at the direct DM search is constructed 
to have maximal sensitivity in the low energy recoil region (near the threshold).
It is unable to register the high-energy (compared with the threshold) 
radiation from the nuclear deexcitation 
caused by the inelastic $\chi A$ interaction and has also bad sensitivity at high  $T_A$.
  Again, it looks quite possible that the "wanted" (spin-dependent) DM interaction can have a noticeable intensity, 
  but a typical DM detector simply cannot detect it.
 \par
 The dominance of the inelastic over the elastic spin-dependent interaction of DM particles with $^{129,131}$Xe nuclei was pointed out in \cite{Baudis:2013bba}, 
 where the inelastic channel was determined by the excitation of one low-lying level, 
 and the expected detectable (scintillation) signal was the sum of the recoil energy 
 and the energy of photons from the nuclear deexcitation.
  \par
 Indeed,  if the transition to the excited state of the nucleus is possible
and took place, i.e.,  the $\chi A$ inelastic interaction occurred
\footnote{This “classical” inelastic approach should not be mixed up with “inelasticity” caused by the transition of the incident $\chi_1$ particle (of dark matter) to a more massive $\chi_2$ particle (also from the dark sector). The nucleus is interacting coherently \cite{Giudice:2017zke,XENON:2020fgj,Zurowski:2020dxe,Baryakhtar:2020rwy,Feng:2021hyz,Filimonova:2022pkj,Bell:2022yxn,Aboubrahim:2022lwb}.},
the nucleus should eventually return to its original (ground) state.
This deexcitation of the nucleus should be accompanied by some release of energy, for example, in the form of $\gamma$ radiation.
The detection of these  $\gamma$-quanta 
was proposed in 1975~\cite{Donnelly:1975ze}.
The energy spectrum of these photons is determined by the structure of nuclear excitation levels
and is fixed for each nucleus.
These photons can produce a detectable signal, which will be correlated 
with the target irradiation time if the source of the external $\chi$ particles is, for example, an accelerator.
According to the figures given above (see, for example,
Fig.~\ref{fig:4DM-1DM-Rates-and-Ratios-vs-TA-eta-mchi-Axial-Iso}), 
the prospects for registration of a noticeable number of events with $\gamma$ quanta 
from incoherent $\chi A$ interaction do not seem futile,
provided that the  $\chi A$ interaction itself has detectable intensity for modern detectors.
\par
Recall
\cite{Bednyakov:2018mjd,Bednyakov:2021ppn,Bednyakov:2021bty,Bednyakov:2022dmc}
that for each target nucleus these photons will be characterized by three important parameters.
First, their energy is often noticeably higher than the nuclear recoil energy 
($T_{\text{$^{127}$I}}\le $ 20 keV in Fig.~\ref{fig:4DM-1DM-Rates-and-Ratios-vs-TA-eta-mchi-Axial-Iso}).
Second, emission of photons with the typical nucleus deexitation energies
will be shifted in time (relative to the beginning of the excitation)
by the deexitation time typical of the excited nuclear level.
Third, the counting rate of these $\gamma$ quanta will be proportional to the ratio
of expected incoherent-to-coherent number of events, 
$N_\text{inc}/N_\text{coh}$, where
$N_\text{inc/coh} = \int dE_\nu \Phi(E_\nu)\int_{dT_A^\text{min}}^{dT_A^\text{max}} dT_A\dfrac{d\sigma_\text{inc/coh}}{dT_A} \varepsilon(T_A),$ and $\varepsilon(T_A)$ is the detector efficiency.
\par
Let us discuss the accuracy of the inelastic $\chi A$ cross section
in the approach \cite{Bednyakov:2018mjd,Bednyakov:2019dbl,Bednyakov:2021ppn,Bednyakov:2022dmc}.
This (combined) inelastic cross section is the upper boundary for contributions to the total measured $\chi A$ cross section from the totality of all allowed inelastic subprocesses (in the given kinematic region). 
This upper limit only follows from the probability conservation rule: the sum of the probabilities of all inelastic processes and the probability of the elastic process is equal to unity.
The degree of “saturation” of this combined inelastic cross section by individual inelastic channels
(contributions from transitions to different allowed levels) depends on the structure of excitation levels of
a particular nucleus and on the incident particle energy. 
Clearly, this saturation cannot be 100\% reproduced numerically. 
\par
Inelastic cross sections for the $\nu(\bar{\nu}) A$ scattering due to weak neutral currents 
were calculated within modern nuclear models for specific excitation levels of particular nuclei, for example, in
 \cite{Divari:2012zz,Divari:2012cj,Lykasov:2007iy}. 
 Recently, similar calculations have been generalized to the 
 neutral DM particle inelastic scattering off nuclei \cite{Sahu:2020kwh,Sahu:2020cfq,Dutta:2022tav}.
In \cite{Sahu:2020kwh,Sahu:2020cfq}, detailed calculations were performed for the 
elastic and inelastic event rates in $^{73}$Ge, $^{127}$I, $^{133}$Cs, $^{133}$Xe 
as well as in the $^{23}$Na and $^{40}$Ar nuclei. 
 It was shown that incoherent (inelastic) WIMP--nucleus processes can be noticeably enhanced. 
Therefore,  the authors of \cite{Sahu:2020kwh,Sahu:2020cfq} confirmed 
importance of the inelastic channel at high recoil energies
first indicated in \cite{Bednyakov:2018mjd,Bednyakov:2019dbl,Bednyakov:2021ppn}.
\par
In \cite{Dutta:2022tav},  inelastic cross sections for scattering of neutrinos and DM particles off
the  $^{40}$Ar, $^{133}$Cs and $^{127}$I nuclei were calculated within the nuclear shell model (NSM). 
The (total) inelastic $\nu A$ cross section \cite{Dutta:2022tav}  at rather low neutrino energies
($E_\nu< $ 20 MeV) was in agreement with other calculations 
(including ones from \cite{Bednyakov:2018mjd}), but at $E_\nu \simeq$  40 MeV 
the cross sections were about an order of magnitude smaller than the relevant upper limits from \cite{Bednyakov:2018mjd}.
Furthermore, from  Fig.~12 
of \cite{Dutta:2022tav}, one can notice that the "simple"\/ Helm nuclear form factor $|F_{\rm H}(T_A)|^2$
used here and in \cite{Bednyakov:2018mjd,Bednyakov:2019dbl,Bednyakov:2021ppn}
coincides with high accuracy with the
more sophisticated form factor $|F_{\rm NSM}(T_A)|^2$ from \cite{Dutta:2022tav}, 
provided both $|F_{\rm NSM/H}(T_A) |^2\ge 0.01$.
\par 
Finally let us stress the following. 
The (individual) inelastic cross sections (and rates) 
calculated within modern nuclear models
are always smaller than the ones obtained here within the formalism  \cite{Bednyakov:2018mjd,Bednyakov:2019dbl,Bednyakov:2021ppn}.
It looks reasonable, because the upper bound for all possible channels
is always larger than a sum of some set of individual channels.
The calculations of inelastic cross sections (and rates) within advanced nuclear models do
not allow the coherence-to-incoherence transition to be quantitatively controlled.
If the "coherency condition"\/ is satisfied "with high precision"\/, 
one can reliably use the "coherent" formulas, otherwise one can reliably 
use the "incoherent"\/ formulas. 
Nobody knows what to do if the  "coherency condition"\/ is satisfied "with bad precision"\/.
There is no any "coherency condition"\/ 
in the approach \cite{Bednyakov:2018mjd,Bednyakov:2019dbl,Bednyakov:2021ppn},  
since the coherence-to-incoherence transition is automatic.

\section{Conclusion} \label{60chiA-Conclusions}
Recall that the formalism \cite{Bednyakov:2018mjd,Bednyakov:2019dbl,Bednyakov:2021ppn}
 proposed to describe the neutrino--nuclear interaction, was generalized in 
\cite{Bednyakov:2022dmc} to the case of non-relativistic
\footnote{The relativistic version of the description is given  in \cite{Bednyakov:2023bbg}.}
weak interaction of a massive neutral particle ($\chi$ lepton) with a nucleus as a composite system.
This interaction was parameterized via effective coupling constants  which determine the probability 
amplitudes by means of  the scalar products of the lepton and nucleon currents.
Regardless of the specific nuclear models, explicit use of the complete set of the nuclear quantum states 
allowed one \cite{Bednyakov:2018mjd,Bednyakov:2019dbl,Bednyakov:2021ppn}
 to obtain the unified description of the elastic (coherent) and inelastic (incoherent) 
scattering of the $\chi$ lepton off nucleus $A$.
The behavior of the elastic and inelastic $\chi A$ cross sections 
 is determined by the nucleon form factors, $|F_{p/n}(\bm{q})|^2$,  averaged over the initial nuclear states. 
 As the nuclear recoil energy $T_A$ increases, one has a controlled transition from the 
dominance of the elastic interaction to the dominance of the inelastic one
 in the nonrelativistic scattering of the {\em massive}\/ $\chi$ particle off the nuclei.
In other words, as $T_A$ increases, the registrable events will change their "origin"\/.
If a DM detector is configured to register only elastic scattering events, it begins to lose the ability to "see" anything, because the elastic processes rather quickly decrease with increasing $T_A$.
At the same time, the expected number of inelastic events increases, 
but the detector is unable to register them, because it cannot see the deexcitation photons, 
or their energies are beyond the detector capabilities, etc.
 \par
Another case is also possible.
If the nuclear recoil energy produced after $\chi A$ scattering is below the threshold of the detector, 
$T_A< T_A^{\min}$, the elastic signal cannot be detected.
With this "invisible"\/ $T_A$,  the only evidence of the $\chi A$ interaction could be from the nuclear deexitation radiation, i.e., the inelastic signal, although its intensity can be an order of magnitude smaller than the intensity of the elastic one.
In general, when only the nuclear recoil energy, $T_A$, can be registered, 
it is impossible to understand whether the elastic or the inelastic interaction took place.
When the inelastic signal "falls"\/ into the registration area of the elastic one, its origin is completely unclear.
\par
In the most critical way, this "phenomenon"\/ can manifest itself in the
direct DM search experiments, the results of which  are usually
 interpreted in terms of the {\em spin-independent}\/ and {\em spin-dependent}\/ cross sections of 
the galactic DM particle interaction with nucleons. 
\par 
In this paper, the main statements of the previous one
\cite{Bednyakov:2022dmc} are transformed into numerical predictions
for the direct DM search experiments. 
For the first time in the formalism \cite{Bednyakov:2018mjd,Bednyakov:2019dbl,Bednyakov:2021ppn,Bednyakov:2022dmc}
the effect of nonzero nuclear excitation energy  was explicitly taken into account
in calculation of the event rates caused by inelastic $\chi A$ interactions.
The resulting rigid kinematic correlation between this excitation energy
and the allowed recoil energy of the excited nucleus appreciably limits the possibility of detecting 
 inelastic events with some set of target nucleus.
\par
In addition to the Standard Halo Model  (SHM) of the DM particle distribution
near the Earth,  the role of other DM distributions 
with higher maximal speeds of DM particles was considered.
There are several mechanisms (proposed, for example,  in \cite{Feng:2021hyz,Bardhan:2022ywd,CDEX:2022fig,Xia:2022tid,Granelli:2022ysi,Wang:2021jic}), 
which allow higher DM-particle velocities at a level of $ |\bm{v}|/c  \simeq 10^{-2} \div 10^{ -1}$.
\par
Three variants of the DM particle interaction with nucleons are considered \cite{Bednyakov:2022dmc}.
The first is the nonrelativistic weak interaction of the Standard Model  
with some suppression parameter $R_0\simeq 10^{-(3\div 4)} R^w_0$ (see
formula (\ref{eq:4DM-R_0})).
The plots for the differential and total expected event rates are
shown in Figs.~\ref{fig:4DM-1DM-Rates-and-Ratios-vs-TA-eta-mchi-Ge73}--%
\ref{fig:4DM-1DM-Rates-Integratedand-I127}.
The incoherent (inelastic) channel does not look much more promising than the usual coherent (elastic) one.
Roughly speaking, in the SM-like $\chi A$ interaction they almost coincide.
This can also be seen for the event rates integrated over $T_A$  in
Figs.~\ref{fig:4DM-1DM-Rates-Integratedand-Ge73}-\ref{fig:4DM-1DM-Rates-Integratedand-I127}.
\par 
The second is the {\em spin-independent}\/, or scalar DM-particle interaction.
When the scalar interaction is the same for the proton and the neutron (iso-scalar), 
the above-mentioned smooth change of the total rate "filling"\/ with increasing $T_A$  is also visible in Fig.~\ref{fig:4DM-1DM-Rates-and-Ratios-vs-TA-eta-mchi-ScalarIsoScalar}.
Nevertheless, the incoherent contribution does not look dominant here.
However, the anti-isoscalar {\em spin-independent}\/ $\chi A$ interaction, 
where the proton and neutron couplings differ only by the sign, 
looks more interesting.
Here (see Fig.~\ref{fig:4DM-1DM-Rates-and-Ratios-vs-TA-eta-mchi-SMScalar})
the smooth change of the total rate "filling"\/ with the growth of $T_A$ is visible very clear.
The inelastic contribution completely dominates at $T_A\ge 50\div 60\, $keV.
This dominance, especially with increasing $m_\chi$, is also clearly seen for the total event rates in
Fig.~\ref{fig:4DM-1DM-Rates-Integratedand-Ge73-SMScalar}.
\par 
The  third is the {\em spin-dependent}\/,   or axial-vector DM particle interaction with nucleons.
In this case, any detection of the DM particles via the coherent (elastic) interaction   
with zero-spin nuclei is not possible.
The only way to observe this interaction is through the inelastic (incoherent) channel. 
Moreover, even if the nucleus has the nonzero spin, the coherent (proportional to the square of the nuclear spin) contribution to the measurable rate is hardly visible above the incoherent
(proportional to the atomic mass of the nucleus $A$) contribution 
for almost all $T_A$ (see Fig.~\ref{fig:4DM-1DM-Rates-and-Ratios-vs-TA-eta-mchi-Axial-SM} and
\ref{fig:4DM-1DM-Rates-and-Ratios-vs-TA-eta-mchi-Axial-Iso}). 
The coherent contribution to the total (coherent plus incoherent and 
integrated over $T_A$) rate is not visible and is negligible
(see Fig.~\ref{fig:4DM-1DM-Rates-Integratedand-Ge73-AxialIso}).
Therefore, in the case of the pure axial-vector $\chi A$ interaction, a 
detector traditionally focused on recording the elastic spin-dependent DM signal is
"doomed"\/ to see nothing, since all potentially "signal-forming"\/ interactions go through the inelastic channel, 
to which this detector is, as a rule, insensitive.
\par 
One meets the situation where the ``wanted'' DM interaction may well have 
a rather noticeable (potentially detectable) intensity,  
but the instrument used to search for the DM is unable to detect it.
Furthermore, since the nature of the DM particle interaction with nucleons is beyond 
the Standard Model and is not yet experimentally determined,
the other combinations of $\chi$--nucleon couplings (in addition to discussed above),  
which can cause the "interesting balance"\/ between the elastic and inelastic rates,  
could be quite interesting  for the direct DM search.
\par  
Therefore, one should plan experiments aimed at 
 direct {\em detection} of dark matter particles in a way, where one can measure two signals. 
The first is the nuclear recoil energy $T_A$ (traditional, elastic), 
and the second is the $\gamma$-quantum with the nuclear deexcitation
energy $\Delta\epsilon_{mn}$ accompanied by the (excited-nucleus) recoil energy $T_{A^*}$, 
which, as a rule,  is {\em higher than} the elastic $T_A$.
This two-fold experiment will give complete information
about the $\chi A$ interaction, if it occurs.
\par
Finally recall, that for reliable registration of dark matter particles from our Galaxy 
by means of direct detection one needs the positive signature 
 \cite{Bednyakov:2015uoa,Bednyakov:2020njj}, 
 which nowadays is only the annual modulation of the observed signal
 \cite{Freese:1987wu}.
In the context of this paper, both elastic (coherent) and inelastic (incoherent) contributions
to the modulation signal are very important
\cite{Sahu:2020kwh,Sahu:2020cfq}.
\par 
The author is grateful to D.V. Naumov, E.A. Yakushev, N.A. Russakovich and I.V. Titkova for 
fruitfull discussions and important remarks.
\enlargethispage{20pt}

\bibliographystyle{JHEP}
{\protect\small  \bibliography{COHERENT,BednyakovCoherence,CommonGeneral,DarkMatter} }
\end{document}